%% file: ms.tex
\documentclass[fleqn,usenatbib]{mnras}
\usepackage{graphicx}
\usepackage{amssymb}
\usepackage{amsmath}
\usepackage{ae,aecompl}
\usepackage{natbib,times}
\usepackage{lscape}

\usepackage{ulem}
\newcommand{\mc}{\multicolumn}

\begin{document}
\title[Cluster partners and mergers] {A catalogue of merging
  clusters of galaxies: cluster partners, merging subclusters, and
  post-collision clusters}

\author[Wen et al.]
{Z. L. Wen$^{1}$\thanks{E-mail: zhonglue@nao.cas.cn}, 
J. L. Han$^{1,2}$\thanks{E-mail: hjl@nao.cas.cn},
and Z. S. Yuan$^{1}$
\\
1. National Astronomical Observatories, Chinese Academy of Sciences, 
20A Datun Road, Chaoyang District, Beijing 100101, China\\
2. School of Astronomy, University of Chinese Academy of Sciences,
           Beijing 100049, China
}

\date{Accepted XXX. Received YYY; in original form ZZZ}

\label{firstpage}
\pagerange{\pageref{firstpage}--\pageref{lastpage}}
\maketitle


\begin{abstract}
Clusters of galaxies are merging during the formation of large-scale
structures in the Universe. Based on optical survey data, we identify
a large sample of pre-mergers of galaxy clusters and merging
subclusters in rich clusters.
We find 39\,382 partners within a velocity difference of 1500
km~s$^{-1}$ and a projected separation of 5\,$r_{500}$ around 33\,126
main clusters, where $r_{500}$ is the radius of the main cluster.
Based on the galaxy distribution inside rich clusters with more than
30 member galaxy candidates, we identify subclusters by modeling the
smoothed optical distribution with a two-component profile, and a
coupling factor is obtained for merging subclusters in 7845 clusters.
In addition, we find 3446 post-collision mergers according to the
deviations of brightest cluster galaxies from other member galaxies,
most of which have been partially validated by using the Chandra and
XMM-Newton X-ray images. Two new bullet-like clusters have been
identified by using the optical and X-ray images.
The large samples of merging clusters of galaxies presented here are
important databases for studying the hierarchical structure formation,
cluster evolution, and the physics of intergalactic medium.

\end{abstract}


\section{Introduction}

The hierarchical formation scenario of the large-scale structure
suggests that larger structures are formed through the continuous
merging of smaller structures \citep{swj+05}. As the largest
virialized systems in the Universe, clusters of galaxies were formed
at the epoch of $z\sim 2$ and grew through merging smaller clusters
\citep{kb12}. The merging process results in obvious substructures or
the non-spherical distribution of member galaxies and/or hot gas
inside a cluster of galaxies, which can be used to reveal the
dynamical state of clusters \citep{wh13} and test models for the
structure formation of the Universe \citep{wod88, jmb+95}.

The merging process of galaxy clusters usually experiences several
stages. Before the merger, two or more clusters have a large
separation with only very weak gravitational interaction. These
clusters may be located in a cosmological filament as suggested by the
distribution of galaxies or diffuse gas \citep{prp+08, zdm+13}, which
are good targets to reveal the missing baryon in the form of warm-hot
intergalactic medium \citep{wfk+08, tpc+16, mwr22}, and to study star
formation history of galaxies around clusters \citep{gwm+04, prp+08,
  pjl+24}.
As two clusters approach but are not connected at the early stage of
mergers, they have a stronger interaction with some gas partly
mixed. Merging clusters at this stage generally have not yet produced
shocks shown in X-ray images \citep[e.g.][]{bps+04,kng+15} except few
cases \citep[e.g.][]{ags+16,gas+19}.
Afterward, two clusters experience mergers with an enormous amount of
energy released. Strong shocks of intracluster hot gas are usually
generated, which can transfer the dynamical energy in the intracluster
medium (ICM) into non-thermal component \citep{vrb+10, vri+12, mmg+11,
  rsf+10}. Radio observations reveal radio relics which are produced
by particles accelerated to relativistic velocity in merging shock
fronts, or radio halo by re-accelerated particles in the turbulence in
the ICM \citep{fgg+12}. The ICM in some merging subclusters can be
stripped by the ram pressure. The extreme case of cluster merger,
e.g. 1E~0657$-$558 \citep[the Bullet cluster, ][]{mgd+02}, shows the
significant offset between the matter and the ICM, which provides
direct evidence of dark matter and constraint on the properties of
dark matter \citep{ mgc+04, cbg+06}.
After merging, the hot gas and member galaxies may experience long
post-merging phase and gradually slow down towards a final relaxed
state which is shown by a bright cool core in X-ray and a dominated
brightest cluster galaxy (BCG) in optical \citep{vmm+05,wh15a,ltl+18}.

Merging clusters are indicated by substructures inside a
cluster. About 40\% -- 70\% of clusters show an obvious signature of
recent mergers \citep{sbr+01, sks+05, wh13}. The dynamical states of
galaxy clusters are direct indications of merging processes inside
clusters and be quantified by the concentration index, the centroid
shift, the power ratio, and the morphology index in X-ray images
\citep{bt95,mef+95,srt+08,yh20}. In optical, it also can be diagnosed
by the 3-D distribution of member galaxies \citep{ds88,yds+18} and the
relaxation parameters based on the 2-D distribution of member galaxies
\citep{wh13}. The above measurements merely give the amount of the
substructure for clusters without the information of merging stages,
and hence are insufficient to describe the dynamical states of
clusters.

Cluster mergers at different stages have been explored previously. A
few tens of paired clusters and pre-merger systems have been
individually studied by optical and X-ray data \citep[e.g.][]{bps+04,
  sp04, planck13b, mcm+19}. Ongoing mergers have been identified for
more than one hundred clusters through X-ray images
\citep[e.g.][]{kd04, mgb+05, sp06, gvc+09, bgb+12} or observations of
radio diffuse halos and relics \citep[e.g.][]{fgg+12,
  vda+19}. \citet{me12} identified 10 clusters that likely have
undergone recent multiple merger events and also 11 systems that are
likely the post-collision of head-on mergers based on the projected
offset between the BCG and the peak of X-ray emission, and also a
visual assessment of the cluster morphology in optical and X-ray. By
using Sloan Digital Sky Survey (SDSS) spectroscopic data release 12
(DR12), \citet{ttk+17} found 498 potential partner systems for the
galaxy groups in the redshift range of $z<0.17$. Recently \citet{op24}
presented an improved search algorithm for interacting galaxy clusters
from the SDSS DR17 and published 160 merging systems and 21
pre-merging/post-merging systems at $z\le0.2$.

In recent years, a large number of galaxy clusters up to $z>1$ have
been identified from the optical survey data, e.g. SDSS
\citep[][]{kma+07b,whl09,whl12,wh15b,rrb+14,ogu14,bsp+18}, Dark
Energy Spectroscopic Instrument (DESI) Legacy Imaging Surveys
\citep{yxh+21,zsx+22,wh24} and Dark Energy Survey \citep[DES;][]{rrh+16,
  wh22}. These cluster samples, supplemented with spectroscopic or
photometric data of galaxies in optical surveys, provide an
unprecedentedly large database to reveal cluster mergers.

Optical imaging data can show cluster mergers projected on the sky
plane and trace the mass distribution inside merging clusters since
galaxies are essentially collisionless during mergers
\citep[e.g.][]{jsd+15, gdw+16, fjg+17}. In this paper, we identify a
large sample of merging clusters at different stages by using the
large optical cluster sample from the DESI Legacy Surveys. In Section
2, we describe the cluster sample and the member galaxies. In Section
3, we obtain a large sample of pre-merger cluster partner systems. In
section 4, we analyze the subclusters within the region of $r_{500}$
for rich clusters and quantify the merging process via a coupling
factor. In section 5, we identify post-collision mergers from rich
clusters according to the deviations of the BCGs from the distribution
of other member galaxies. A summary is presented in Section 6.

Throughout this paper, we assume a flat Lambda cold dark matter
($\Lambda$CDM) cosmology taking $H_0=70$ km~s$^{-1}$ Mpc$^{-1}$,
$\Omega_m=0.3$ and $\Omega_{\Lambda}=0.7$.

\begin{table*}
\setlength{\tabcolsep}{3pt}
\small 
\caption[]{The cluster partners (CP) for 33\,126 clusters}
\begin{center}
\begin{tabular}{lrccccccc}
\hline
\mc{1}{c}{Name of partners}
&\mc{1}{c}{Name of cluster}
&\mc{1}{c}{R.A.} & \mc{1}{c}{Dec.} & \mc{1}{c}{$z$} & \mc{1}{c}{$z_{\rm m,BCG}$} &
\mc{1}{c}{$r_{500}$}  & \mc{1}{c}{$M_{500}$} &\mc{1}{c}{$r_p$} \\
\mc{1}{c}{(1)} & \mc{1}{c}{(2)} & \mc{1}{c}{(3)} & \mc{1}{c}{(4)} & \mc{1}{c}{(5)} &
\mc{1}{c}{(6)} & \mc{1}{c}{(7)} & \mc{1}{c}{(8)} & \mc{1}{c}{(9)} \\
\hline
 WHY-CP00001:M & WH24-J000000.5$+$021911 & 0.00201 & 2.31980 & 0.4282& 18.043 & 0.612 & 1.07 & 0.000\\     
 WHY-CP00001:p1& WH24-J000008.5$+$022658 & 0.03525 & 2.44944 & 0.4279& 18.215 & 0.672 & 0.94 & 2.695\\[1mm]
 WHY-CP00002:M & WH24-J000002.3$+$051718 & 0.00959 & 5.28825 & 0.1694& 15.412 & 0.708 & 1.32 & 0.000\\     
 WHY-CP00002:p1& WH24-J000037.1$+$053308 & 0.15451 & 5.55236 & 0.1701& 16.228 & 0.643 & 0.89 & 3.132\\[1mm]
 WHY-CP00003:M & WH24-J000004.2$+$021941 & 0.01741 & 2.32799 & 0.6443& 18.314 & 0.627 & 1.32 & 0.000\\     
 WHY-CP00003:p1& WH24-J000011.9$+$021704 & 0.04961 & 2.28446 & 0.6439& 18.473 & 0.538 & 1.04 & 1.344\\[1mm]
 WHY-CP00004:M & WH24-J000009.8$+$081846 & 0.04067 & 8.31275 & 0.4862& 17.604 & 0.900 & 3.13 & 0.000\\     
 WHY-CP00004:p1& WH24-J000008.8$+$082858 & 0.03665 & 8.48269 & 0.4849& 18.019 & 0.497 & 0.73 & 3.677\\[1mm]
 WHY-CP00005:M & WH24-J000014.2$+$014424 & 0.05902 & 1.73990 & 0.7909& 19.603 & 0.585 & 1.04 & 0.000\\     
 WHY-CP00005:p1& WH24-J000031.3$+$014241 & 0.13023 & 1.71140 & 0.7933& 19.997 & 0.487 & 0.55 & 2.061\\     
  ... & ... & ...&  ... & ... &  ... & ... & ... & .   \\
\hline
\end{tabular}
\end{center}
{Note.
Column (1): name of cluster partners: `-M' stands for the main cluster, `-p1' or `-p2' and so on for partners; 
Column (2): name of cluster; 
Column (3): RA (J2000) of a cluster (degree); 
Column (4): Decl. (J2000) of a cluster (degree);
Column (5): redshift of cluster; 
Column (6): $z$-band magnitude of BCG; 
Column (7): cluster radius, $r_{500}$, in Mpc; 
Column (8): derived cluster mass, in $10^{14}~M_{\odot}$; 
Column (9): projected distance to the main cluster, in Mpc. 
\\
(This table is available in its entirety in a machine-readable form.)
}
\label{tab1}
\end{table*}

\begin{figure*}
\includegraphics[width = 0.3\textwidth]{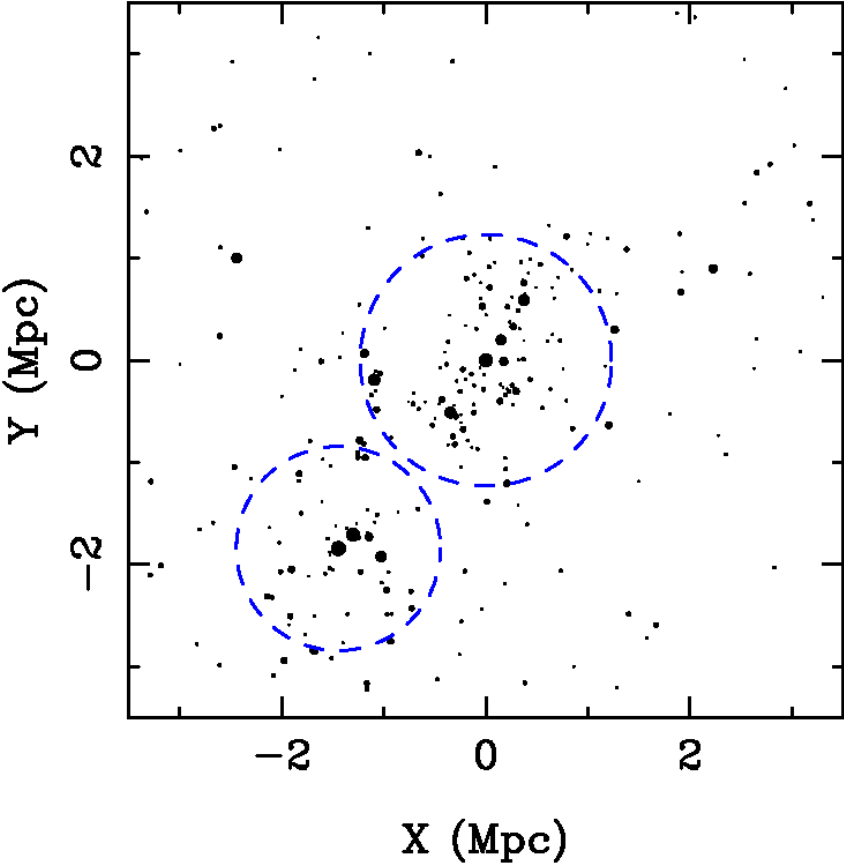}
\includegraphics[width = 0.3\textwidth]{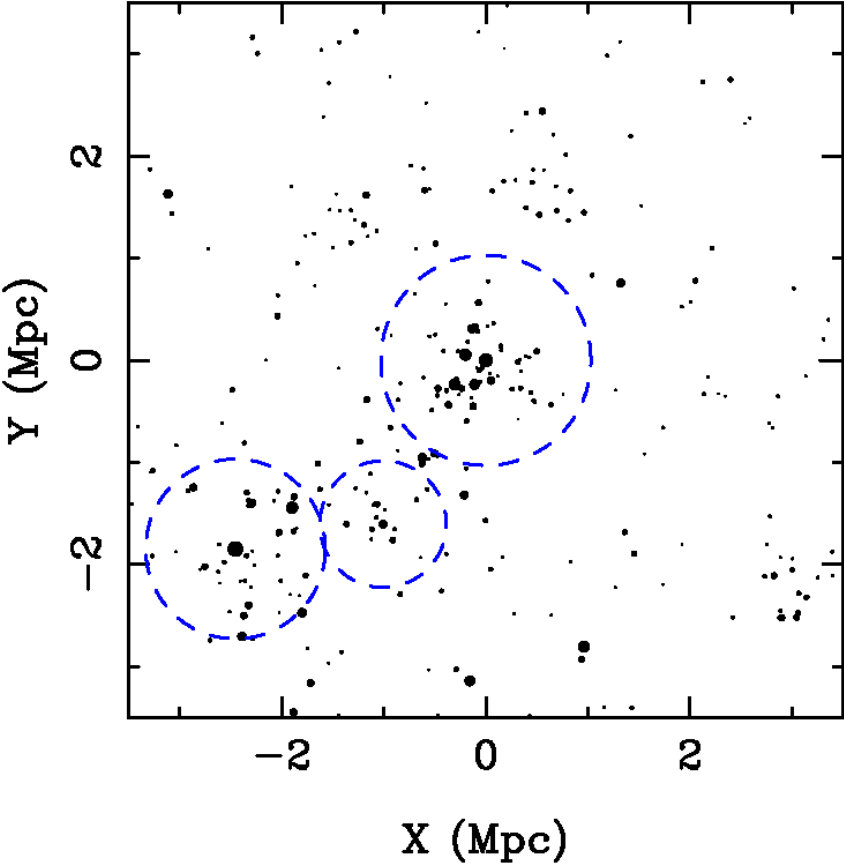}
\includegraphics[width = 0.3\textwidth]{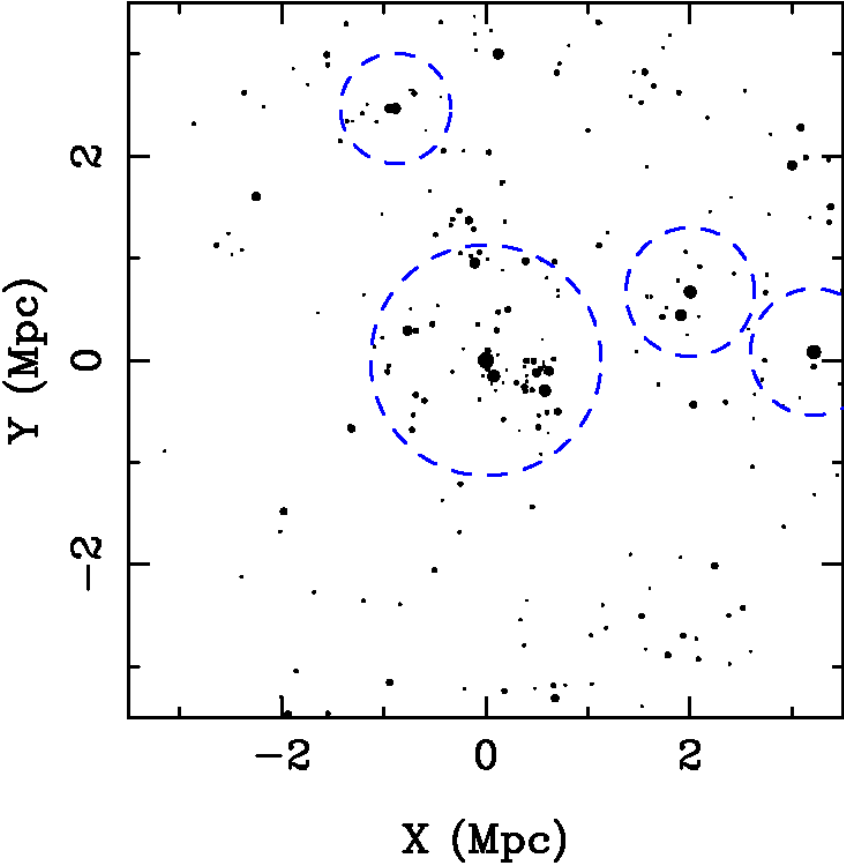}
\caption{Projected distribution of member galaxies (indicated by dots
  with a size proportional to the root square of stellar masses) on
  the sky plane for three examples of cluster partner systems, with
  one, two, and three partner clusters, respectively. The dashed
  circles indicate the region of $r_{500}$ for these clusters.}
\label{exam_comp}
\end{figure*}

\section{Galaxy clusters and member galaxies}

This study uses the catalogue of 1.58 million galaxy clusters in the
redshift range of $z<1.5$ identified by \citet{wh24} from the DESI
Legacy Imaging Surveys \citep{dsl+19} DR9 and new data from DR10,
covering a total sky area of $\sim$24,000 deg$^2$. Our algorithm first
selects a sample of massive BCG-like galaxies with a stellar mass of
$M_{\star}\ge10^{11} M_{\odot}$, a high optical/infrared luminosity
and a red colour defined by known BCGs. Then clusters are searched as
being the overdensity peaks of galaxy stellar masses in the redshift
slices around the massive BCG-like galaxies. The galaxies within a
given redshift slice and a given projected separation from the
BCG-like galaxies were taken as member galaxy candidates of clusters
to-be-identified. The cluster redshifts were then determined from the
photometric redshifts and available spectroscopic redshifts of member
galaxy candidates. The cluster center is defined as being the position
of the BCGs. The cluster radius ($r_{500}$, the radius within which
the mean density of a cluster is 500 times the critical density of the
universe) and mass ($M_{500}$, the cluster mass within $r_{500}$) were
estimated from the total stellar masses of member galaxy candidates
via the calibrated scaling relations. The clusters have a mass
$M_{500}$ in the range of $(0.47- 12)\times10^{14}~M_{\odot}$ with an
uncertainty of 0.2 dex. Among the 1.58 million clusters of galaxies,
338\,841 clusters have spectroscopic redshifts, and the others have
photometric redshifts with an uncertainty of about
0.01. Cross-matching with known clusters suggests that our cluster
catalogue has a high purity of $\sim$97\% at $z<0.9$ and contains
$\sim$90\% of X-ray clusters with
$M_{500}>0.5\times10^{14}~M_{\odot}$. See details in \citet{wh24}.

In this paper, we use the 338\,841 clusters that have spectroscopic
redshifts among the 1.58 million clusters for finding cluster partner
systems at the pre-merger stage. The spectroscopic redshifts are
mostly taken for the member galaxies from the 2MASS Redshift Survey at
low redshifts and from the SDSS up to the intermediate redshifts.

From 27\,685 rich clusters in the catalogue of 1.58 million clusters,
which have more than 30 member galaxy candidates within $r_{500}$, we
find merging subclusters and post-collision mergers. These clusters
have a mass $M_{500}$ in the range of
$(1.2-12)\times10^{14}~M_{\odot}$ with the median mass of
$2.87\times10^{14}~M_{\odot}$. For this work, only massive galaxies
with a stellar mass of $M_\star\ge 10^{10}~M_{\odot}$ within a
projected radius of 3 Mpc are taken as member galaxy (candidates), if
they have spectroscopic redshifts within a velocity difference of
$\Delta v<2500$ km~s$^{-1}$ from the cluster redshifts or in a
photometric redshift slice according to the uncertainty of their
photometric redshifts \citep{wh21}. The spectroscopic data from the
GAMA survey \citep{dbr+22} have been used to test the accuracy and
completeness of so-detected bright member galaxies. We found that 83\%
of the member galaxy candidates from photometric redshifts can be
verified well by spectroscopic redshifts, and only about 8\% of true
members are missing by our algorithm.

\begin{figure}
\centering
\includegraphics[width = 0.45\textwidth]{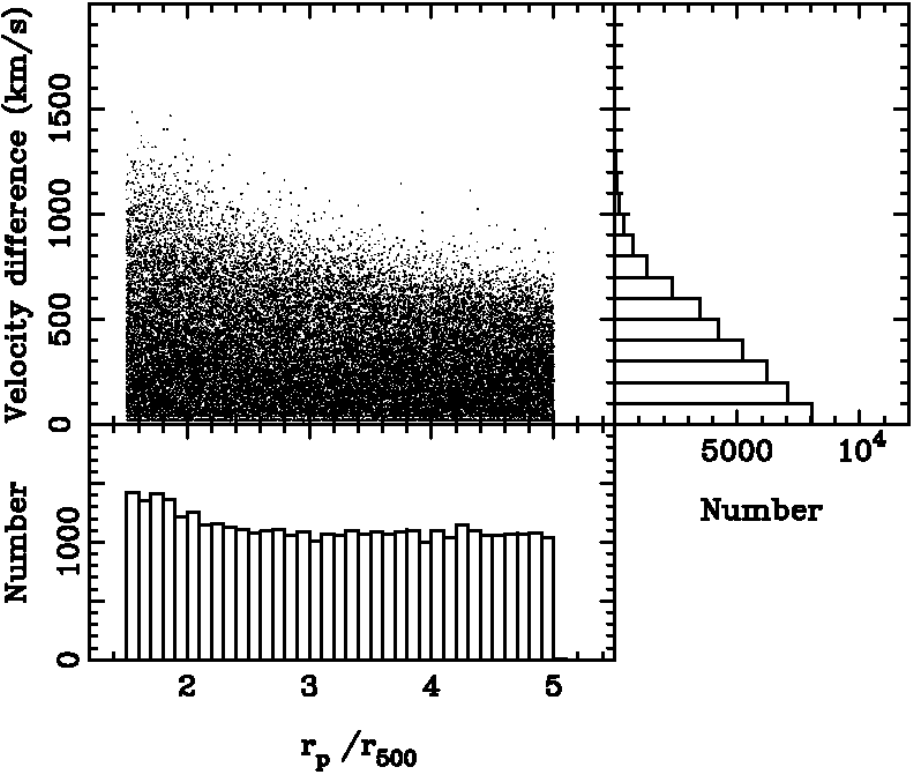}
\caption{The distributions of velocity difference and projected
  separation ($r_p/r_{500}$) of cluster partners from the main clusters.}
\label{dis_double}
\end{figure}

\begin{figure}
\centering
\includegraphics[width = 0.35\textwidth]{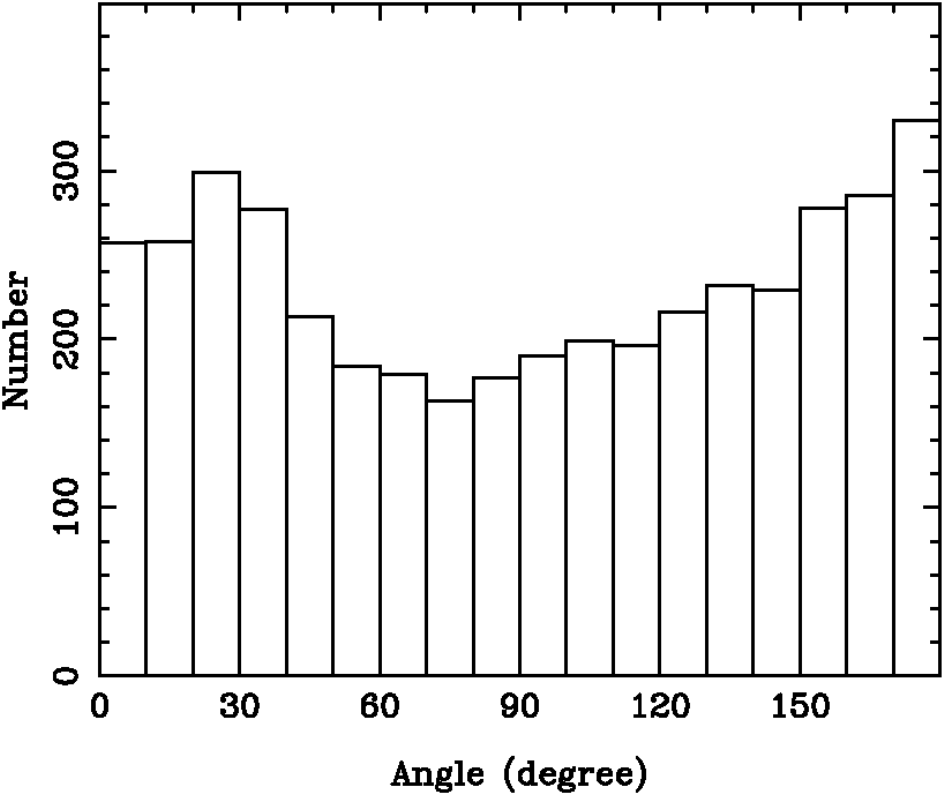}
\caption{Distribution of the angle between two partners to the main cluster.}
\label{orien}
\end{figure}

\begin{figure}
\centering
\includegraphics[width = 0.35\textwidth]{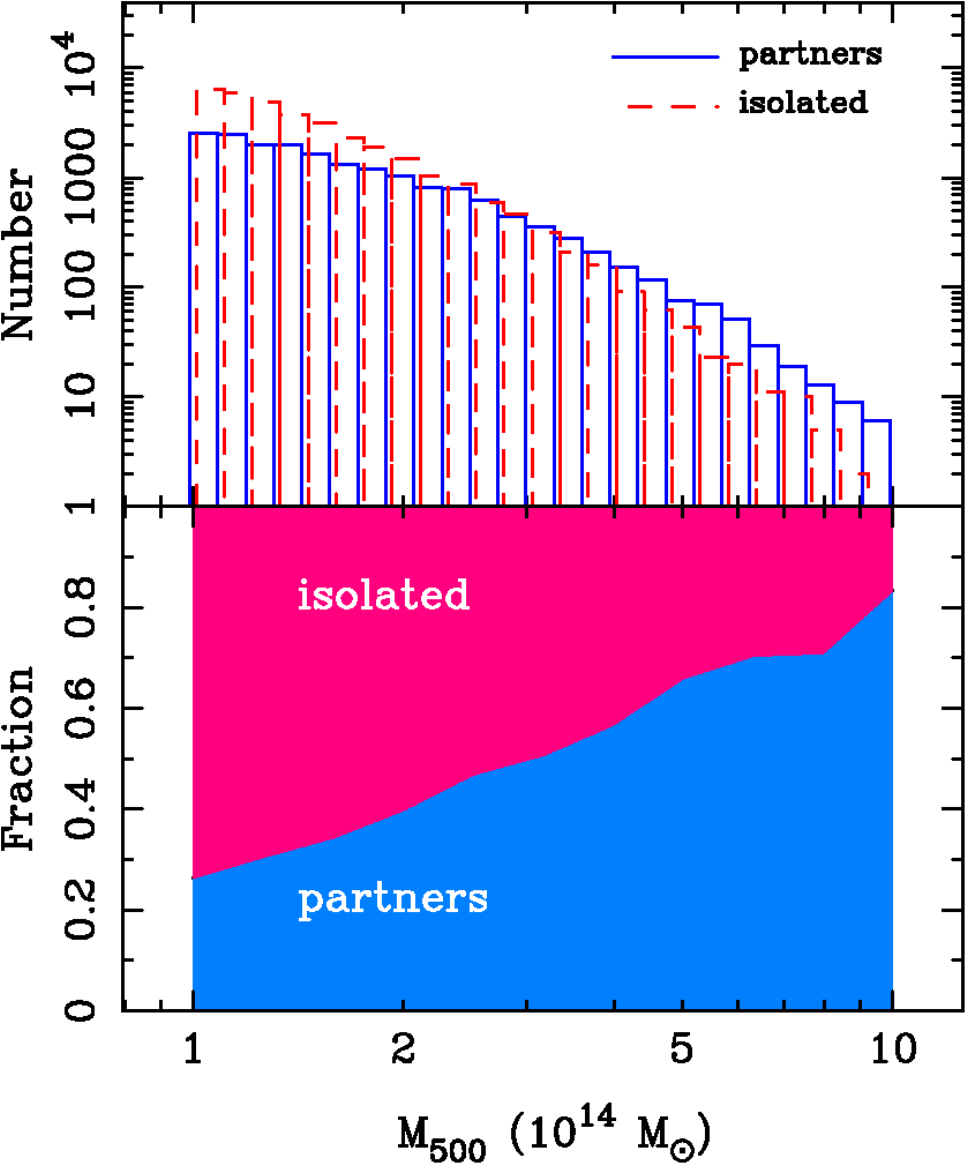}
\caption{Comparison of the mass distribution of clusters in the
  partner systems and that of isolated clusters ({\it upper panel}),
  and the fraction of two kinds of clusters in the mass range ({\it
    lower panel}).}
\label{dis_mass}
\end{figure}

\begin{figure*}
\includegraphics[width = 0.32\textwidth]{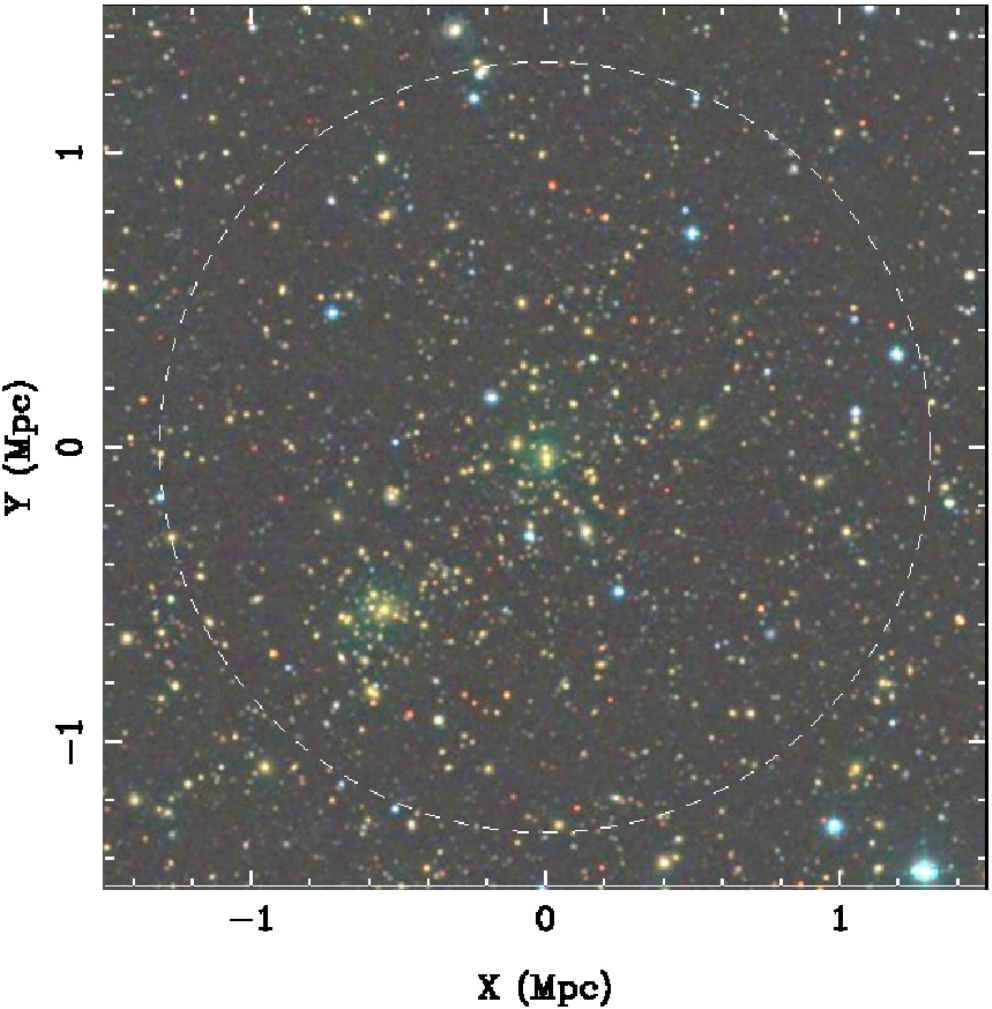}
\includegraphics[width = 0.32\textwidth]{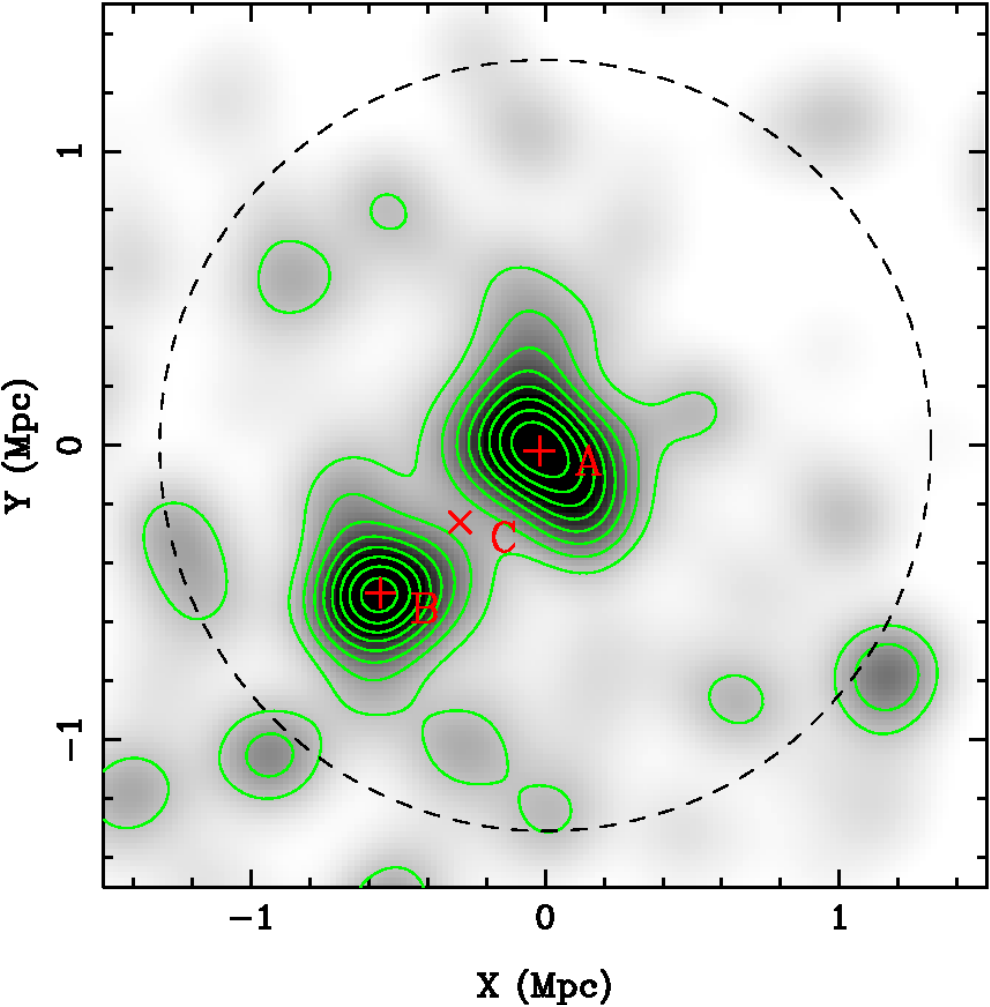}
\includegraphics[width = 0.32\textwidth]{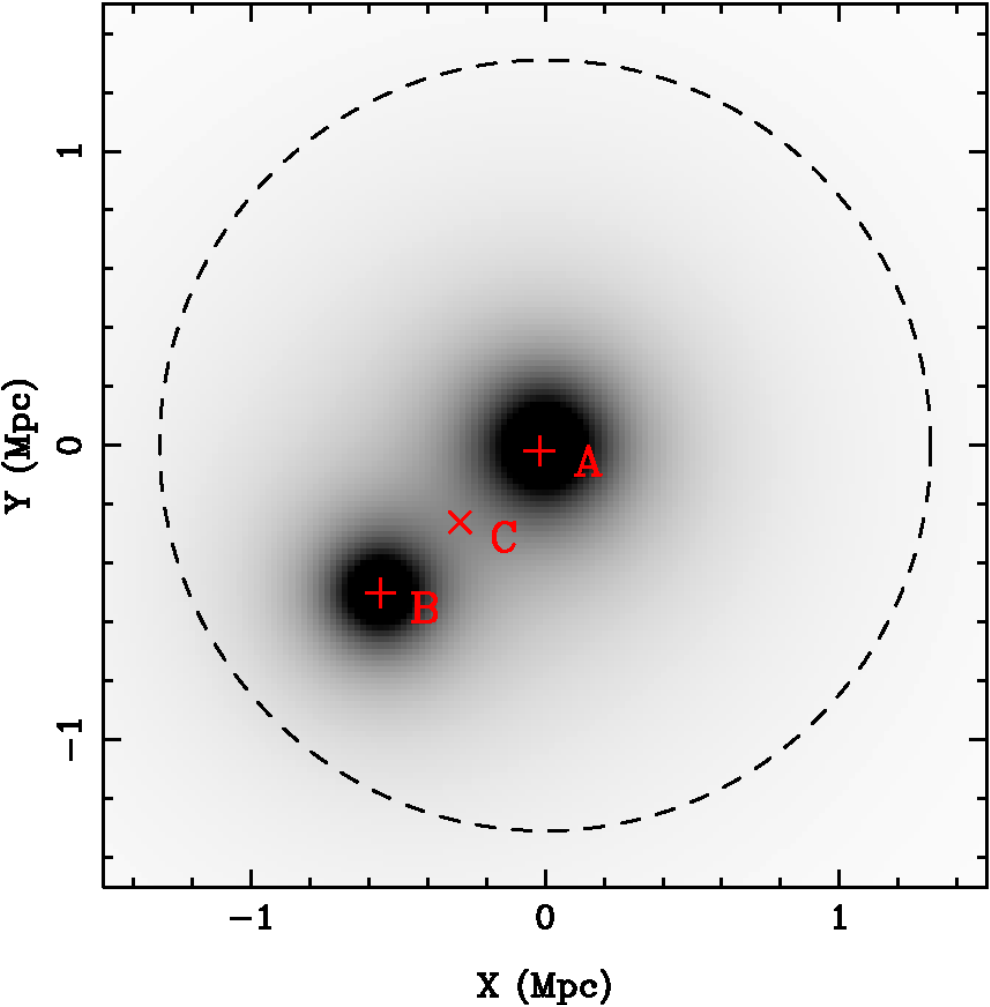}
\caption{An illustration for decoupling distinct subclusters. {\it The
    left panel} is the DESI colour image for A1758. {\it The middle
    panel} is the smoothed distribution of member galaxies, with the
  peaks for the central subcluster at 'A' in the center and the
  secondary subcluster at 'B', both marked by `+'. The minimum of the
  stellar mass distribution is located at at 'C' marked by $\times$
  between the two peaks.  The dashed circles indicate the radius of
  $r_{500}$. {\it The right panel} is for the best fit by the
  dual-King model.}
\label{example2}
\end{figure*}

\section{Cluster partners}

We first find cluster partner systems before merging, which contain
two or more nearby clusters that are probably gravitationally bound
together. The more massive one is called the main cluster, and the
other(s) with a less mass are partners. To avoid any
misidentifications, we find the cluster partner systems from the
338\,841 clusters with spectroscopic redshifts.

Clusters in the partnership are identified with criteria of a small
separation on the sky plane and a small velocity difference indicated
by spectroscopic redshifts. We set a maximum projected separation of
5\,$r_{500}$ of the main cluster, which is about the half turn-around
radius of a cluster \citep{rg89,rd06,hhl+20}. The velocity difference
between the two partners is temporarily set to be less than 1500
km~s$^{-1}$, and then we use the two-body model for gravitational
binding \citep{bgh82} to further constrain the velocity difference by,
\begin{equation}
v^2_rr_p\le 2GM\sin^2\alpha\cos\alpha,
\label{bind}
\end{equation}
where $v_r$ is the relative velocity along the line of sight, $r_p$ is
the projected separation, $M$ is the total virial mass of the system,
and $\alpha$ is the angle between the line connecting two clusters and
the plane of the sky. The right term of Eq.~\ref{bind} has a maximum
value of $4\sqrt{3}GM/9$, hence the partners with $v^2_rr_p >
4\sqrt{3}GM/9$ are discarded.

From the 338\,841 clusters, we find 39\,382 partners for 33\,126 main
clusters as listed in Table~\ref{tab1}, which satisfy the criteria for
the partnership. Among them, 28\,041, 4161, 737, 146, 29, 7, 4, and 1
main clusters have one, two, three, four, five, six, seven, and even
nine partners, respectively.
The projected distributions of member galaxies for three examples of
partnerships with one, two, and three partners, respectively, are
shown in Fig.~\ref{exam_comp}.
Not surprisingly, galaxies are concentrated inside each cluster or in
the region between clusters. The distributions for the projected
separation and the velocity difference of 39\,382 partners from the
main clusters are shown in Fig.~\ref{dis_double}, together with the
histogram number distributions. Under the gravitational binding
condition, most ($\sim$97\%) systems have a velocity difference of
less than 1000 km~s$^{-1}$, and a smaller velocity difference is
desired at a larger projected separation.

The sample of our identified partner systems includes the previously
well-known double clusters, such as A21 -- IVZw015
\citep{planck13b,pap+11}, A1095W -- A1095E \citep{gwt+16}, A1560A --
A1560B \citep{uc82}, 400dJ1359 -- ZwCl1358.1 \citep{planck13b,pap+11},
A1882A -- A1882B \citep{obb+13}, and A2029 -- A2033 \citep{gdo+18}.
However, some previously proposed double clusters, such as A2061 --
2067 \citep{pbb+14} and A1758N -- A1758S \citep{dk04}, cannot satisfy
the criterion for the velocity difference for the gravitation binding
(Eq.~\ref{bind}), and therefore are not listed in Table~\ref{tab1} as
being cluster partner systems.
Some close double clusters, e.g. A98N -- A98S \citep{prb+14} and
A1750N -- A1750C \citep{brb+16}, are not in our sample since they have
a projected separation smaller than $r_{500}$. They are considered as
merging subclusters inside massive clusters in this work (see
Section~\ref{mergers}).

\begin{table*}
\footnotesize
\setlength{\tabcolsep}{4pt}
\caption[]{Parameters of the subclusters in 7845 rich merging clusters.}
\begin{center}
\begin{tabular}{rrrccccrrrccc}
\hline
\mc{1}{c}{ID}&\mc{1}{c}{R.A.} & \mc{1}{c}{Dec.} & \mc{1}{c}{$z$} & \mc{1}{c}{flag$_{z}$} & \mc{1}{c}{$z_{\rm m,BCG}$} &
\mc{1}{c}{$r_{500}$}  & \mc{1}{c}{$M_{500}$} & \mc{1}{c}{R.A.$_2$} & \mc{1}{c}{Dec.$_2$} &
\mc{1}{c}{$r_p$} & \mc{1}{c}{$\mu$ 
} & \mc{1}{c}{$\gamma$}  \\
\mc{1}{c}{(1)} & \mc{1}{c}{(2)} & \mc{1}{c}{(3)} & \mc{1}{c}{(4)} & \mc{1}{c}{(5)} & 
\mc{1}{c}{(6)} & \mc{1}{c}{(7)} & \mc{1}{c}{(8)} & \mc{1}{c}{(9)} & \mc{1}{c}{(10)} & 
\mc{1}{c}{(11)} & \mc{1}{c}{(12)} & \mc{1}{c}{(13)} \\
\hline
WH24-J000007.6$+$155003 & 0.03175 & $ 15.83424$ & 0.1528 & 1& 14.918& 0.892 &  2.25 & 0.02523 & $ 15.85307$ & 0.19 &  1.79 &  0.71 \\
WH24-J000017.5$-$090235 & 0.07307 & $ -9.04308$ & 0.6564 & 1& 19.245& 0.717 &  1.94 & 0.06336 & $ -9.04548$ & 0.25 &  0.58 &  0.39 \\
WH24-J000026.3$+$215405 & 0.10957 & $ 21.90142$ & 0.1790 & 0& 15.418& 1.027 &  3.44 & 0.14325 & $ 21.95473$ & 0.67 &  1.66 &  0.02 \\
WH24-J000037.8$-$681727 & 0.15750 & $-68.29070$ & 0.6070 & 0& 18.379& 0.996 &  5.02 & 0.09934 & $-68.31801$ & 0.84 &  1.63 &  0.17 \\
WH24-J000044.0$-$085223 & 0.18353 & $ -8.87319$ & 0.8185 & 0& 18.923& 0.807 &  3.12 & 0.18279 & $ -8.88496$ & 0.32 &  1.08 &  0.31 \\
WH24-J000045.5$-$460728 & 0.18979 & $-46.12441$ & 0.8455 & 0& 19.979& 0.701 &  2.26 & 0.19294 & $-46.11713$ & 0.21 &  1.37 &  0.52 \\
WH24-J000102.7$-$051311 & 0.26138 & $ -5.21968$ & 0.7013 & 0& 19.154& 0.752 &  2.41 & 0.26529 & $ -5.20723$ & 0.34 &  1.19 &  0.09 \\
WH24-J000113.8$-$332203 & 0.30742 & $-33.36749$ & 0.2828 & 0& 17.142& 0.902 &  2.63 & 0.32454 & $-33.31289$ & 0.87 &  1.84 &  0.00 \\
WH24-J000117.2$-$031648 & 0.32184 & $ -3.28000$ & 0.2932 & 1& 16.448& 1.009 &  3.62 & 0.35485 & $ -3.29901$ & 0.60 &  1.11 &  0.11 \\
WH24-J000117.5$-$051944 & 0.32308 & $ -5.32900$ & 0.2633 & 1& 16.727& 0.884 &  2.47 & 0.32858 & $ -5.34952$ & 0.31 &  0.93 &  0.28 \\
\hline
\end{tabular}
\end{center}
{Note.
Column (1): Name of merging cluster;
Column (2): RA (J2000) of BCG (degree);
Column (3): Decl. (J2000) of BCG (degree);
Column (4): redshift of cluster;
Column (5): flag for cluster redshift. '1' indicates spectroscopic redshift and '0'
indicates photometric redshift;
Column (6): $z$-band magnitude of the BCG;
Column (7): cluster radius, $r_{500}$, in Mpc;
Column (8): derived mass in $10^{14}~M_{\odot}$; 
Column (9)-(10): RA (J2000) and Decl. (2000) of the secondary subcluster;
Column (11): projected distance between the central subcluster and the secondary subcluster, in Mpc;
Column (12): mass ratio between the central subcluster and the secondary subcluster;
Column (13): coupling factor between the central cluster and the secondary subcluster.
\\
(This table is available in its entirety in a machine-readable form.)
}
\label{tab2}
\end{table*}

Clusters of galaxies are preferably located at the knots of
filamentary large-scale structures in the Universe, and are preferably
aligned together \citep{acc+06,smb+12}. Here we investigate the
alignment of the partner positions to the main clusters using the 4161
clusters with two partners, and find the preferential alignment. The
distribution of the angles of two partners to the main cluster is
shown in Fig.~\ref{orien}, showing peaks at about 20 -- 30 degrees for
similar directions of two partners and at 170 -- 180 degrees for the
opposite. The partners at around 70-80 degrees are 30\% below the
average.

We noticed that more massive clusters have a higher fraction with
partners (see Fig.~\ref{dis_mass}). Considering the completeness of
cluster spectroscopic redshifts, for statistics, we use only these
clusters in the region of SDSS Baryon Oscillation Spectroscopic Survey
with a redshift of $z<0.5$ and a mass $M_{500}\ge 1\times10^{14}
~M_{\odot}$ for this statistics because about 90\% of such clusters in
the catalogue of 1.58 million clusters have spectroscopic
redshifts. Among the 338\,841 clusters with spectroscopic redshifts in
this specific region, we get 20\,145 clusters in the partner systems
and 40\,038 isolated clusters. It is understandable that in the
hierarchical cluster formation scenario massive clusters are
preferably formed in a denser environment and are experiencing more
merging processes than isolated clusters.

\section{Merging subclusters inside massive clusters}
\label{mergers}

Clusters of galaxies in merging processing usually contain distinct
subclusters, which are not dynamically relaxed. Here we work on
27\,685 rich clusters with more than 30 member galaxies, decompose the
subclusters, and finally obtain a coupling factor for overlapping
subclusters in 7845 rich clusters.

\subsection{Subcluster decomposition based on optical data}

To reveal the mass distribution of subclusters inside a massive
cluster, weak lensing analysis is a very reliable approach but can be
done only for a small number of massive clusters
\citep[e.g.][]{jsd+15,jds+16,fjg+17,kjh+21}. The ICM in subclusters
can be stripped and shocked in merging processing. Hence the X-ray
images can show substructures directly but are not accurate for the
mass distribution of merging clusters. The member galaxies, however,
can be regarded as collisionless objects, and their distribution can
be an excellent indicator for subclusters and the mass distribution
projected on the sky plane.

Because member galaxies are discretely distributed, to reduce the
Poisson uncertainty, we smooth the optical images by putting the
member galaxies with a Gaussian kernel weighted by the stellar mass of
each galaxy following \citet{wh13}, as shown in
Fig.~\ref{example2}. For example, the cluster A1758 shows two distinct
merging subclusters from the distribution of galaxies in the DESI
colour image\footnote{https://www.legacysurvey.org/viewer}. The
positions (RA, Dec.) of all member galaxies are transformed into the
Cartesian coordinates ($x$, $y$) centering around the BCG. To get a
smoothed optical map, the surrounding area of 4~Mpc $\times$ 4~Mpc is
divided into 200 pixels $\times$ 200 pixels. The stellar mass value of
each pixel ($x_i$,$y_j$) can then be calculated by adding the weighted
Gaussian kernels for all member galaxies:
\begin{equation}
I(x_i,x_j)=\sum_{k=1}^{N_{\rm gal}}M_{\star,k}g(x_i-x_k,y_j-y_k,\sigma_k),
\label{smoothmap}
\end{equation}
where $M_{\star,k}$ is the stellar mass of the $k$th member galaxy,
$x_k$ and $y_k$ are the coordinates of the $k$th member galaxy,
$N_{\rm gal}$ is the total number of member galaxies within the region
of 4~Mpc $\times$~4 Mpc and $g(x,y,\sigma)$ is a two-dimensional
Gaussian function with a smooth scale of $\sigma$,
\begin{equation}
g(x,y,\sigma)=\frac{1}{2\pi \sigma^2}\exp\Big(-\frac{x^2+y^2}{2\sigma^2}\Big).
\label{gauss}
\end{equation}
The structures shown on the smoothed optical map depend on the
smoothing scale of the Gaussian function. The smooth scale should be
adopted to match the diffuse distributions between member galaxies and
the underlying matter in clusters. A smaller scale leads to more
discrete false subclusters, but a larger scale could smear out more
real subclusters. \citet{djs+15} carried out the Merging Cluster
Collaboration (MC$^2$) project and performed a kernel density
estimation (KDE) to model the light distribution of ongoing merging
clusters. A smooth scale is obtained by maximizing the likelihood of
fit between the KDE model and the data, which is in the range of
0.07--0.15\,$r_{500}$ for several massive clusters \citep{jsd+15,
  jds+16, gdw+16, gvd+17, fjg+17}. Here we adopt a smooth scale of
$0.08\,r_{500}$ to construct the smoothed optical map (see
Appendix~\ref{scale}). The background and fluctuations are obtained
from the region outside of 1.5\,$r_{500}$.

\begin{figure}
\centering
\includegraphics[width = 0.35\textwidth]{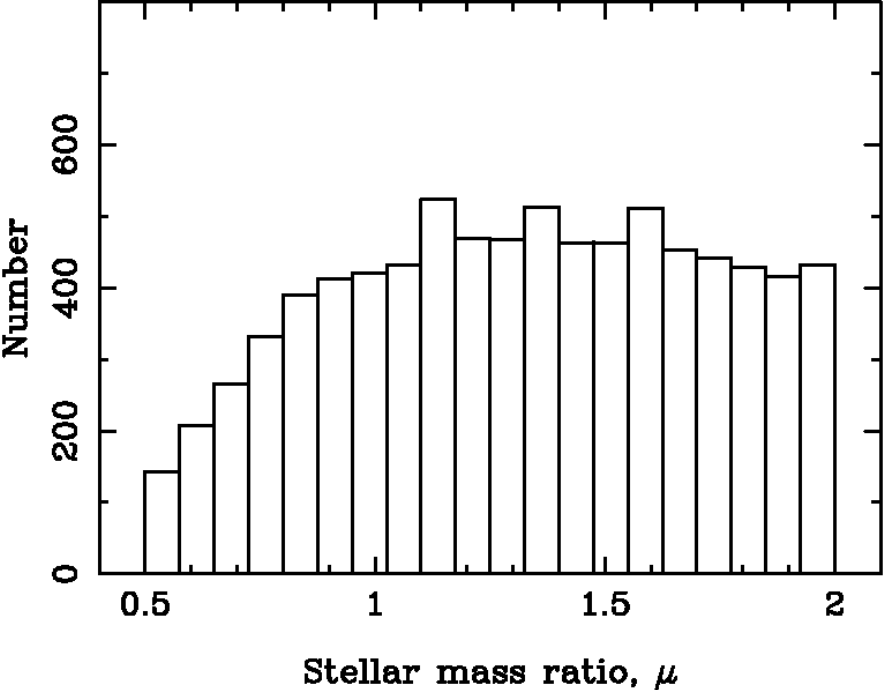} \\[2mm]
\includegraphics[width = 0.35\textwidth]{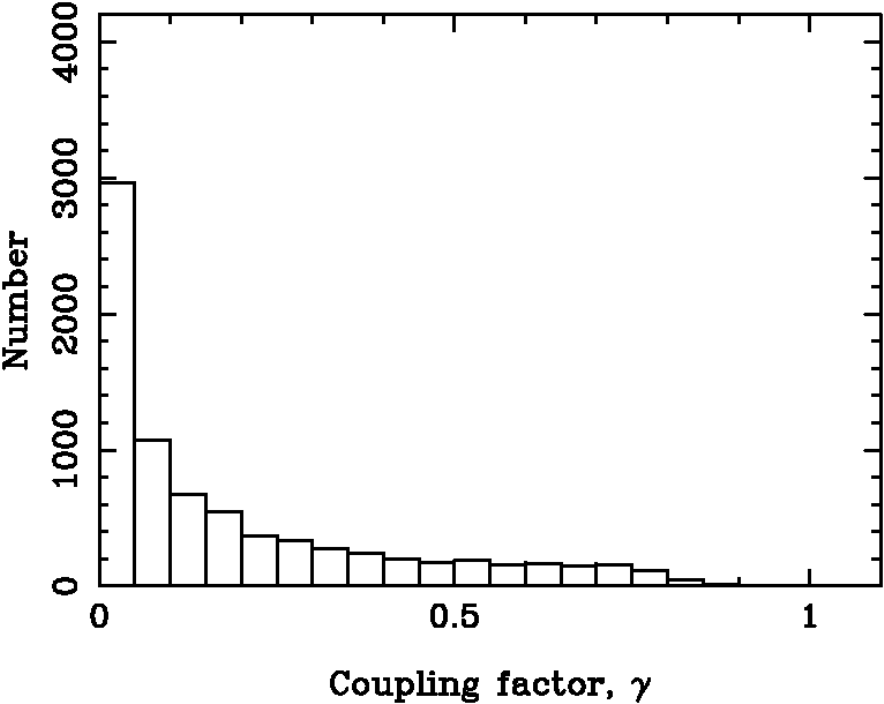}
\caption{Distribution of the stellar mass ratio $\mu$ ({\it upper
    panel}) and the coupling factor $\gamma$ ({\it lower panel}) of
  two merging subclusters in 7845 rich clusters.}
\label{dist_mu_gamma}
\end{figure}

We then decompose the smoothed optical map to recognize
subclusters. From the limited number of member galaxies, the smoothed
map always contains many clumps. We work on the two dominant
subclusters from the highest peaks on the smoothed map within
$r_{500}$ and with a signal-to-noise greater than 10. The central
subcluster contains the BCG, and the secondary is located aside, see
Fig.~\ref{example2} for example. Afterward, the smoothed map is fitted
to a dual-King model:
\begin{eqnarray}
  I_{\rm model}(x,y)&=&\frac{I_1}{1+(x^2+y^2)/r_1^2}+\nonumber\\
  &&\frac{I_2}{1+[(x-x_2)^2+(y-y_2)^2]/r_2^2}.
\end{eqnarray}
The first term is for the central subcluster and the second term is
for the secondary. $I_1$ and $I_2$ are the stellar mass values at the
centers of two components, $r_1$ and $r_2$ are their core radii,
respectively, and $(x_2,y_2)$ is the relative position of the
secondary subcluster to the central one.
The total stellar mass of each subcluster is then obtained by the
best-fitted King model, denoted as $M^{\star}_{\rm central}$ for the
central subcluster, and $M^{\star}_{\rm secondary}$ for the
secondary. In most clusters, the central subcluster is more
massive. Nevertheless, in some cases the secondary subcluster could be
more massive.
The mass ratio between two subclusters is then calculated as being
\begin{equation}
\mu=M^{\star}_{\rm central}/M^{\star}_{\rm secondary}.
\label{massratio}
\end{equation}
If the central subcluster is more massive, the mass ratio is
$\mu>1$. Otherwise, $\mu<1$.  A cluster showing two distinct
subclusters with a comparable stellar mass should have a mass ratio
near 1.0. The mass ratio for our selected rich clusters has a
distribution shown in Fig.~\ref{dist_mu_gamma}. We limit the merging
clusters to those showing subclusters with a mass ratio of $0.5\le \mu
\le 2$ for reliability of subcluster identification, so we have 7845
clusters listed in Table~\ref{tab2}.

\subsection{The coupling factor $\gamma$ between subclusters}

Subclusters are merging at various stages. At the early stage of a
merger, they are separated. As the two subclusters approach and merge,
they partly overlap together more and more, so that more galaxies
between the subclusters could be mixed in the superposition area
between them, Here we define a coupling factor for the two subclusters
to show the merging stage based on the smoothed optical map. As shown
in the middle panel of Fig.~\ref{example2}, we first find the minimum
stellar mass $I(x_C, y_C)$ on the connection line between two peaks,
and then the coupling factor is defined by
\begin{equation}
\gamma=\frac{2\,I(x_C, y_C)} {I(x_A,y_A)+I(x_B,y_B)},
\end{equation}
here $I(x_A,y_A)$ and $I(x_B,y_B)$ are the peak values of two
subclusters, respectively. A larger value of $\gamma$ indicates a more
significant overlapping between the central and the secondary
subclusters. The coupling factors for 7845 massive clusters are listed
in Table~\ref{tab2}, and the distribution of $\gamma$ is shown in
Fig.~\ref{dist_mu_gamma}.

\begin{figure*}
\centering
\includegraphics[width = 0.45\textwidth]{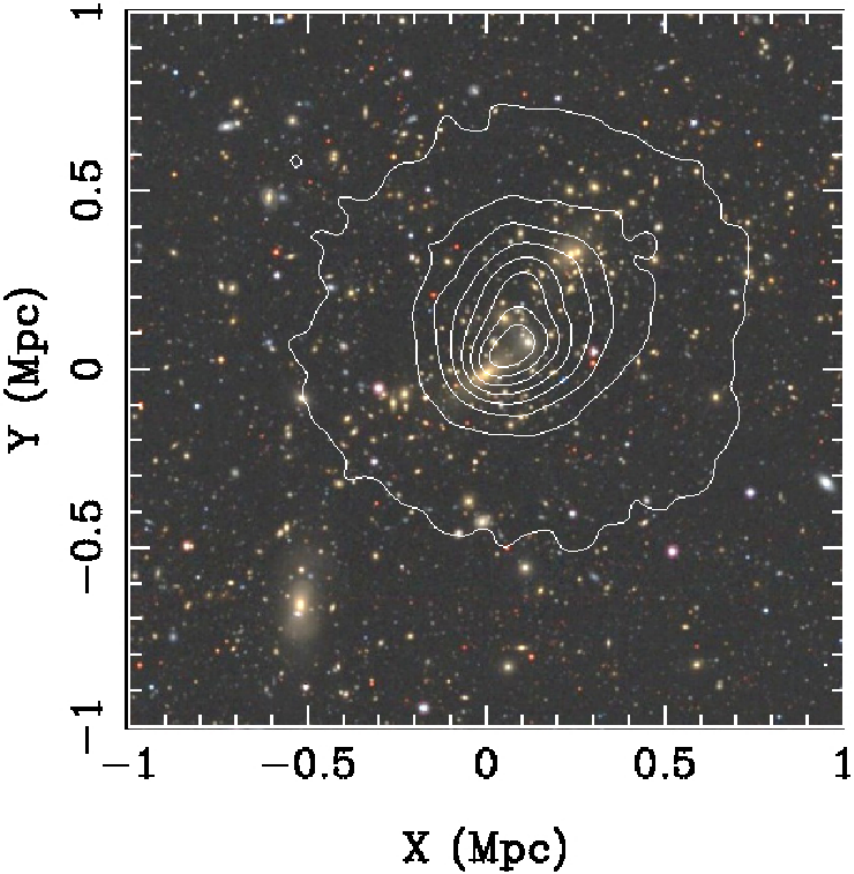}
\includegraphics[width = 0.45\textwidth]{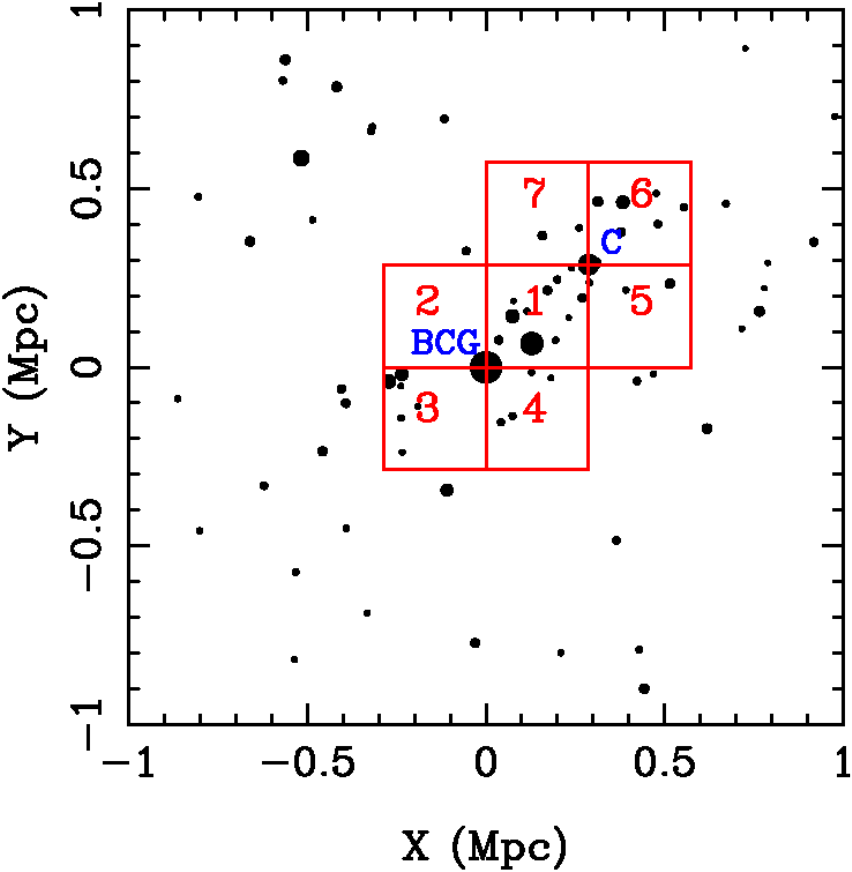}
\caption{An illustration of identification for post-collision mergers.
  {\it The left panel} shows the DESI colour image of an example
  cluster, A1553, overlaid with the contour for the X-ray emission
  observed from {\it XMM-Newton}.  {\it The right panel} shows the
  rotated projected distribution of member galaxies with `C' galaxy in
  the northwest direction. The 7 subregions around the BCG and the 'C'
  galaxy are chosen according to their positions, and have been used
  for mass calculations and the overdensity factor estimation. The
  figures have a size of 2~Mpc$\times$2~Mpc.}
\label{postmerger}
\end{figure*}

\subsection{Different merging signatures shown by galaxy distribution and hot gas}

Both the member galaxies and hot intracluster gas can trace the
gravitational potential inside clusters. However, member galaxies
often have a different distribution from that of the hot gas
\citep{yhb+23}.  We here compare the subclusters seen from our optical
analysis with those shown in the X-ray images obtained by \citet{yh20}
and \cite{yhw22} from observations by the {\it Chandra} \citep{wtv+00}
and {\it XMM-Newton} \citep{sbd+01}.  The X-ray images have been
smoothed by a Gaussian function with a scale of 30 kpc. We cross-match
the 7845 rich clusters are listed in Table~\ref{tab2} with the X-ray
cluster sample, and get 153 X-ray clusters, 35 with Chandra images and
118 with XMM-Newton images. In Fig.~\ref{coupling} in the Appendix, we
show the smoothed optical map of member galaxies overlaid with X-ray
contours for these clusters sorted by the coupling factor.

For clusters with a small coupling factor of $\gamma<0.1$, the
identified subclusters have a large separation, e.g. A1750, A689, RXC
J1003.0+3254, A750, A2933, A2051, A1035, A2028, A2507, A98 and
MCXC J1419.8+0634, which show double X-ray components from each optical
subcluster, without interacting features between subclusters. They
should be at the early stage of merging, as claimed for the clusters,
e.g. A1750 \citep{bps+04} and A98 \citep{prb+14}.

Subclusters overlap more significantly with a larger coupling factor
for a deeper merging, showing more features from the mixed ICM in
X-ray images.  During subcluster merging, the intracluster gas can be
stripped from one subcluster. Most clusters with $\gamma>0.4$ have
only one main disturbed X-ray component, indicating that they are at
the later merging stage. Subcluster collisions may cause the
displacements between the member galaxies (and underlying dark matter)
and gas \citep{iws+15, obm+11, rms+12, omg04, zss+21}.
The comparison between optical distribution for subclusters and X-ray
images for intracluster gas shows that the X-ray emission of one
subcluster is sometimes completely absent or very weak compared to the
secondary in the clusters, e.g. A3771, ACT J0102-4915,
SPTCL J0354-5904, MCXC J1311.7+2201, MCXC J0336.8-2804, A3921, APMCC772,
A2069, ZwCl1742.1+3306, MS2215.7-0404, A665, A1443, A1240, A2146,
A961, A3984, A85, MCXC J0510.7-0801, MACSJ0417.5-1154, A2163, A223, and
A1201, indicating ongoing mergers responsible for the gas stripping,
with obviously disturbed X-ray morphologies.

The most interesting among them is A223, very special for the weaker
X-ray emission from the more massive central subcluster
\citep{dsc+05}. Similar cases we find are A523, A2069, MS2215.7-0404,
A3984, PSZ2G225.93-19.99, PSZ2G239.27-26.01, A141 and A2163.

Some clusters, e.g. A168, A2255, A2443, PSZ1G295.60-51.95 and
PSZ2G314.26-55.35, show an elongated X-ray morphology or a stripped
gas tail, indicating that they are probably at the stage of
post-collision on the sky plane, as we will discuss below in details.

\begin{table*}
\caption[]{3446 clusters with post-collision merging features.}
\begin{center}
\begin{tabular}{rrrrccccrrrrc}
\hline
\mc{1}{c}{Name}&\mc{1}{c}{R.A.} & \mc{1}{c}{Dec.} & \mc{1}{c}{$z$} & \mc{1}{c}{flag$_{z}$}& \mc{1}{c}{$z_{\rm m,BCG}$} &
\mc{1}{c}{$r_{500}$}  & \mc{1}{c}{$M_{500}$} & \mc{1}{c}{R.A.$_{\rm C}$} & \mc{1}{c}{Dec.$_{\rm C}$} &
\mc{1}{c}{$\delta_1$} & \mc{1}{c}{$\delta_2$} & \mc{1}{c}{Note}  \\
\mc{1}{c}{(1)} & \mc{1}{c}{(2)} & \mc{1}{c}{(3)} & \mc{1}{c}{(4)} & \mc{1}{c}{(5)} & 
\mc{1}{c}{(6)} & \mc{1}{c}{(7)} & \mc{1}{c}{(8)} & \mc{1}{c}{(9)} & \mc{1}{c}{(10)} & 
\mc{1}{c}{(11)} & \mc{1}{c}{(12)} & \mc{1}{c}{(13)} \\
\hline
WH24-J000003.1$-$033245 & 0.01275 & $ -3.54574$ & 0.5973 & 1& 18.010& 0.851 &  3.20 & 0.03489 & $ -3.56529$ &  4.6 &  6.2 & 0\\
WH24-J000007.6$+$155003 & 0.03175 & $ 15.83424$ & 0.1528 & 1& 14.918& 0.892 &  2.25 & 0.03161 & $ 15.90206$ &  7.7 & 25.9 & 0\\
WH24-J000023.4$-$280612 & 0.09770 & $-28.10340$ & 0.2822 & 0& 15.744& 1.039 &  3.92 & 0.14379 & $-28.14625$ &  2.7 &  3.0 & 0\\
WH24-J000037.8$-$681727 & 0.15750 & $-68.29070$ & 0.6070 & 0& 18.379& 0.996 &  5.02 & 0.09718 & $-68.31879$ &  3.1 &  7.4 & 0\\
WH24-J000055.7$-$071235 & 0.23225 & $ -7.20973$ & 0.5515 & 1& 18.338& 0.811 &  2.29 & 0.22204 & $ -7.18830$ &  4.5 & 15.1 & 0\\
WH24-J000113.8$-$332203 & 0.30742 & $-33.36749$ & 0.2828 & 0& 17.142& 0.902 &  2.63 & 0.25904 & $-33.37627$ & 10.2 &  3.2 & 0\\
WH24-J000117.2$-$031648 & 0.32184 & $ -3.28000$ & 0.2932 & 1& 16.448& 1.009 &  3.62 & 0.36384 & $ -3.29274$ &  4.9 &  3.3 & 0\\
WH24-J000126.3$-$000143 & 0.35969 & $ -0.02863$ & 0.2490 & 1& 15.798& 1.073 &  4.47 & 0.29256 & $ -0.04964$ &  2.9 &  9.9 & 0\\
WH24-J000156.4$-$560937 & 0.48500 & $-56.16028$ & 0.3041 & 0& 16.366& 1.077 &  4.69 & 0.49819 & $-56.12814$ &  3.0 &  7.4 & 0\\
WH24-J000212.4$-$032509 & 0.55172 & $ -3.41914$ & 0.7247 & 1& 19.115& 0.764 &  2.54 & 0.55274 & $ -3.40491$ &  6.5 &  8.4 & 0\\
\hline
\end{tabular}
\end{center}
{Note.
Column (1): Cluster name;
Column (2): RA (J2000) of BCG (degree);
Column (3): Decl. (J2000) of BCG (degree);
Column (4): redshift of the cluster;
Column (5): flag for cluster redshift. '1' indicates spectroscopic redshift and '0'
indicates photometric redshift;
Column (6): $z$-band magnitude of the BCG;
Column (7): cluster radius, $r_{500}$, in Mpc;
Column (8): derived mass, in $10^{14}~M_{\odot}$;
Column (9)-(10): RA and Decl. of the `C' galaxy as being the subcluster BCG;
Column (11): overdensity factor $\delta_1$ around the cluster BCG;
Column (12): overdensity factor $\delta_2$ around the `C' galaxy;
Column (13): It is '1' if the post-collision merger has X-ray data from {\it{Chandra} and} {\it XMM-Newton}, otherwise '0'.
\\
(This table is available in its entirety in a machine-readable form.)
}
\label{tab3}
\end{table*}

\section{Post-collision mergers of clusters}

Cluster merging at the post-collision stage is most intriguing for
many aspects, e.g. dark matter distribution, electron acceleration,
synchrotron radiation and the ICM heating \citep{mgc+04,fgg+12}. Some
post-collision clusters have been recognized previously from X-ray
images based on highly elongated features, the shock front with a
sharp edge of X-ray surface brightness, and the stripped gas tails
\citep{mv07, rsf+10, mhs+12, jsd+15}. Here we identify post-collision
merging clusters based on the two-dimensional distribution of member
galaxies on the sky plane.

\subsection{Identification of post-collision mergers}

When two clusters collide, the mass distributions of the two clusters
are reshaped. The member galaxies can be stripped out from the
original clusters under gravitational force and probably also have an
elongated spatial distribution in a new merging cluster. The BCG
location should also deviate from the number density peak of the
member galaxies \citep{bcg+06, gvd+17, jhm+14, mcf+08}, and the member
galaxies of two merging subclusters are not a simple superposition of
two groups of galaxies. After the collision of two clusters, galaxies
should have an excessive density in the region between the BCGs of two
original clusters. We therefore identify the post-collision merging
cluster from the above mentioned 27\,685 rich clusters by finding the
excessive density via the following steps (see Fig.~\ref{postmerger}):

1. Find the potential BCGs of two original clusters which are the
colliding subclusters inside a rich cluster. We try all BCG-like
galaxies \citep[see][for details]{wh24} within $r_{500}$ as being the potential BCGs of
the colliding subclusters. The BCG-like galaxies are selected to have
a stellar mass of $M_{\star}\ge10^{11} M_{\odot}$, a high
optical/infrared luminosity and a red colour which are defined by
known BCGs from Abell clusters \citep{lps+14}, WHL and WH22 clusters
\citep{whl12,wh15b,wh22}. We mark the potential BCGs of the colliding
subclusters by 'C', and do the following tests. If a rich cluster does
not have such a 'C' galaxy in addition to the recognized central BCG,
it is skipped for the identification of post-collision mergers.

2. Divide the sky region near the BCG and the 'C' galaxy and find the
galaxy mass distribution. For a cluster with the BCG and the 'C'
galaxy, we rotate the positions of member galaxies so that the 'C'
galaxy is located in the northwest direction with a position angle of
$-45$ degree. As shown in Fig.~\ref{postmerger}, we then divide the
local region near the BCG and the 'C' galaxy into seven square
sub-regions with the same size on each side, with the diagonal length
being the distance between them. The subregion No.1 is the common
region between the BCG and the 'C' galaxy. The sub-regions No.2, 3,
and 4 are aside from the BCG, and the sub-regions No.5, 6, and 7 are
aside from the 'C' galaxy.

If in the sub-region No.1, there are less than 8 recognized member
galaxies, i.e. $N_{\rm gal}(1) <8$, the cluster is skipped for the
post-collision mergers to avoid large Poisson noise. If there are more
than 8 member galaxies, we get the total stellar mass of member
galaxies in these seven sub-regions, denoted as being $M_{\rm
  \star}(i)$ ($i=$1, 2, 3, 4, 5, 6 and 7). Here the BCG and the 'C'
galaxy are not included in these mass calculations.

3. Find rich clusters with over-dense subregion No.1. The
post-collision mergers in principle contain more galaxies in the
sub-region No.1 than other sub-regions. We need to compare its total
stellar mass with those of three sub-regions near the BCG
(i.e. regions No.2, No.3 and No.4 in Fig.~\ref{postmerger}), and then
with those of three sub-regions No. 5 ,6 and 7 near the 'C' galaxy. We
define the overdensity factor as
\begin{equation}
\delta_1=\frac{3\,M_{\rm \star}(1)}{M_{\rm \star}(2)+M_{\rm \star}(3)+M_{\rm \star}(4)},
\end{equation}
\begin{equation}
\delta_2=\frac{3\,M_{\rm \star}(1)}{M_{\rm \star}(5)+M_{\rm \star}(6)+M_{\rm \star}(7)}.
\end{equation}
If two symmetric clusters with a similar mass are simply overlapped,
we would get $\delta_1=2$ and $\delta_2=2$. Here the post-collision
mergers are recognized with the criteria of both $\delta_1\ge2.5$ and
$\delta_2\ge2.5$, which ensures that the stellar mass density from
member galaxies between the two original BCGs is more enhanced than
the simple sum of the two original clusters.

4. Finally, as we test for all bright galaxies as potential BCGs of
subclusters, we get the final catalogue of post-collision mergers. For
a cluster with more than two 'C' galaxies satisfying the above
criteria, the 'C' galaxy with the largest $\delta_1$ is adopted. We
finally get 3446 rich clusters with post-collision merging features,
as listed in Table~\ref{tab3}.

\begin{figure*}
\includegraphics[width = 0.33\textwidth]{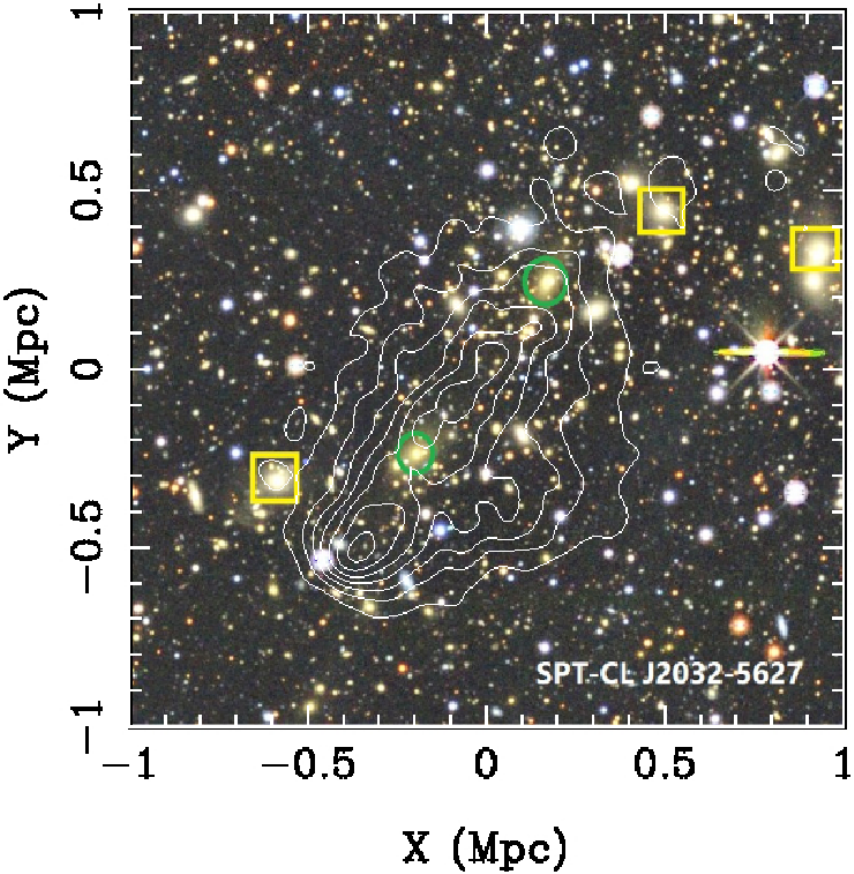}
\includegraphics[width = 0.33\textwidth]{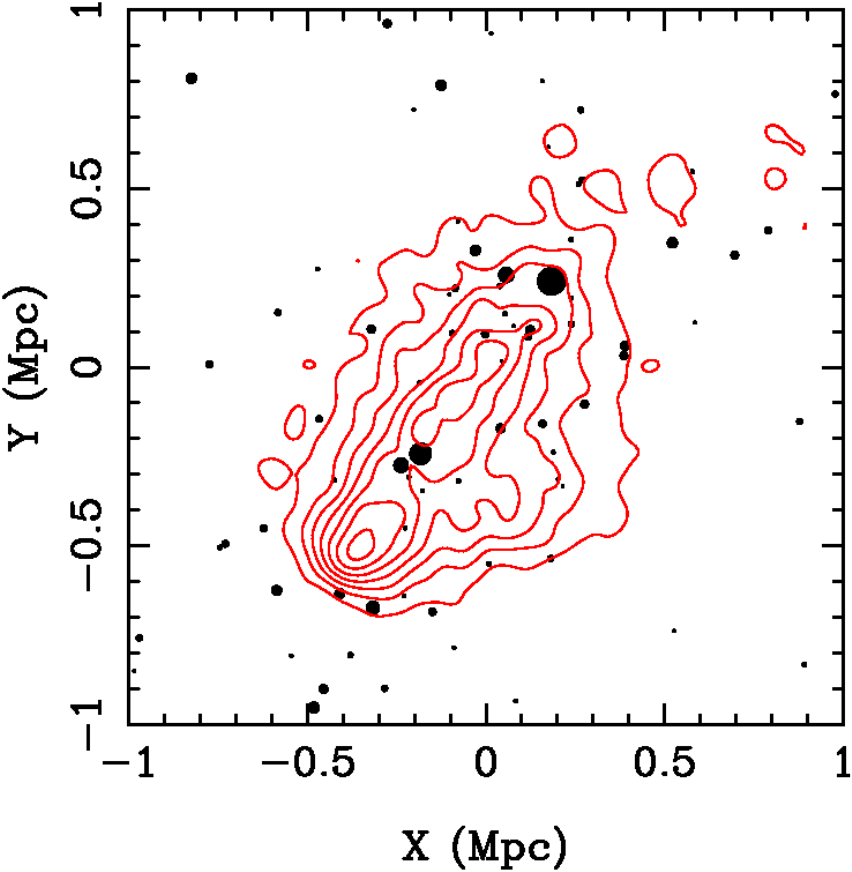}
\includegraphics[width = 0.33\textwidth]{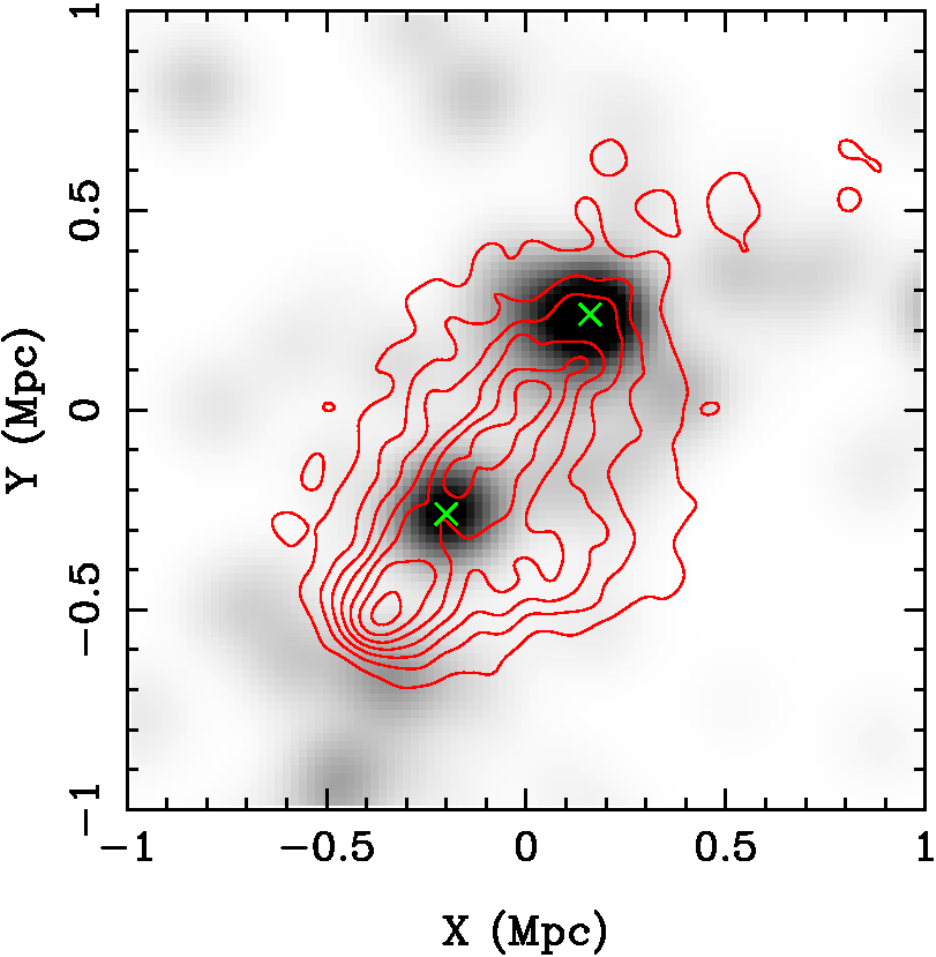}
\caption{An example of post-collision clusters for the bullet-like
  structure identified by \citet{djb+21}. Left: X-ray contours
  overlaid on DESI colour image for the bullet-like cluster, SPT-CL
  J2032-5627. The green circles mark the BCGs in the colliding
  subclusters. The yellow squares indicate bright member galaxies in a
  foreground cluster, A3685. Middle and right: X-ray contours are
  overlaid on the distribution of member galaxies and the smoothed
  map, respectively. In the right panel, the crosses are the locations
  of peaks for the colliding subclusters.}
\label{J2032}
\end{figure*}

\begin{figure*}
\centering
\includegraphics[width = 0.47\textwidth]{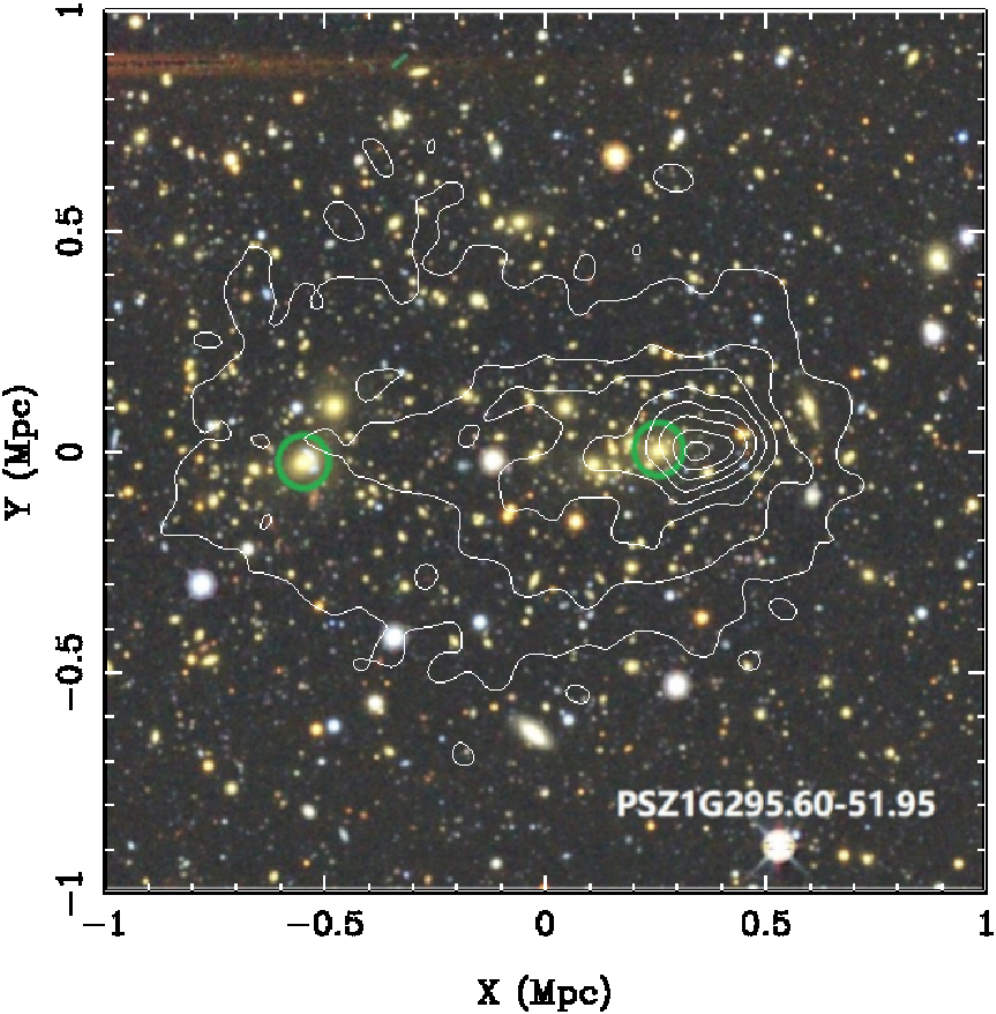}
\includegraphics[width = 0.47\textwidth]{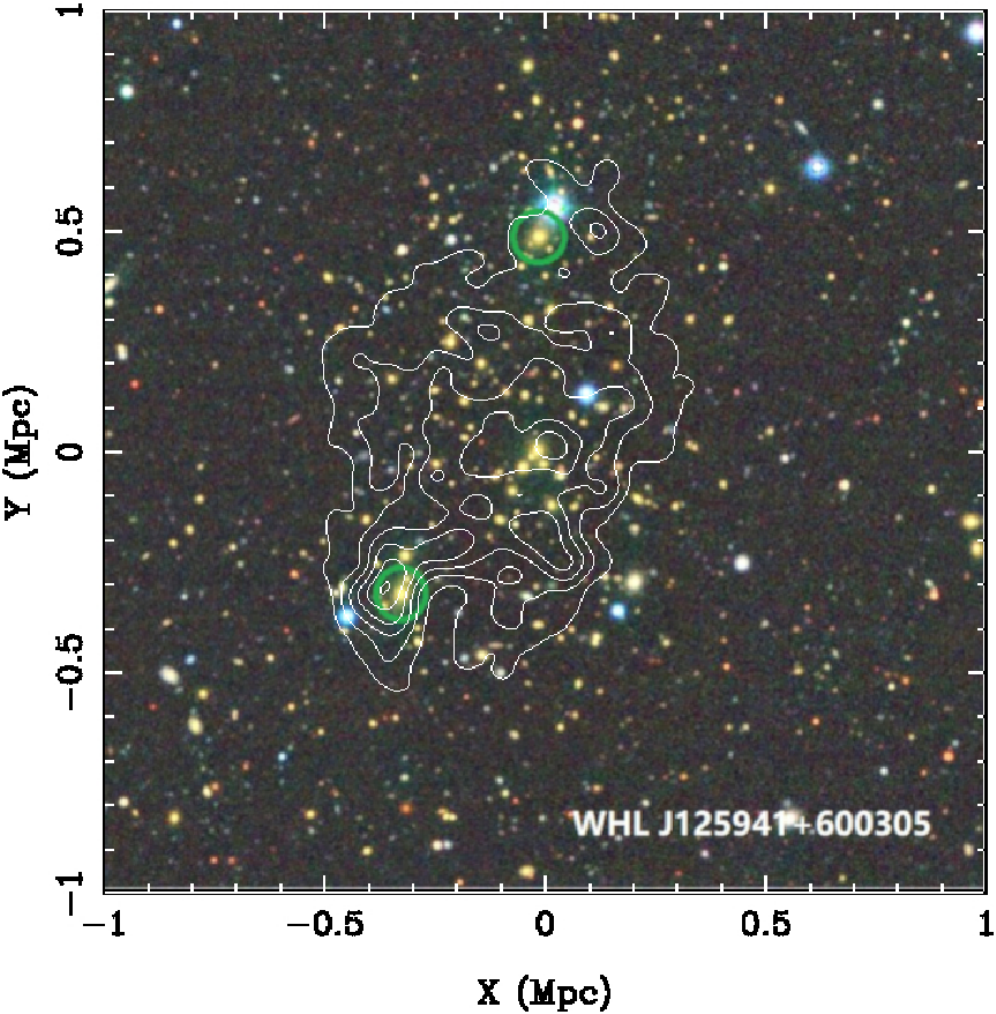}
\caption{Two new bullet-like clusters identified from Post-collision
  mergers, PSZ1 G295.60$-$51.95 at $z=0.3342$ and WHL J125941$+$600305
  at $z=0.3436$. The X-ray contours are measured by the XMM-Newton
  ({\it the left panel}) and the Chandra ({\it the right panel})
  overlaid on DESI colour images. Circles mark the BCGs of the
  colliding subclusters.}
\label{2bullet}
\end{figure*}

\subsection{Verification of post-collision mergers by X-ray images}

We validate the post-collision mergers in Table~\ref{tab3} by checking
X-ray images, as done in Section 4.3. Cross-matching with the X-ray
cluster sample in \citet{yhw22}, we get the {\it Chandra} and {\it
  XMM-Newton} images for 152 clusters (39 from the Chandra and 113
from the XMM-Newton) as listed in Table~\ref{tab3}, and all of them
are shown in Fig.~\ref{post}. About 90\% of clusters show
post-collision features in X-ray images, suggesting that our method
efficiently recognizes post-collision mergers of clusters, with many
interesting cases and some exceptions as discussed below.

Some clusters, e.g. A141, A168, A384, A520, A1240, A1553, A1914,
A2034, A2443, A2645, A521, the Bullet, MCXC J0510.7-0801, MACS
J1006.9+3200, PSZ1G295.60-51.95, SPTCLJ2032-5627, WHL J125941+600305
and WHL J120735.9+525459, have distinct offsets between their BCGs and
the X-ray emission peaks, indicating the separations of the ICM and
dark matter during the cluster collision. The sharp edges of the X-ray
images indicate the front shock of post-collision mergers.

Some clusters, e.g. A68, A222, A1201, A2345, A2410, A2507, A329,
A3399, MCXC J0244.1-2611, WHL J122657.4+3343, PSZ2G314.26-55.35,
PSZ2G207.88+81.31 and 2MASXJ101007+3253, have X-ray peaks coincident
with the BCG or the second BCG. The tail feature of the X-ray emission
indicates a bulk flow of the ICM, which is the indication of clusters
in an off-axis collision.

Diffuse radio halos or relics have been detected from some of these
clusters, e.g. A209, A520, A521, A1240, A1351, A1914, A2034, A2254,
A2255, A2443, the Bullet and ZwCl 2341.1+0000 \citep{fgg+12}, A141
\citep{djw+21}, A168 \citep{dpk+18}, A2645 \citep{kcr+22}, AS1063
\citep{rrd+21}, PSZ2G057.61+34.93 \citep{lrv+20} and SPT-CL J2032-5627
\citep{djb+21}, which are signatures for the ongoing mergers.

We noticed, however, that some clusters, e.g. PSZ2G204.10+16.51,
A2051, AS974, A611, A1084, A2055, and A2261, show a regular and
concentrated X-ray emission around one subcluster as if in a relaxed
dynamical state. Our optical galaxy luminosity distribution analysis
suggests that they are in the post-merger stage. Possibly these
clusters are experiencing an off-axis minor merger, which is not
powerful enough to destroy the bright X-ray core of one cluster, as in
A2261 shown from X-ray and radio data \citep{sbi+17}. Otherwise, they
are probably clusters in the sky due to the projection effect.

All in all, these 3446 rich clusters with post-collision features in
the galaxy distribution are interesting objects worth follow-up.

\subsection{Two newly identified bullet-like clusters}

The Bullet cluster \citep{mgd+02} is an extremely great example of
post-collision mergers, which has a bullet-like subcluster moving away
from the more massive subcluster with a high speed indicated by a
prominent bow-like shock in X-ray. According to simulations
\citep{hw06, ks15}, the bullet-like clusters are rare in the
Universe. Up to now, only a few such cases have been found, including
A2146 \citep{rsf+10}, Abell 3376 \citep{bdn+06}, ACT-CL J0102-4915
``El Gordo'' \citep{mhs+12}, ZwCl 0008.8+5215 \citep{gvd+17}, RXC
J2359.3-6042 \citep{cb15} and SPT-CL J2032-5627 \citep{djb+21}.

We get the plots of SPT-CL J2032$-$5627 in
Fig.~\ref{J2032}. \citet{djb+21} have showed the bullet-like structure
in the X-ray image and detected the double radio relics at the
southeast and northwest of this cluster. We find that SPT-CL
J2032$-$5627 has an elongated distribution of member galaxies (middle
panel of in Fig.~\ref{J2032}) extracted from the optical data, and the
smoothed map shows two distinct subclusters (right panel of in
Fig.~\ref{J2032}). The northwest subcluster hosts the first BCG and is
more massive than the southeast subcluster with a mass ratio of
2.07. The optical and X-ray images suggest that this cluster is a very
unusual merger. In a post-collision merger, the X-ray emission peak is
expected to be located between the merging subclusters
\citep{cbg+06,mcv+18}. However, the {\it XMM-Newton} image of this
cluster shows that the X-ray emission peak is leading both
subclusters. The projected separation between the X-ray peak and the
optical peak of the smaller subcluster is as large as 304 kpc,
indicating a very significant offset between ICM and dark matter,
similar to the Bullet cluster \citep{cbg+06}.

With such post-collision features, we find two new bullet-like
clusters among our galaxy clusters, PSZ1 G295.60$-$51.95 at $z=0.3327$
and WHL J125941$+$600305 at $z=0.3513$, see Fig.~\ref{2bullet} for the
optical images overlaid with the X-ray emission detected by the {\it
  XMM-Newton} and the {\it Chandra} as processed by \citet{yh20} and
\citet{yhw22}.

The optical colour image shows that PSZ1 G295.60$-$51.95 contains
several luminous galaxies. The {\it XMM-Newton} X-ray data show a
head-tail morphology with a sharp brightness edge. The first BCG (east
circle) is located in the tail with very faint X-ray emission. The
west subcluster is associated with the X-ray emission peak which is
ahead of the second BCG. The offset between the closer second BCG and
the X-ray peak is 79.7 kpc. These features indicate that the ICM is
rapidly moving from the east to the west, and the ICM in the east
subcluster has been stripped during the cluster merger. Hence, PSZ1
G295.60-51.95 is a bullet-like cluster.

WHL J125941$+$600305 also has a head-tail morphology in X-ray emission
detected by the {\it Chandra}, and shows two luminous BCGs in the
optical image. The X-ray peak is located at the edge of the contours
with a sharp brightness in the southeast, and it has an offset of 55.5
kpc from the first BCG. The X-ray emission around the other BCG is
much weaker. These features suggest that WHL J125941$+$600305 is also
a bullet-like cluster.

\section{Summary}

We present large samples of cluster mergers according to the optical
properties of a large sample of galaxy clusters identified from the
DESI Legacy Surveys data. By searching nearby clusters, we identify a
sample of 39\,382 partner clusters for 33\,126 main clusters within a
projected distance of 5\,$r_{500}$. About 97\% partners have a
velocity difference of 1000 km~s$^{-1}$ from the main clusters. For
rich clusters with more than 30 member galaxies, we analyze the
subclusters by smoothing the optical stellar mass distribution and
then fit them with the dual-King model. We define the coupling factor
to quantify the merging status for 7845 rich clusters. Based on the
projected distribution of member galaxies, we further develop a new
approach to recognize 3446 rich clusters with post-collision features,
some of which have been validated by using X-ray images.

Such large samples of merging clusters of galaxies are important
databases for studying the hierarchical structure formation, cluster
evolution, and the physics of intergalactic medium. The partner
systems are objects for the study of the process in the pre-merger
stage, the material between clusters, and the structure formation. The
mergers in the rich clusters can help to understand the violent
process of cluster formation, diffuse radio emission, and properties
of dark matter.

\section*{Acknowledgements}

We thank the referee for valuable comments that helped to improve the
paper. The authors are supported by the National Natural Science
Foundation of China (Grant Numbers 11988101, 11833009 and 12073036),
the Key Research Program of the Chinese Academy of Sciences (Grant
Number QYZDJ-SSW-SLH021). We also acknowledge the support by the
science research grants from the China Manned Space Project with
Numbers CMS-CSST-2021-A01 and CMS-CSST-2021-B01.

The DESI Legacy Imaging Surveys consist of three individual and
complementary projects: the Dark Energy Camera Legacy Survey (DECaLS),
the Beijing-Arizona Sky Survey (BASS), and the Mayall z-band Legacy
Survey (MzLS). DECaLS, BASS and MzLS together include data obtained,
respectively, at the Blanco telescope, Cerro Tololo Inter-American
Observatory, NSF’s NOIRLab; the Bok telescope, Steward Observatory,
University of Arizona; and the Mayall telescope, Kitt Peak National
Observatory, NOIRLab. NOIRLab is operated by the Association of
Universities for Research in Astronomy (AURA) under a cooperative
agreement with the National Science Foundation. Pipeline processing
and analyses of the data were supported by NOIRLab and the Lawrence
Berkeley National Laboratory (LBNL). Legacy Surveys also uses data
products from the Near-Earth Object Wide-field Infrared Survey
Explorer (NEOWISE), a project of the Jet Propulsion
Laboratory/California Institute of Technology, funded by the National
Aeronautics and Space Administration. Legacy Surveys was supported by:
the Director, Office of Science, Office of High Energy Physics of the
U.S. Department of Energy; the National Energy Research Scientific
Computing Center, a DOE Office of Science User Facility; the
U.S. National Science Foundation, Division of Astronomical Sciences;
the National Astronomical Observatories of China, the Chinese Academy
of Sciences and the Chinese National Natural Science Foundation. LBNL
is managed by the Regents of the University of California under
contract to the U.S. Department of Energy. The complete
acknowledgments can be found at
https://www.legacysurvey.org/acknowledgment/.

Funding for the Sloan Digital Sky Survey IV has been provided by the
Alfred P. Sloan Foundation, the U.S. Department of Energy Office of
Science, and the Participating Institutions. SDSS acknowledges support
and resources from the Center for High-Performance Computing at the
University of Utah. The SDSS web site is www.sdss4.org.
SDSS is managed by the Astrophysical Research Consortium for the
Participating Institutions of the SDSS Collaboration including the
Brazilian Participation Group, the Carnegie Institution for Science,
Carnegie Mellon University, Center for Astrophysics | Harvard \&
Smithsonian (CfA), the Chilean Participation Group, the French
Participation Group, Instituto de Astrof{\'i}sica de Canarias, The
Johns Hopkins University, Kavli Institute for the Physics and
Mathematics of the Universe (IPMU) / University of Tokyo, the Korean
Participation Group, Lawrence Berkeley National Laboratory, Leibniz
Institut f{\"u}r Astrophysik Potsdam (AIP), Max-Planck-Institut
f{\"u}r Astronomie (MPIA Heidelberg), Max-Planck-Institut f{\"u}r
Astrophysik (MPA Garching), Max-Planck-Institut f{\"u}r
Extraterrestrische Physik (MPE), National Astronomical Observatories
of China, New Mexico State University, New York University, University
of Notre Dame, Observat{\'o}rio Nacional / MCTI, The Ohio State
University, Pennsylvania State University, Shanghai Astronomical
Observatory, United Kingdom Participation Group, Universidad Nacional
Aut{\'o}noma de M{\'e}xico, University of Arizona, University of
Colorado Boulder, University of Oxford, University of Portsmouth,
University of Utah, University of Virginia, University of Washington,
University of Wisconsin, Vanderbilt University, and Yale University.

This work also uses observations obtained with XMM–Newton, an ESA
science mission with instruments and contributions directly fundedby
the ESA Member States and the National Aeronautics and Space
Administration (NASA). This research has used data obtained from the
Chandra Data Archive and the Chandra Source Catalogue, and software
provided by the Chandra X-ray Center (CXC) in the application packages
CIAO, CHIPS and SHERPA.

\section*{Data availability}

We publish the full tables for 33\,126 cluster partner systems, 7845
clusters with distinct subclusters and 3446 post-collision
mergers. They are also publicly available at
http://zmtt.bao.ac.cn/galaxy\_clusters/.

\bibliographystyle{mnras}
\bibliography{wise}

\begin{appendix}

\section{The best scale for smoothed optical images}
\label{scale}

\begin{figure}
\centering
\includegraphics[width = 0.35\textwidth]{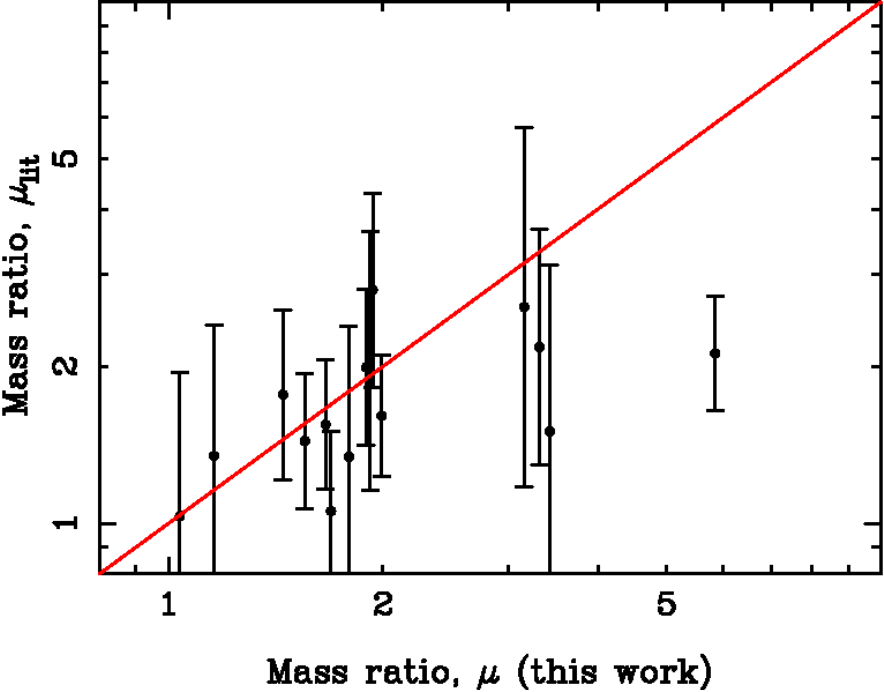}
\caption{Comparison of the mass ratio of subclusters obtained from our
  smoothed optical maps with those from the weak lensing or dynamical
  methods in the literature. The solid line represents $Y=X$. This is
  the best matches at the smooth scale of $0.08r_{500}$. In general,
  they are consistent given their uncertainty.}
\label{comp_ratio}
\end{figure}

\begin{figure}
\centering
\includegraphics[width = 0.35\textwidth]{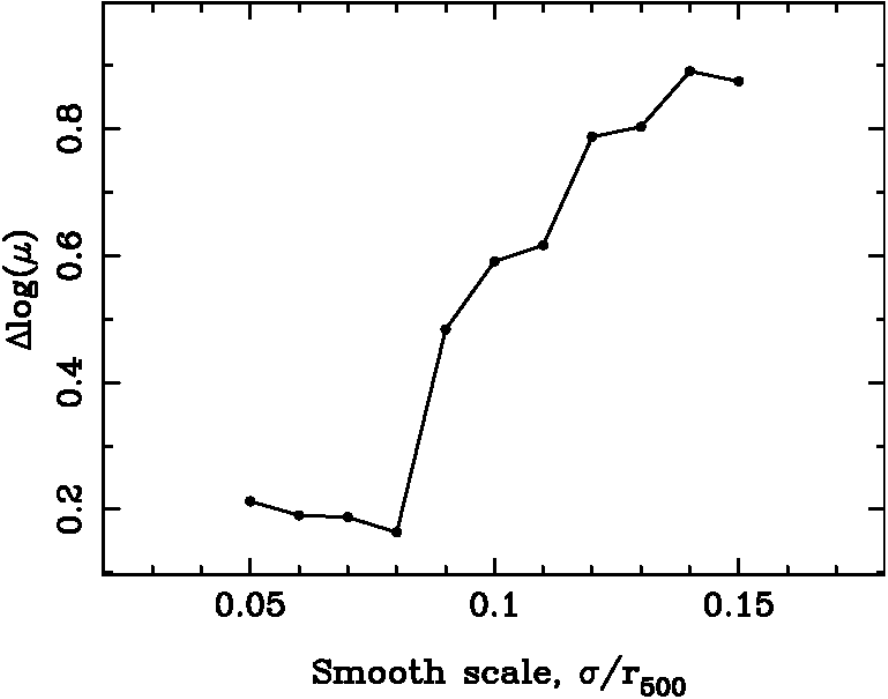}
\caption{Scatter of mass ratios between the values from our smoothed
  optical maps and those values in the literature varies with the
  smooth scale.}
\label{scatter}
\end{figure}

We have to find a proper smoothing scale to construct a smoothed
optical map for the stellar mass distribution of galaxies, from which
merging subclusters can be recognized. We test a series of smooth
scales from 0.05\,$r_{500}$ to 0.15\,$r_{500}$ which have been
proposed for some merging clusters in literature, e.g. for clusters of
CIZA J2242.8+5301, RX J0603.3+4214, MACS J1149.5+2223, ZwCl
0008.8+5215 and PLCK G287.0+32.9 \citep{jsd+15, jds+16, gdw+16,
  gvd+17, fjg+17}.  We also compile the merging clusters with
subclusters recognized and the masses of subclusters derived by weak
lensing and dynamical methods, including A1758, MACS J1149.5+2223,
ZwCl 2341.1+0000, Abell 56, A115, A1240, A2034, A1750, A1995, A2440,
SPT-CL J2023-5535, ACT-CL J0256.5, RXCJ1230.7+3439, RM
J150822.0+575515.2 and HSC J085024+001536 \citep{mgw96, hl09, bgb+09,
  bbc22, bgb12, kib+16, gdw+16, mcm+17, bwg+17, mcv+18, kjf+19,
  hjr+20, tfo+21, wsf+23, swf+23}. For each smooth scale, we recognize
the two most prominent subclusters and their mass ratios from our
smoothed optical maps are then compared with the values in the
literature, as shown in Fig.~\ref{comp_ratio}.  The data scatter of
the mass-ratio offset from the equivalent line is shown in
Fig.~\ref{scatter}, which varies with the smooth scale.  The best
consistency for the $\mu$ is found when we adopt
$\sigma=0.08\,r_{500}$.  At a larger smooth scale, the subclusters in
some clusters can not be resolved, and then the scatter increases
significantly.

\section{Overlaid X-ray images onto merging clusters and the post-collision mergers}

We show in Fig.~\ref{coupling} the smoothed optical map for the 153
merging clusters showing subcluster features with the X-ray contours
from the {\it Chandra} \citep{yh20} and the {\it XMM-Newton}
observations \citep{yhw22}, sorted by the coupling factor. Such
overlapped images should be helpful for readers for their future
works.

We present in Fig.~\ref{post} the projected distribution of member
galaxies for the 152 clusters of post-collision mergers overlapped
with the X-ray contours from the {\it Chandra} \citep{yh20} and the
{\it XMM-Newton} \citep{yhw22}.
 
\input{coupling.tex}
\input{post.tex}

\end{appendix}

\label{lastpage}
\end{document}

%% file: coupling.tex
\begin{figure*}
\centering
\includegraphics[width=0.95\textwidth]{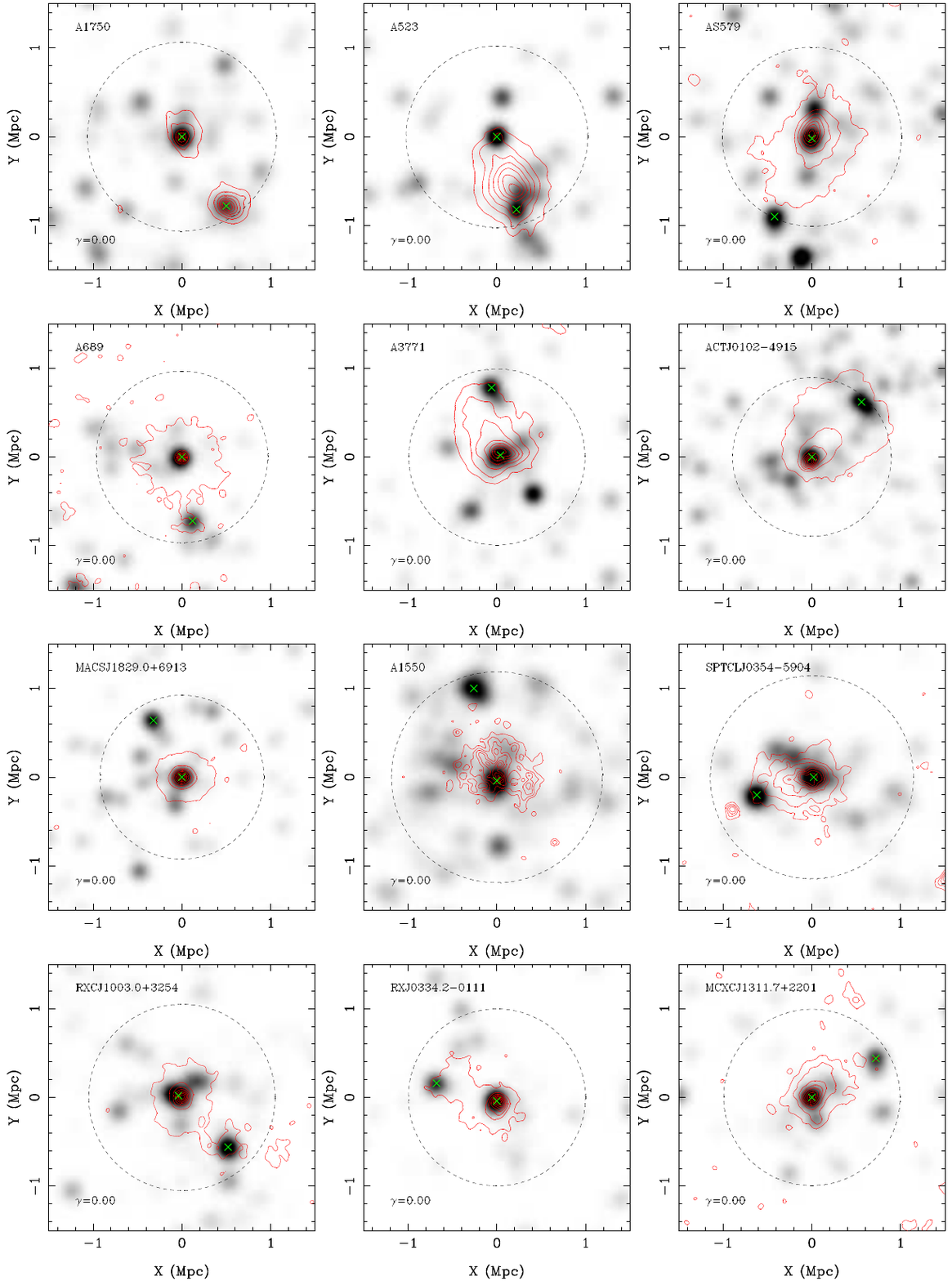}~%
\caption{The contours of X-ray emission from the {\it XMM-Newton} and {\it Chandra} observations overlaid on the smoothed optical map for 153 merging clusters showing subclusters. The big circle indicates the region with a radius of $r_{500}$.  The crosses are the locations of optical peaks
  for the central and the secondary subclusters.
\label{coupling}}
\end{figure*}

\addtocounter{figure}{-1}  \begin{figure*}
\centering
\includegraphics[width=0.95\textwidth]{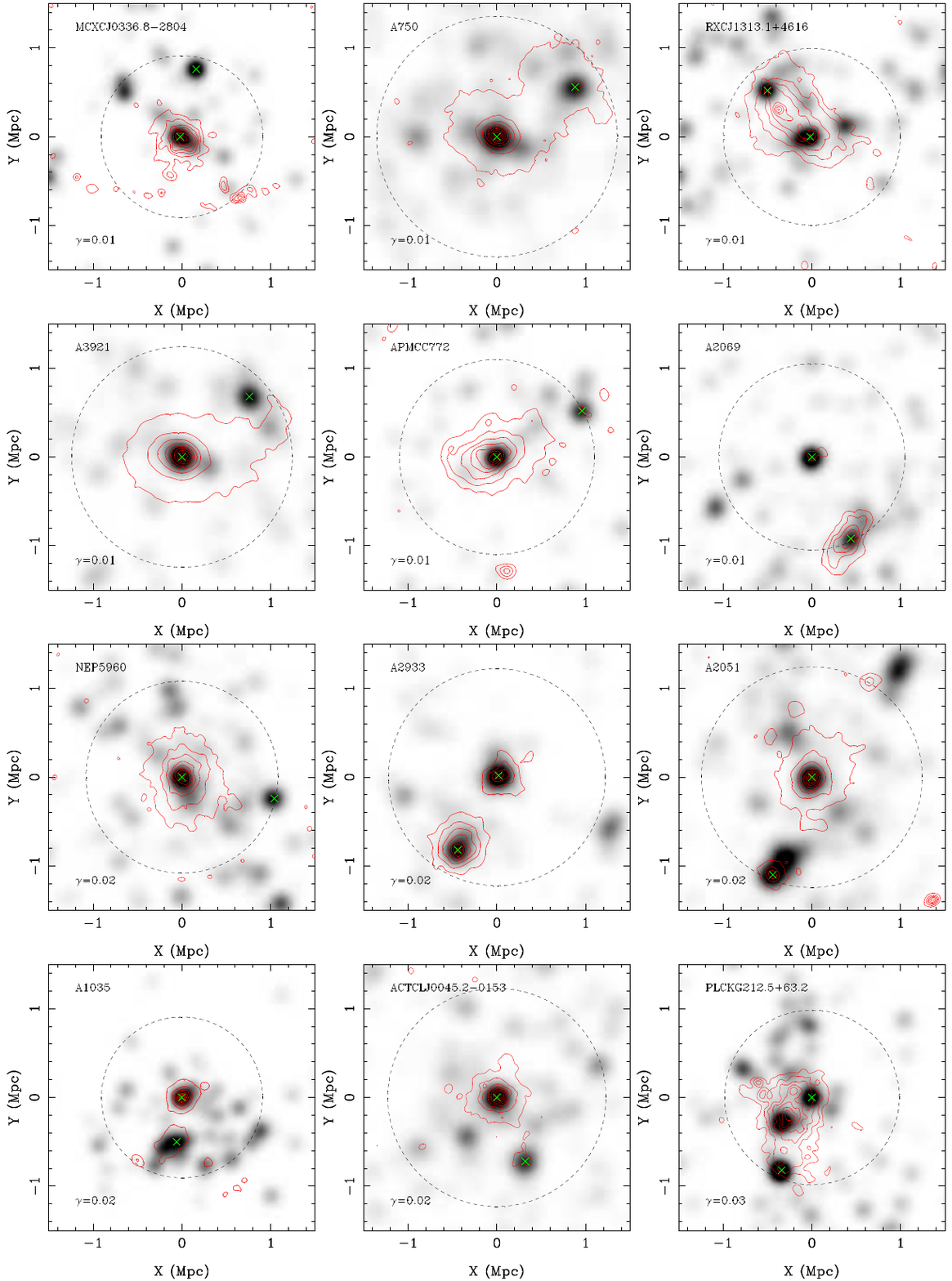}~%
\caption{{\it --- continued} }
\end{figure*}

\addtocounter{figure}{-1}  \begin{figure*}
\centering
\includegraphics[width=0.95\textwidth]{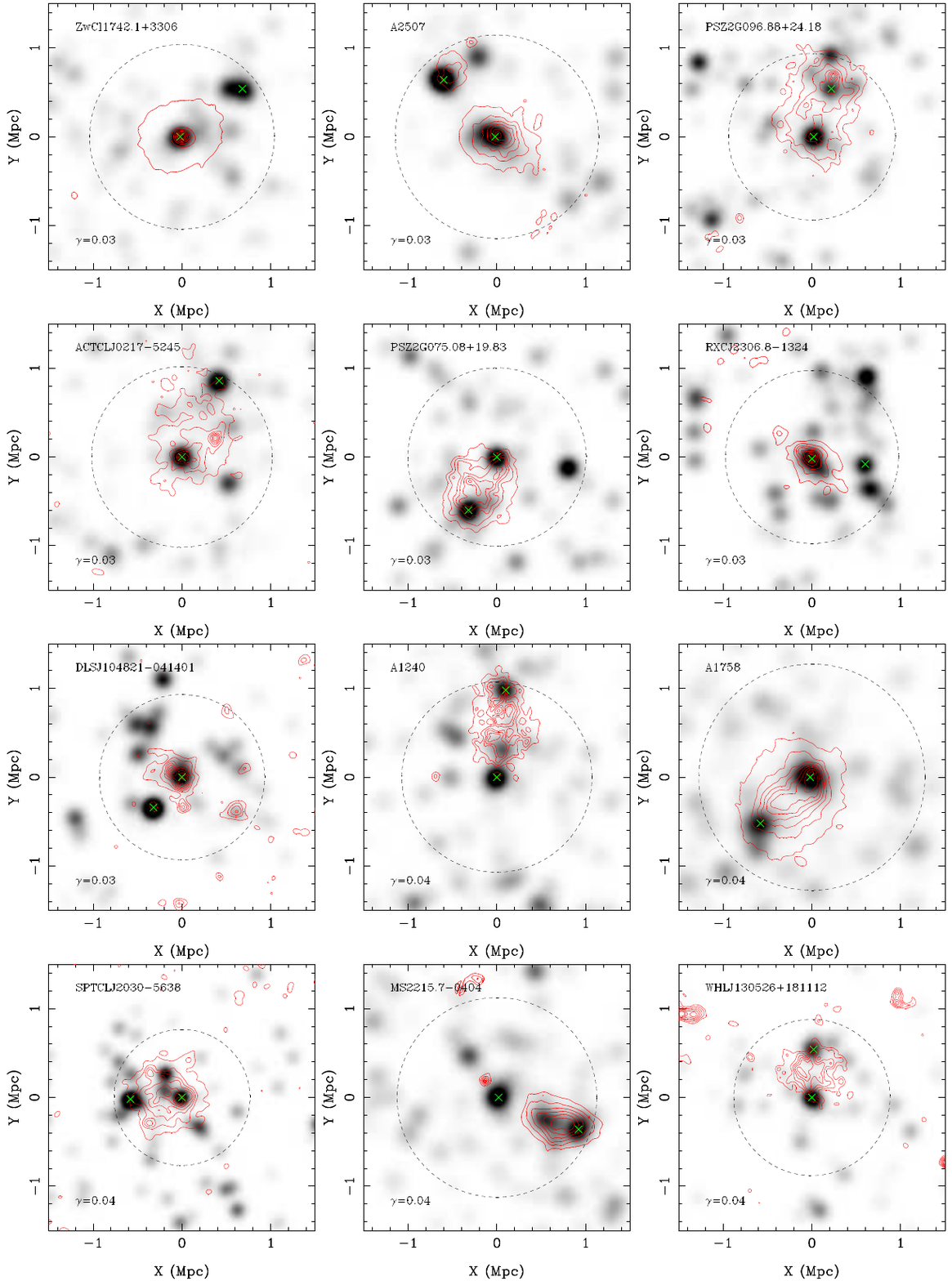}~%
\caption{{\it --- continued} }
\end{figure*}

\addtocounter{figure}{-1}  \begin{figure*}
\centering
\includegraphics[width=0.95\textwidth]{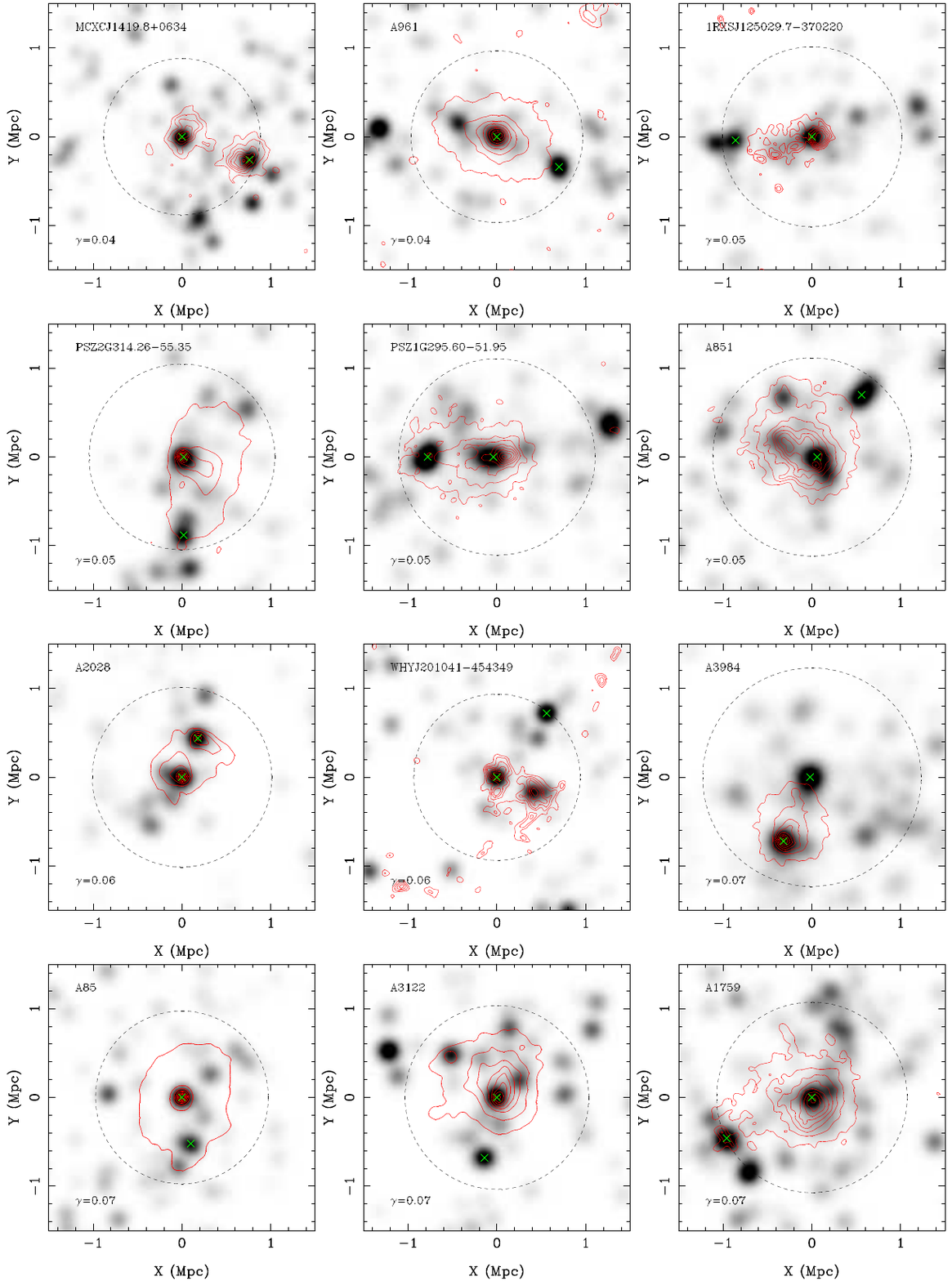}~%
\caption{{\it --- continued} }
\end{figure*}

\addtocounter{figure}{-1}  \begin{figure*}
\centering
\includegraphics[width=0.95\textwidth]{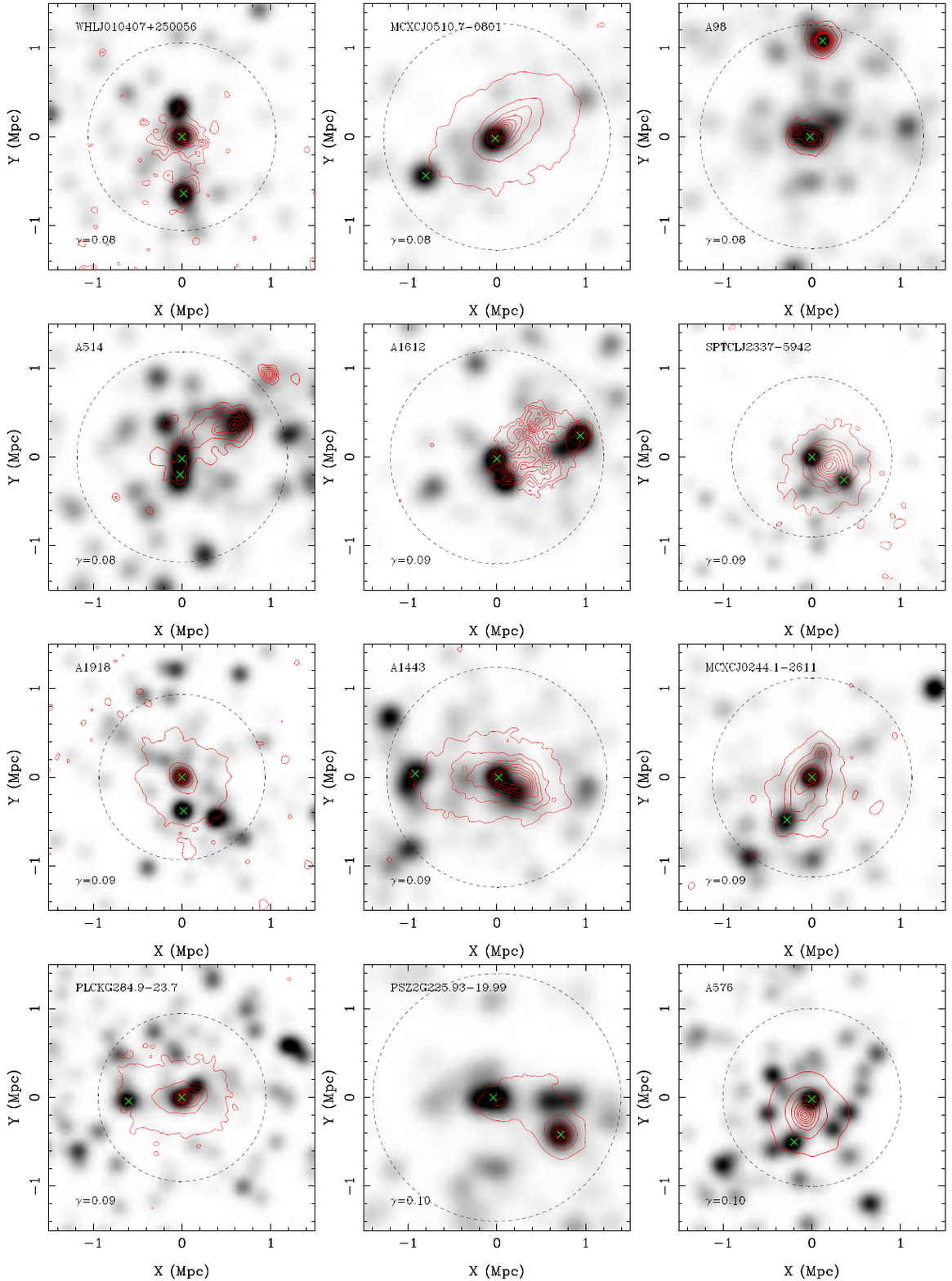}~%
\caption{{\it --- continued} }
\end{figure*}

\addtocounter{figure}{-1}  \begin{figure*}
\centering
\includegraphics[width=0.95\textwidth]{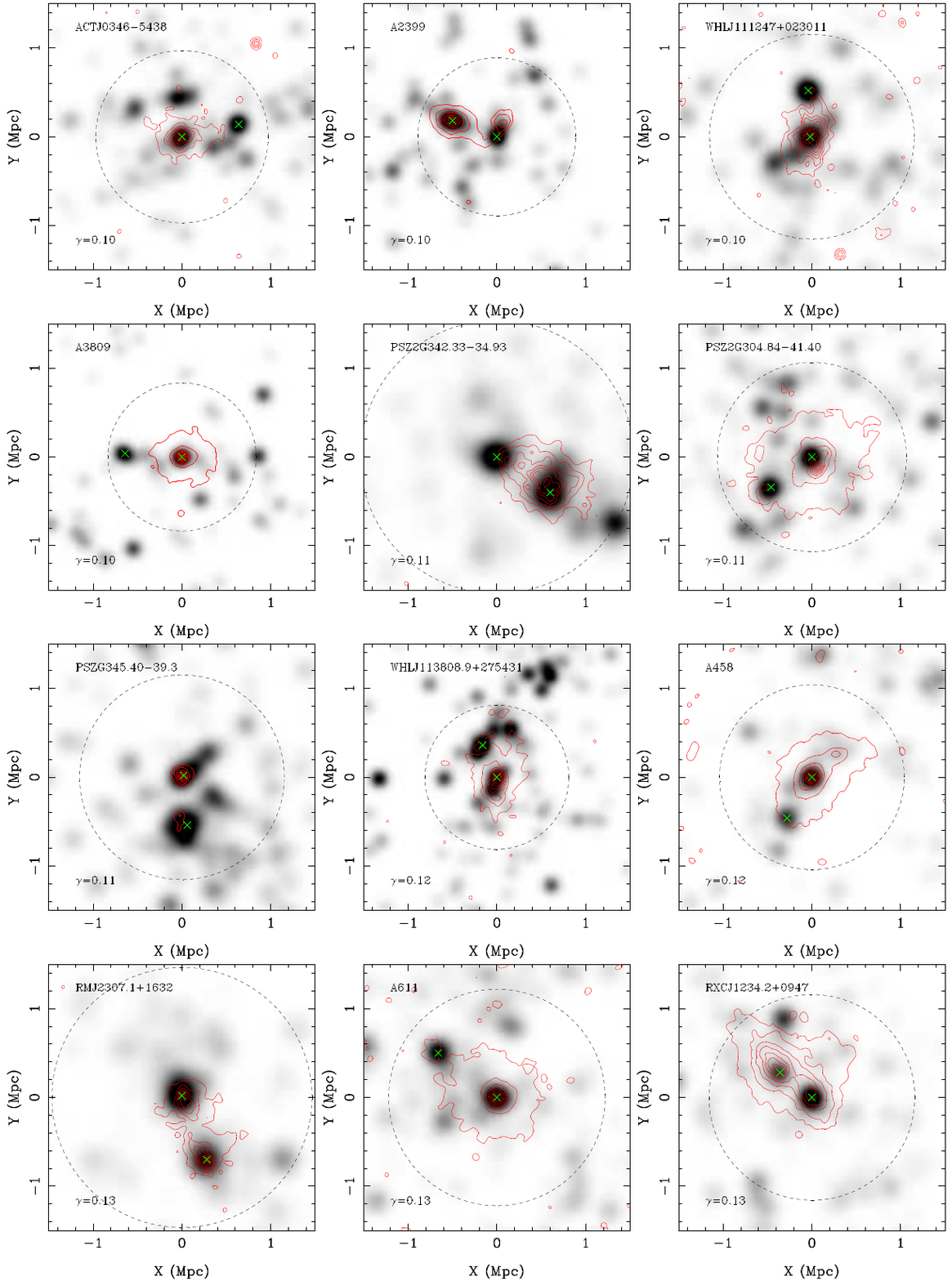}~%
\caption{{\it --- continued} }
\end{figure*}

\addtocounter{figure}{-1}  \begin{figure*}
\centering
\includegraphics[width=0.95\textwidth]{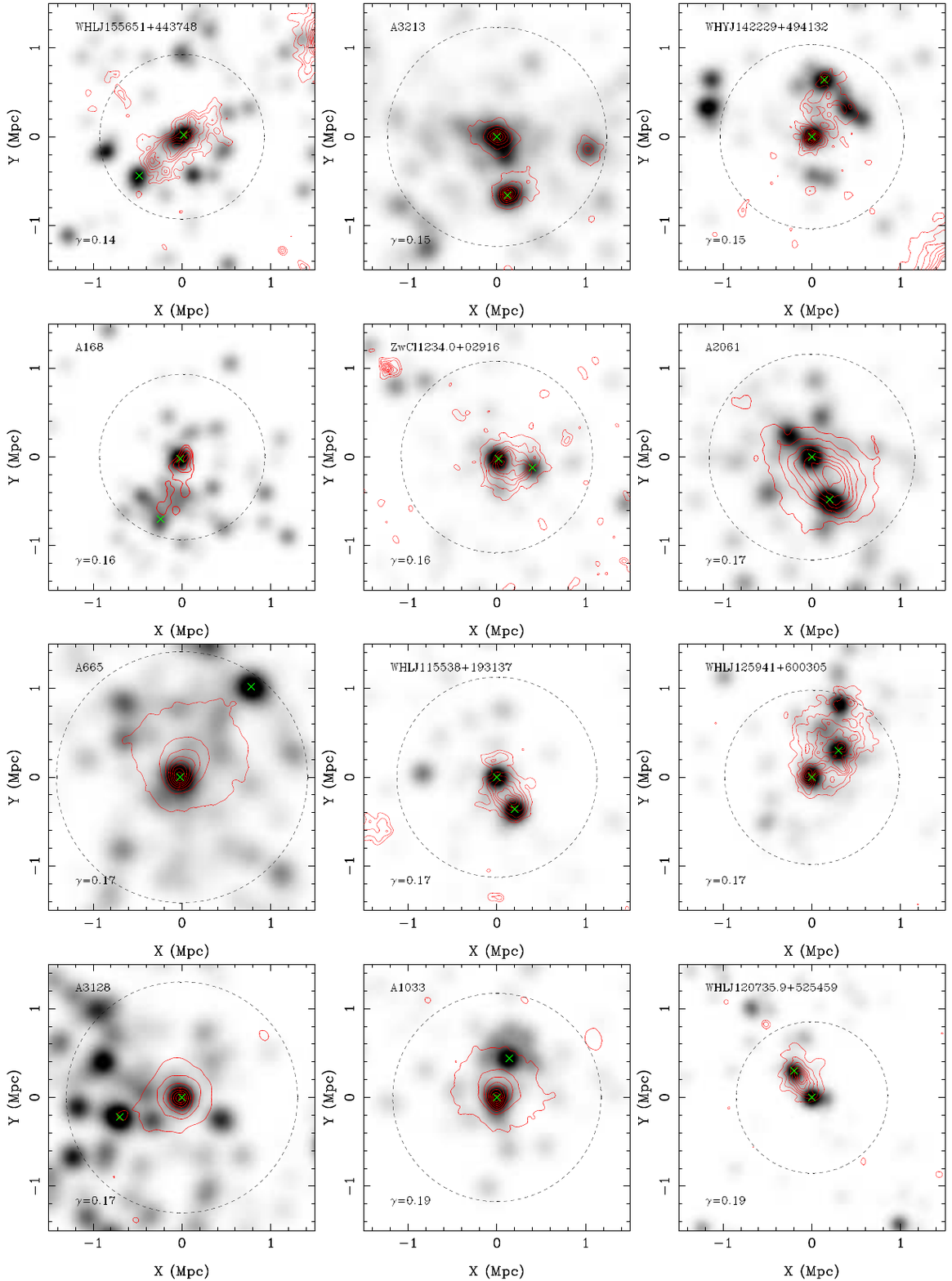}~%
\caption{{\it --- continued} }
\end{figure*}

\addtocounter{figure}{-1}  \begin{figure*}
\centering
\includegraphics[width=0.95\textwidth]{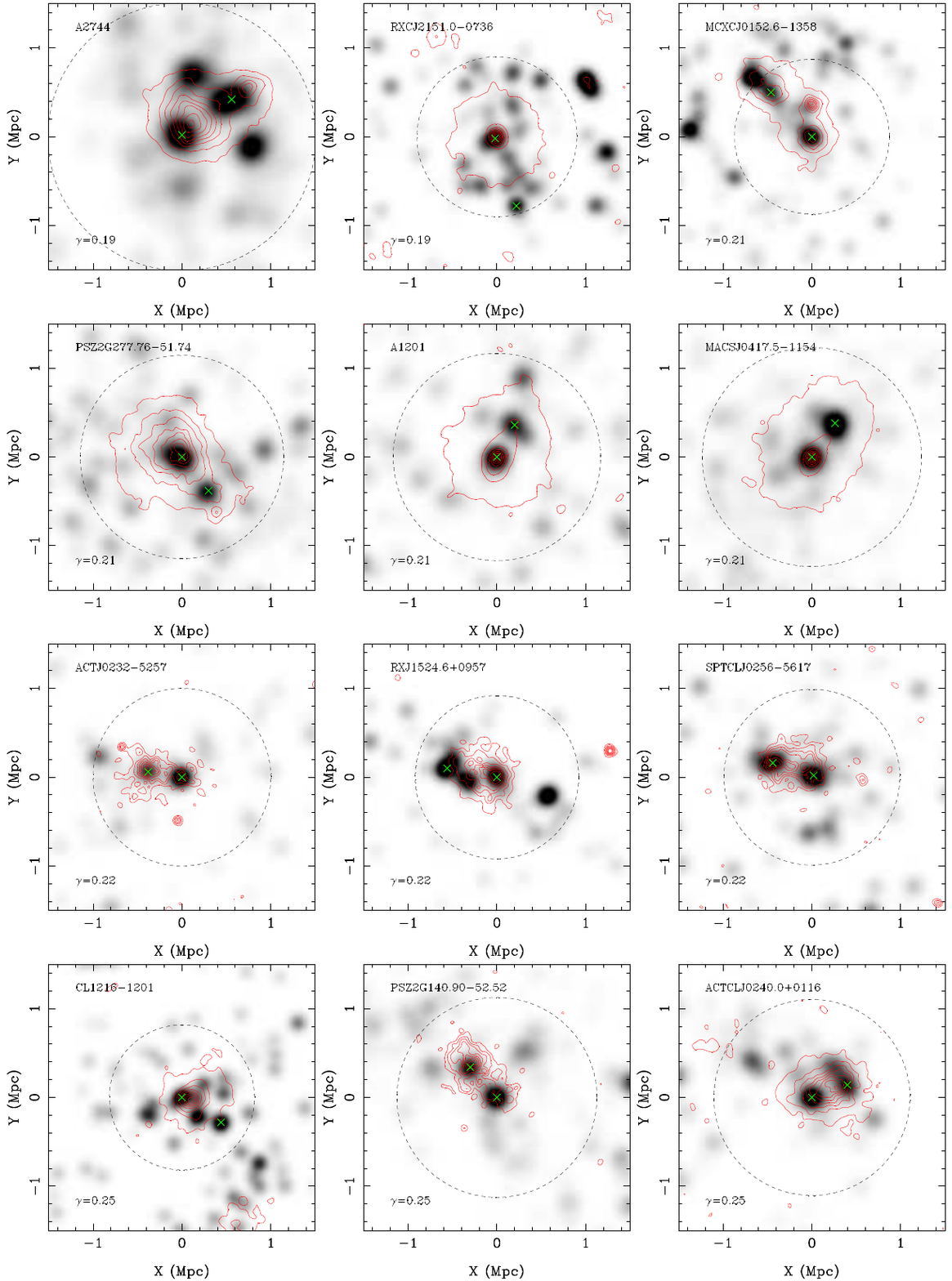}~%
\caption{{\it --- continued} }
\end{figure*}

\addtocounter{figure}{-1}  \begin{figure*}
\centering
\includegraphics[width=0.95\textwidth]{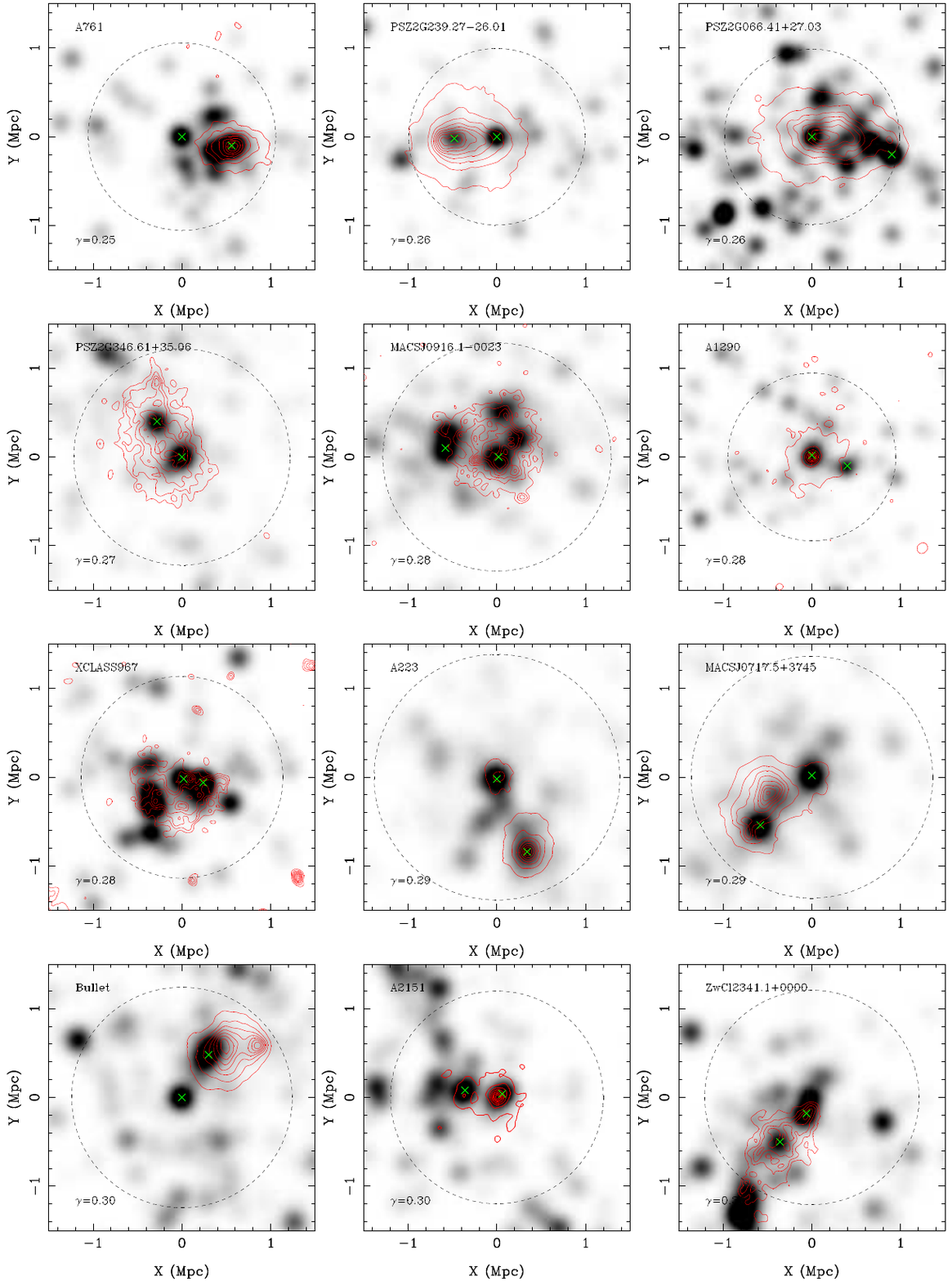}~%
\caption{{\it --- continued} }
\end{figure*}

\addtocounter{figure}{-1}  \begin{figure*}
\centering
\includegraphics[width=0.95\textwidth]{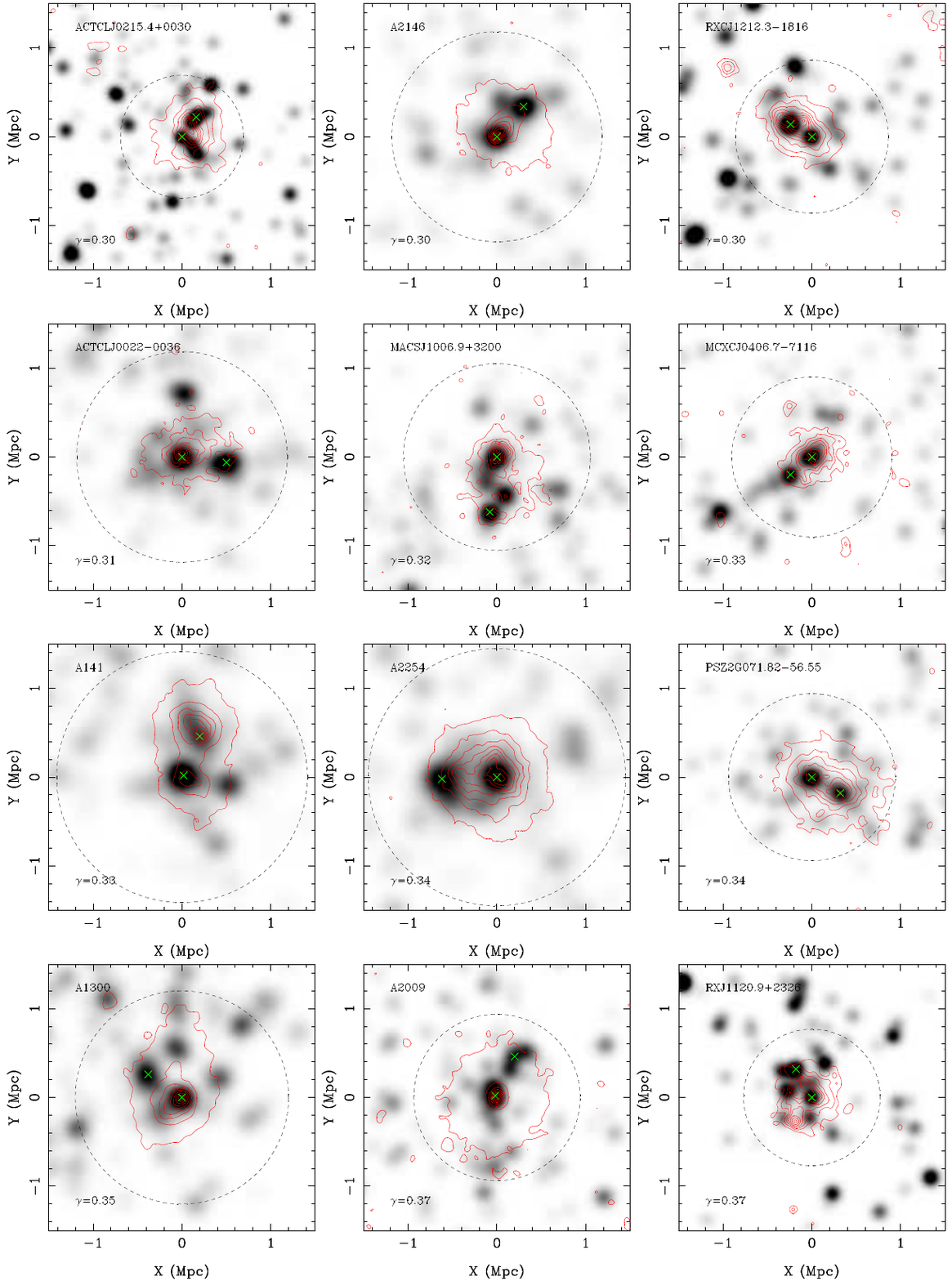}~%
\caption{{\it --- continued} }
\end{figure*}

\addtocounter{figure}{-1}  \begin{figure*}
\centering
\includegraphics[width=0.95\textwidth]{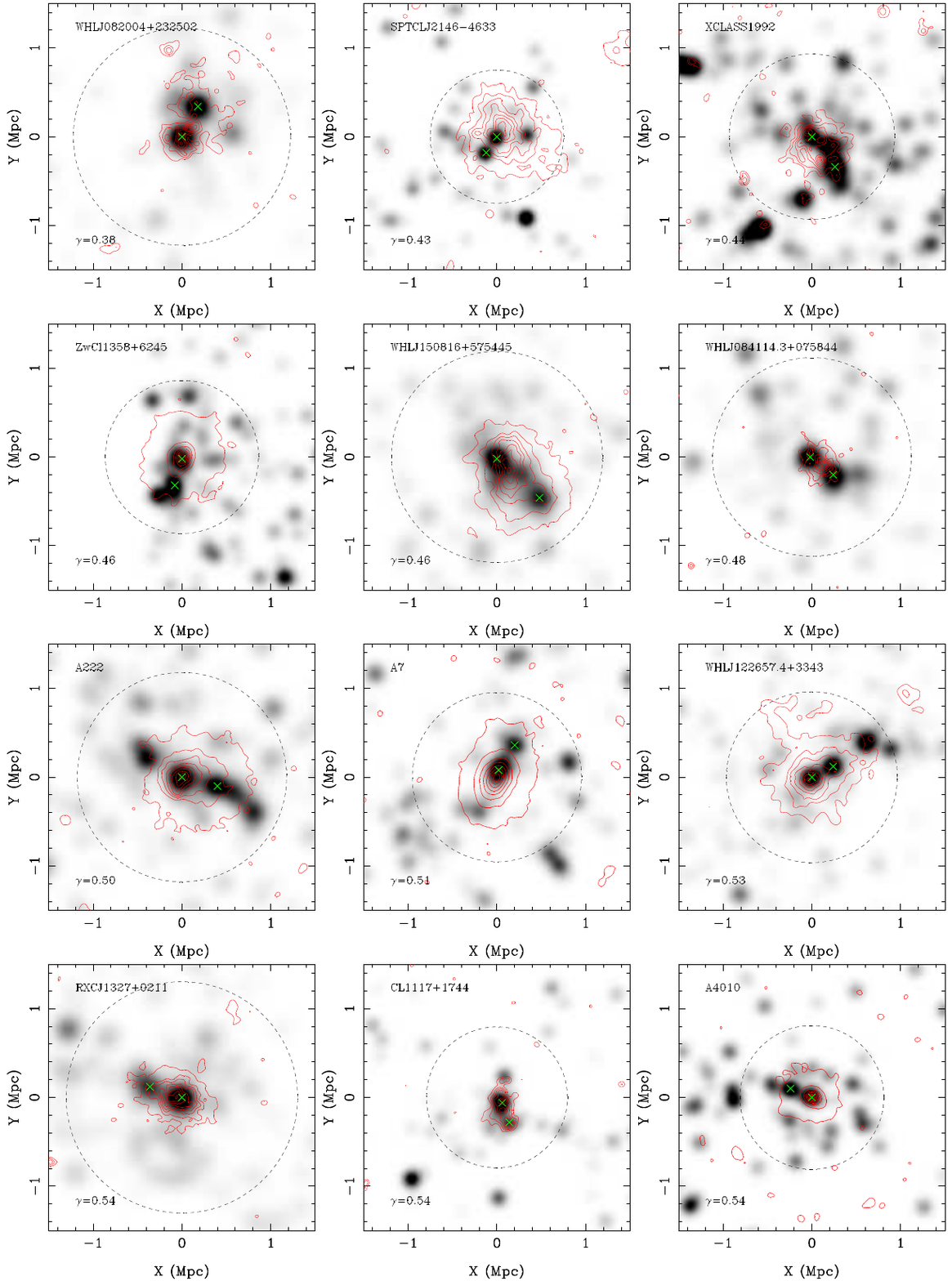}~%
\caption{{\it --- continued} }
\end{figure*}

\addtocounter{figure}{-1}  \begin{figure*}
\centering
\includegraphics[width=0.95\textwidth]{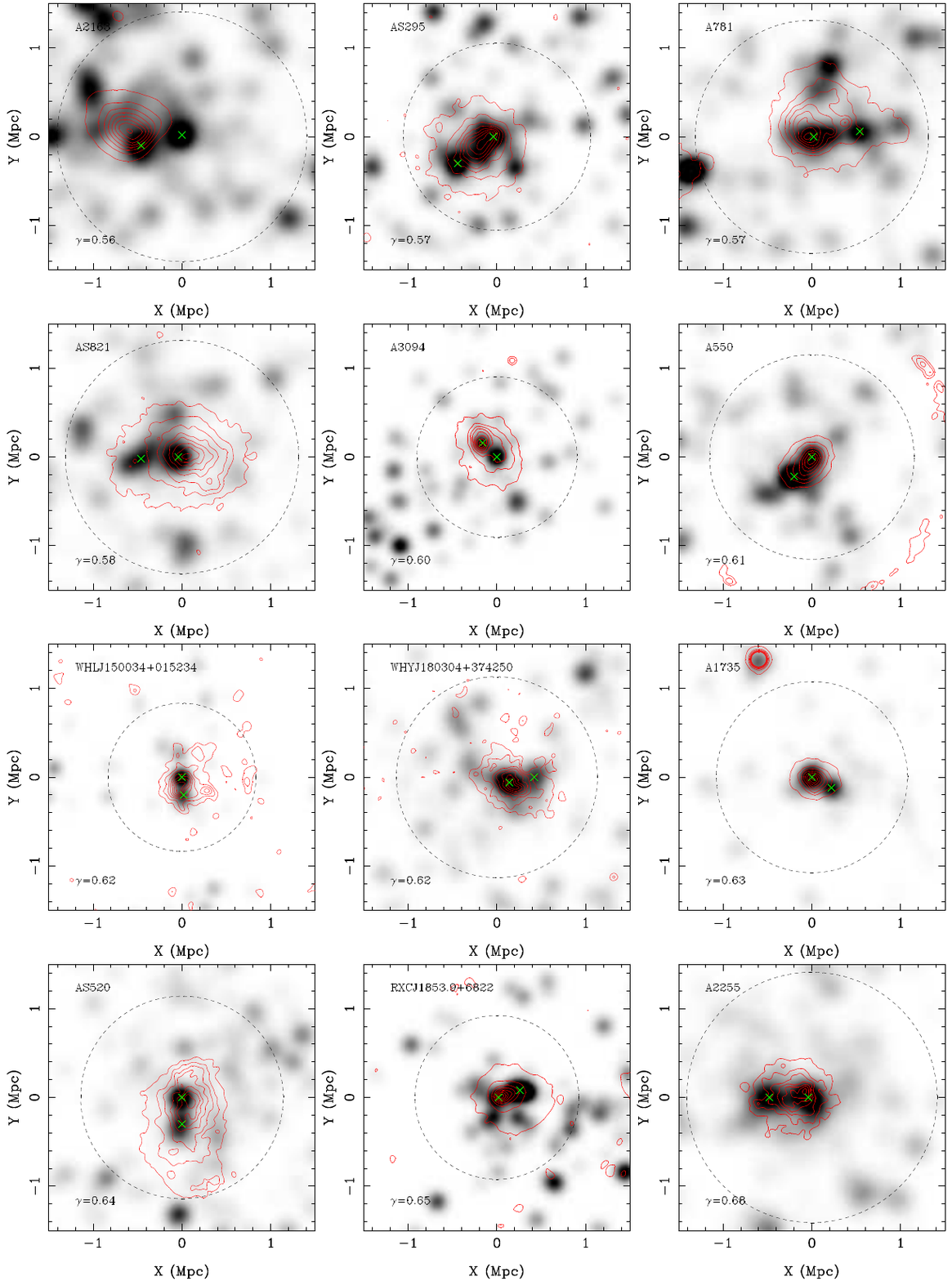}~%
\caption{{\it --- continued} }
\end{figure*}

\addtocounter{figure}{-1}  \begin{figure*}
\centering
\includegraphics[width=0.95\textwidth]{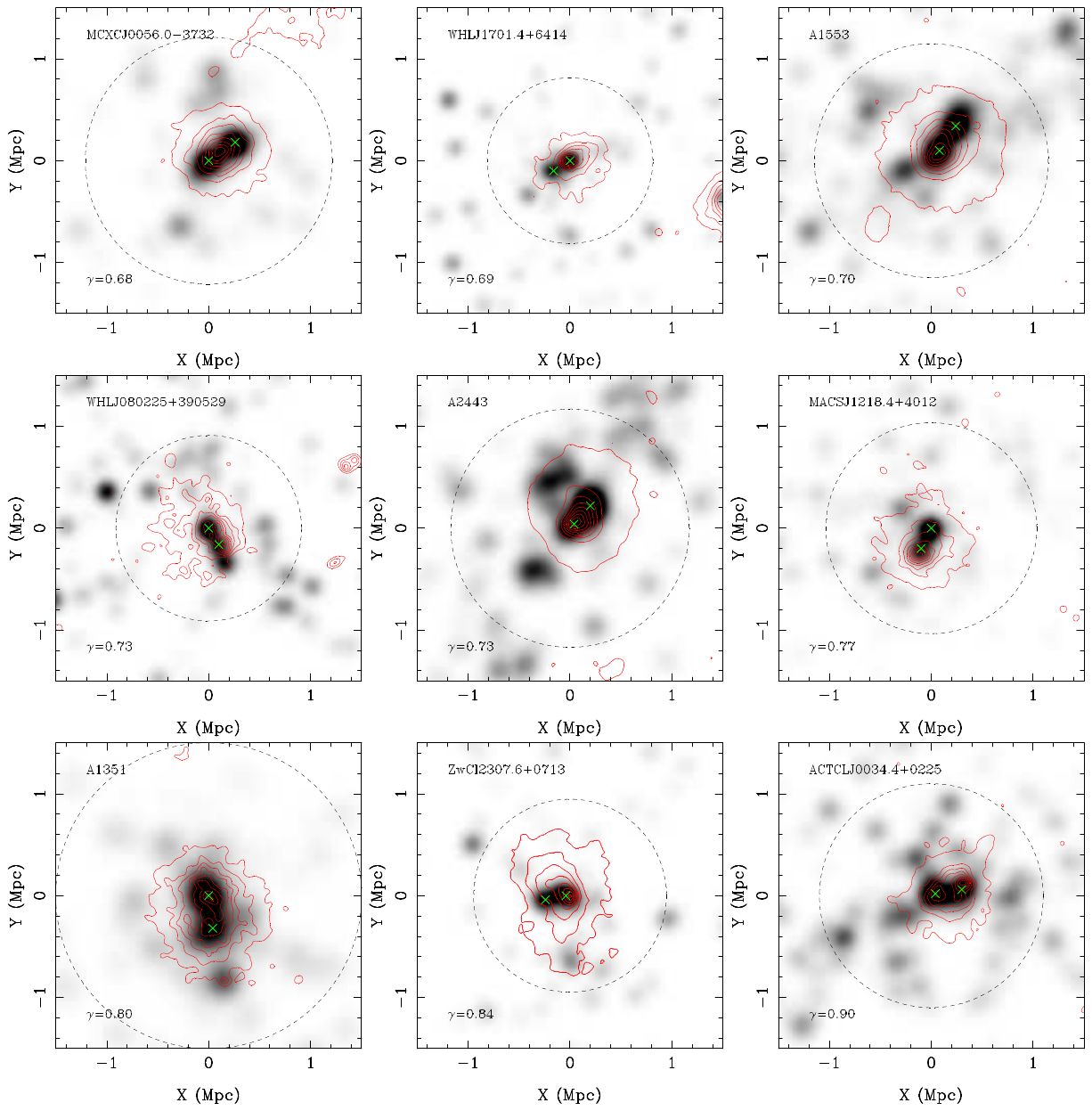}~%
\caption{{\it --- continued} }
\end{figure*}

%% file: post.tex
\begin{figure*}
\centering
\includegraphics[width=0.95\textwidth]{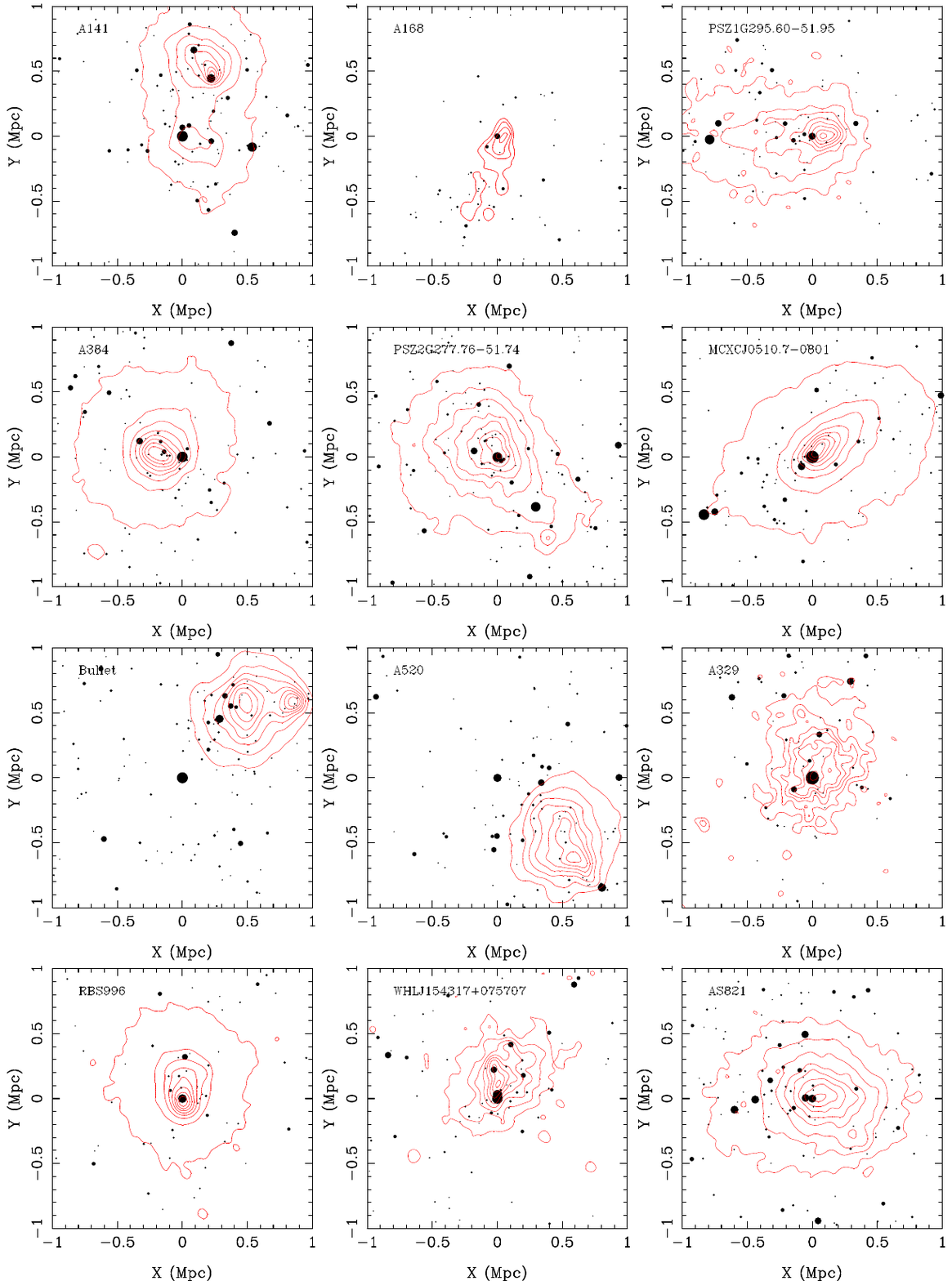}~%
\caption{The contours of X-ray emission from the {\it Chandra} \citep{yh20} and the {\it XMM-Newton} \citep{yhw22} are overlaid on the
 projected distribution of member galaxies for 152 clusters showing post-collision feature.}
\label{post}
\end{figure*}

\addtocounter{figure}{-1}
\begin{figure*}
\centering
\includegraphics[width=0.95\textwidth]{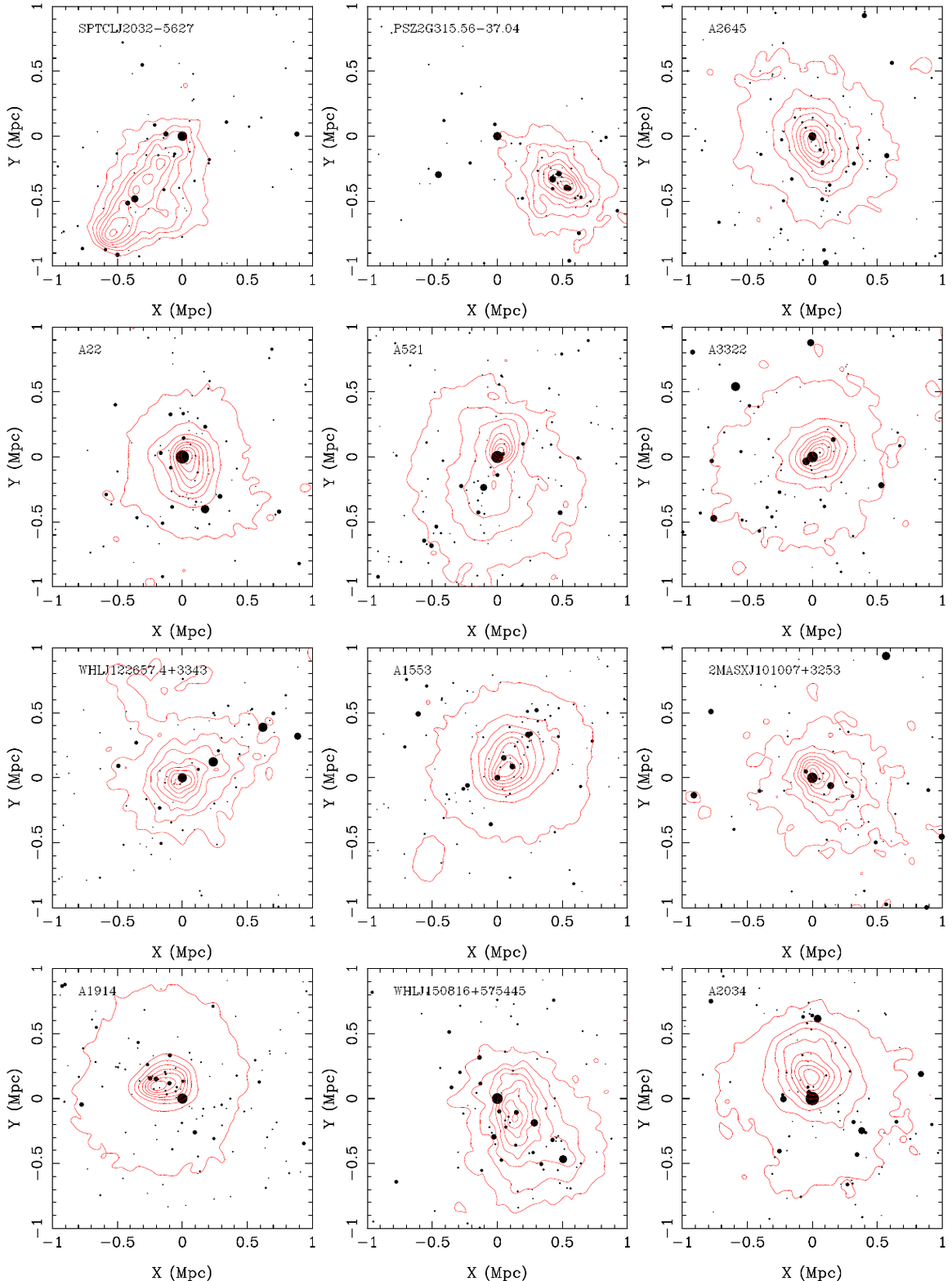}~%
\caption{{\it --- continued}} 
\end{figure*}

\addtocounter{figure}{-1}
\begin{figure*}
\centering
\includegraphics[width=0.95\textwidth]{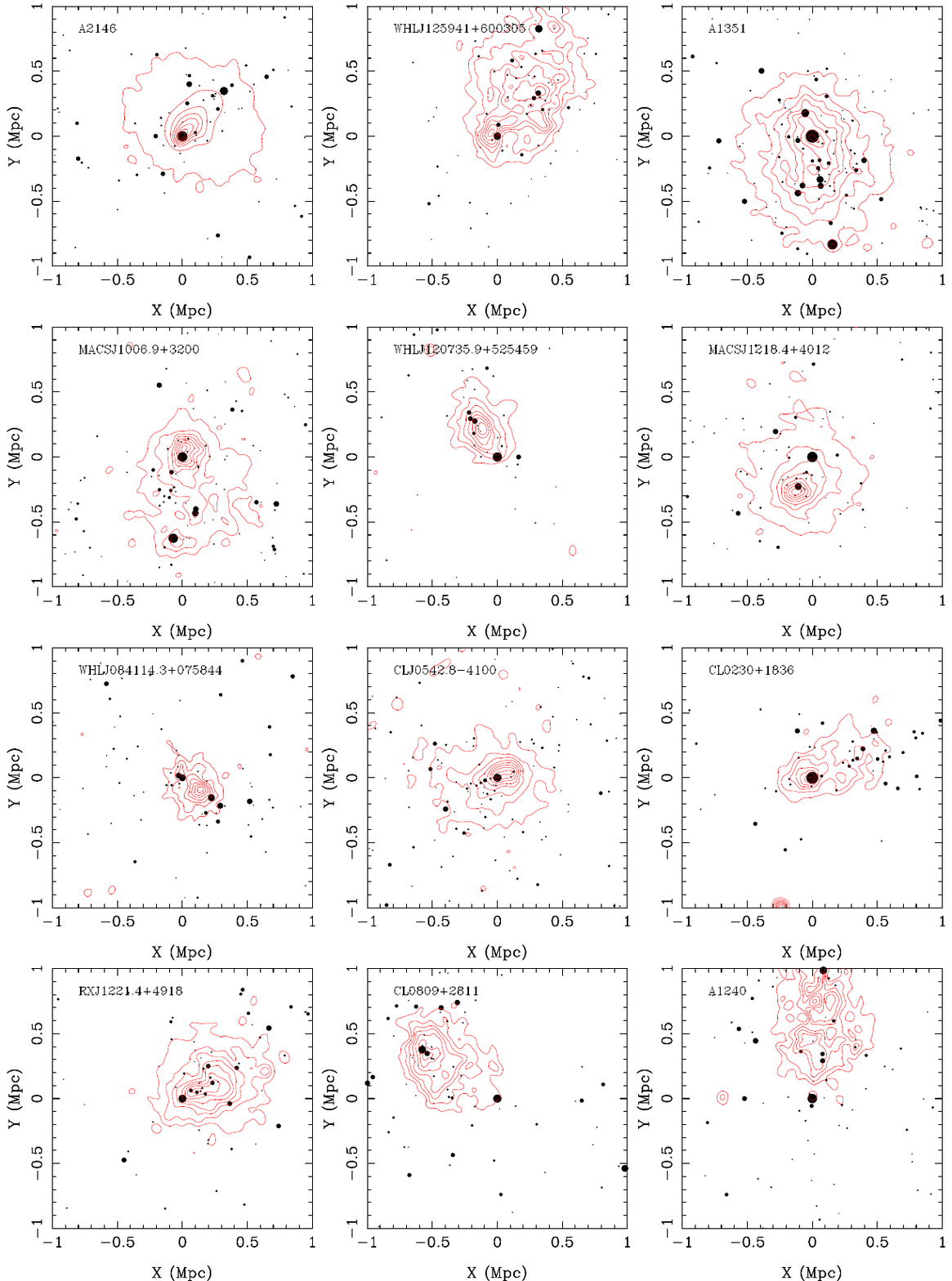}~%
\caption{{\it --- continued}} 
\end{figure*}

\addtocounter{figure}{-1}
\begin{figure*}
\centering
\includegraphics[width=0.95\textwidth]{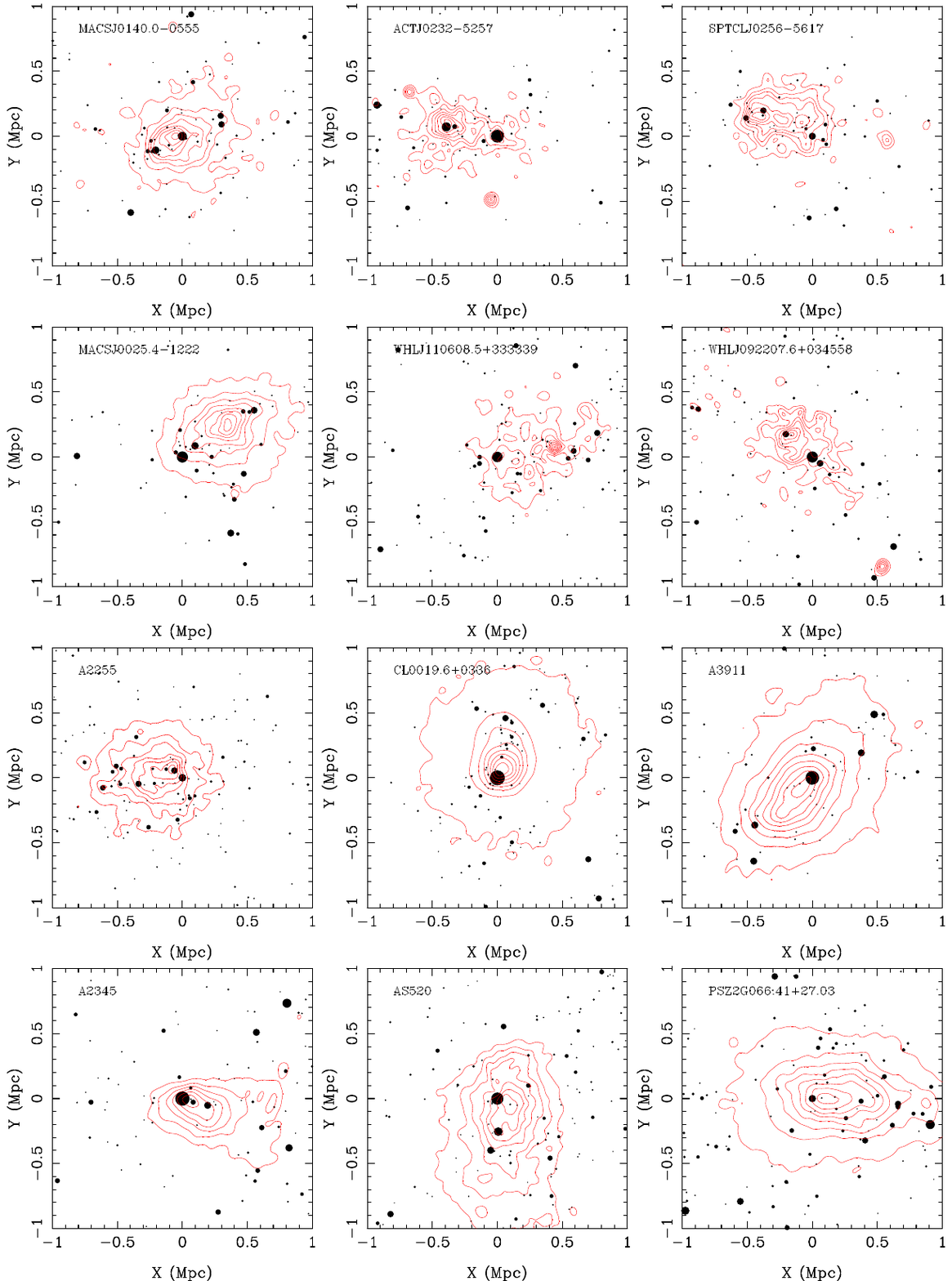}~%
\caption{{\it --- continued}} 
\end{figure*}

\addtocounter{figure}{-1}
\begin{figure*}
\centering
\includegraphics[width=0.95\textwidth]{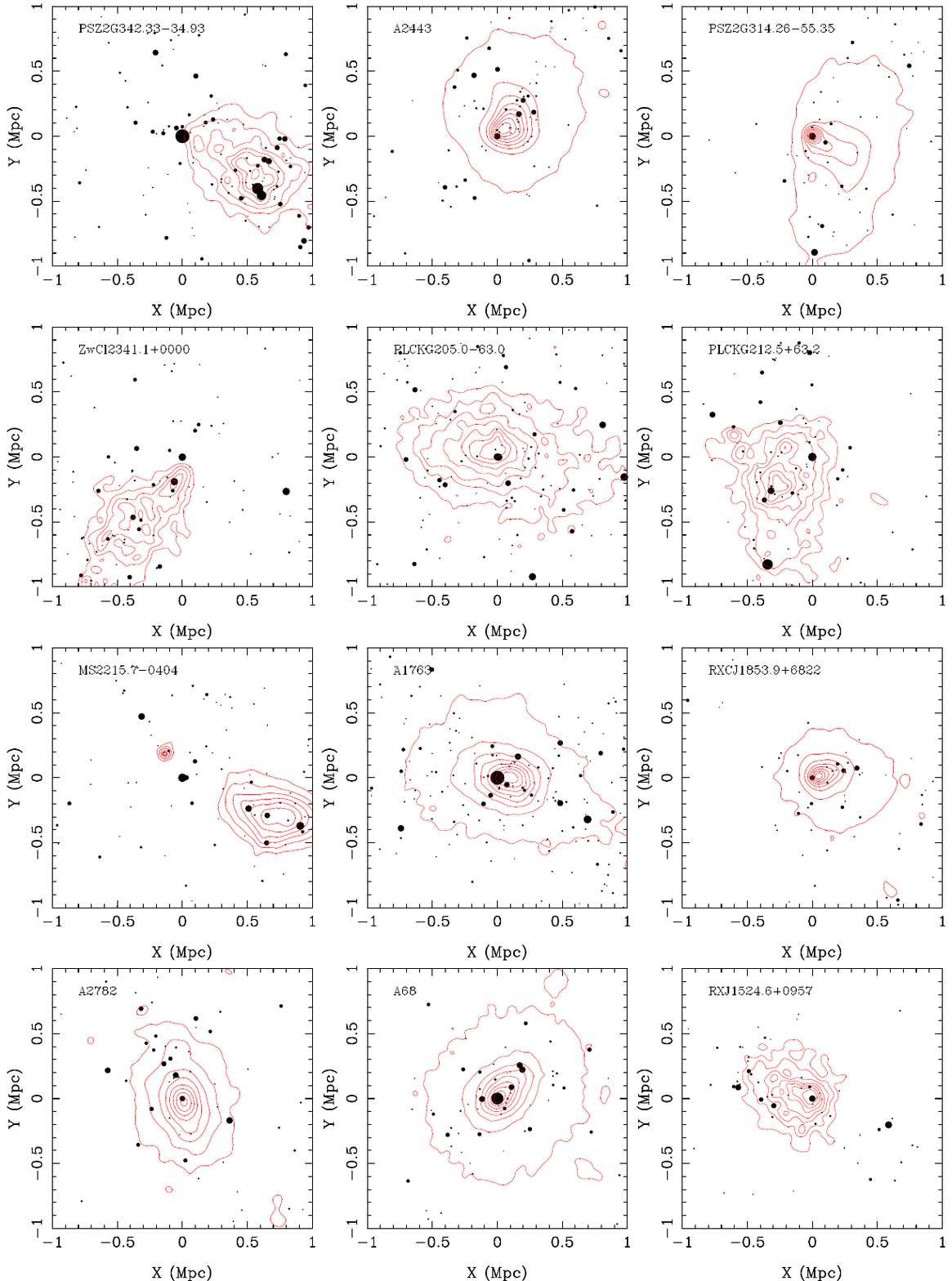}~%
\caption{{\it --- continued}} 
\end{figure*}

\addtocounter{figure}{-1}
\begin{figure*}
\centering
\includegraphics[width=0.95\textwidth]{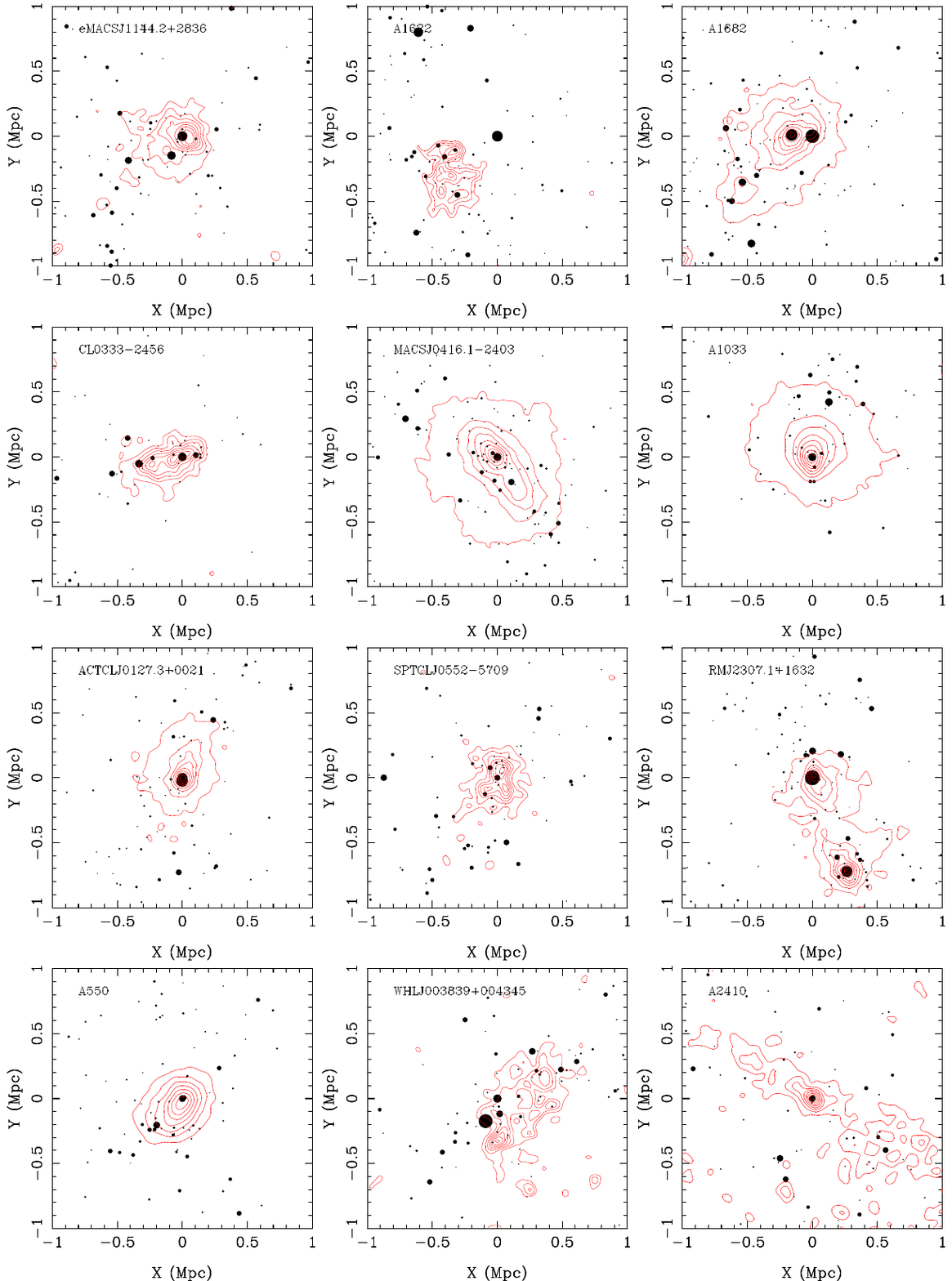}~%
\caption{{\it --- continued}} 
\end{figure*}

\addtocounter{figure}{-1}
\begin{figure*}
\centering
\includegraphics[width=0.95\textwidth]{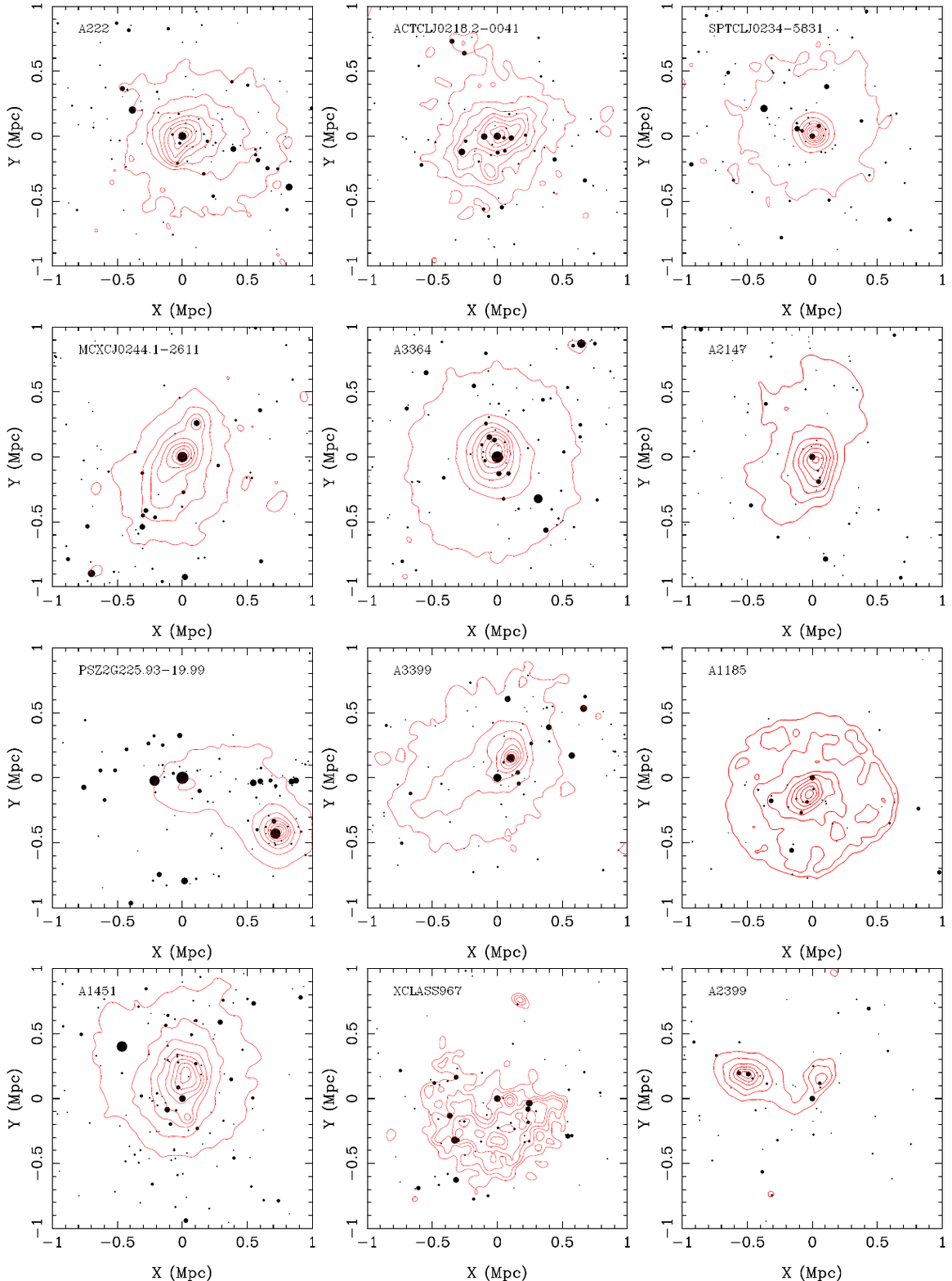}~%
\caption{{\it --- continued}} 
\end{figure*}

\addtocounter{figure}{-1}
\begin{figure*}
\centering
\includegraphics[width=0.95\textwidth]{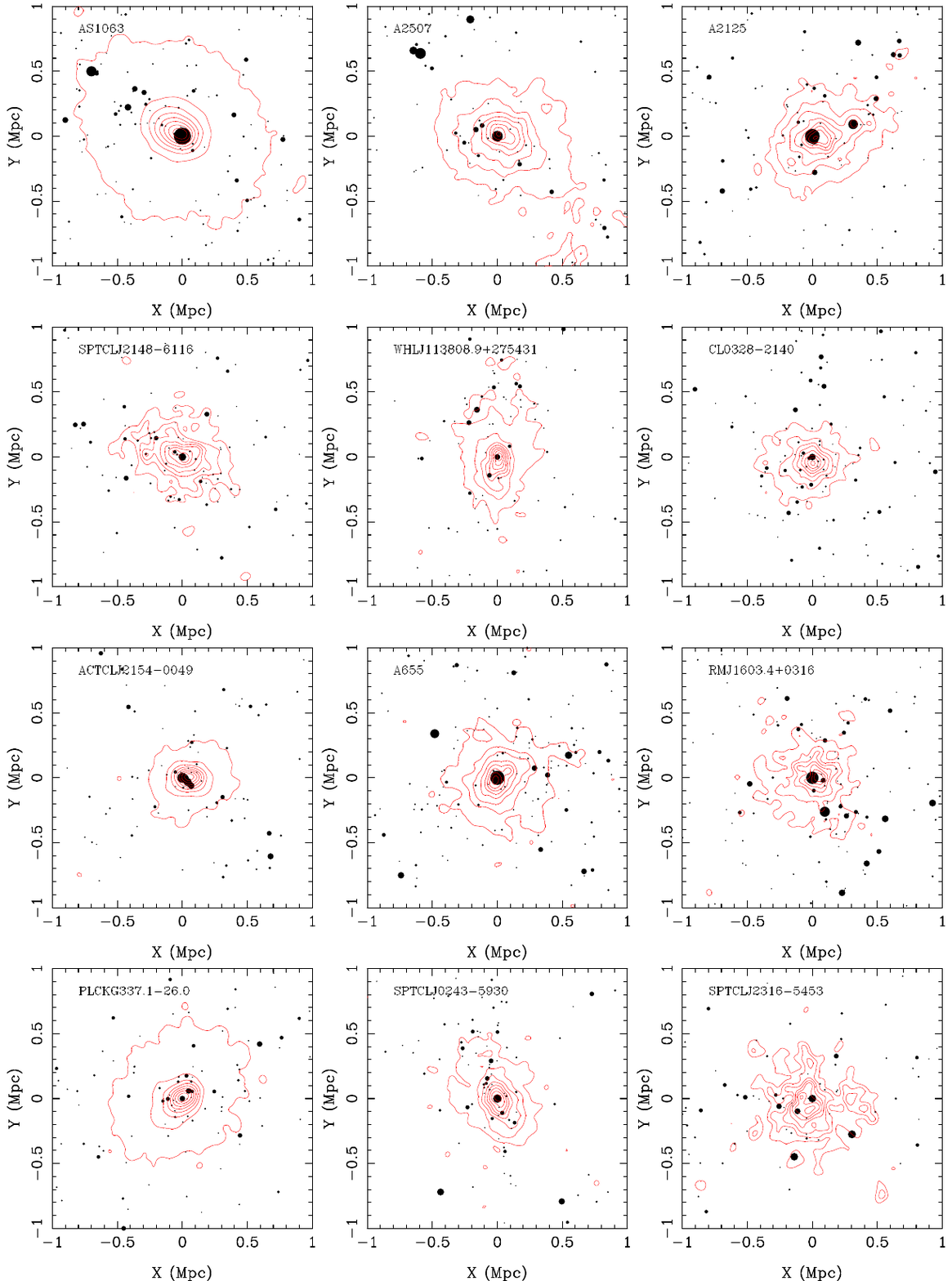}~%
\caption{{\it --- continued}} 
\end{figure*}

\addtocounter{figure}{-1}
\begin{figure*}
\centering
\includegraphics[width=0.95\textwidth]{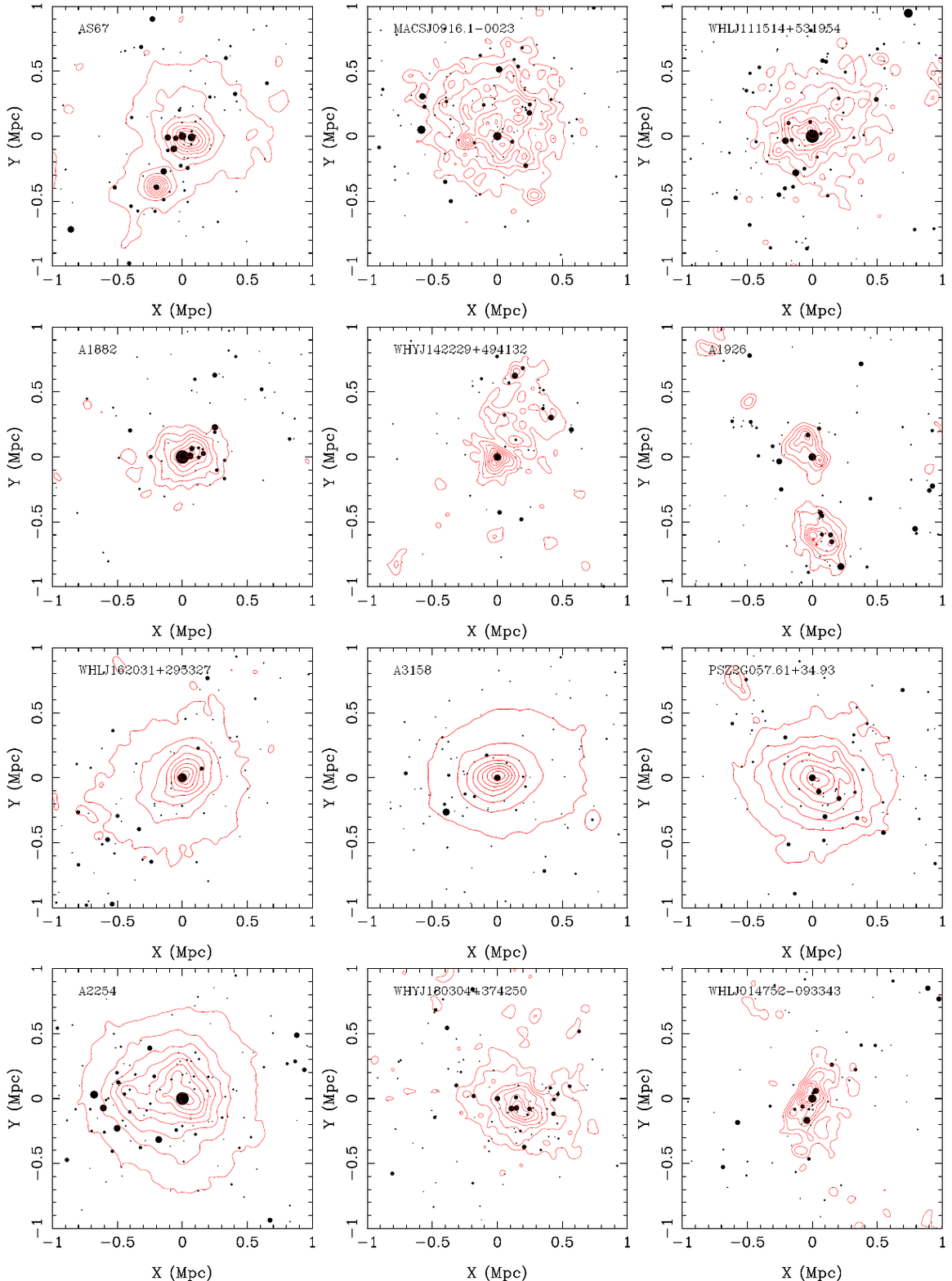}~%
\caption{{\it --- continued}} 
\end{figure*}

\addtocounter{figure}{-1}
\begin{figure*}
\centering
\includegraphics[width=0.95\textwidth]{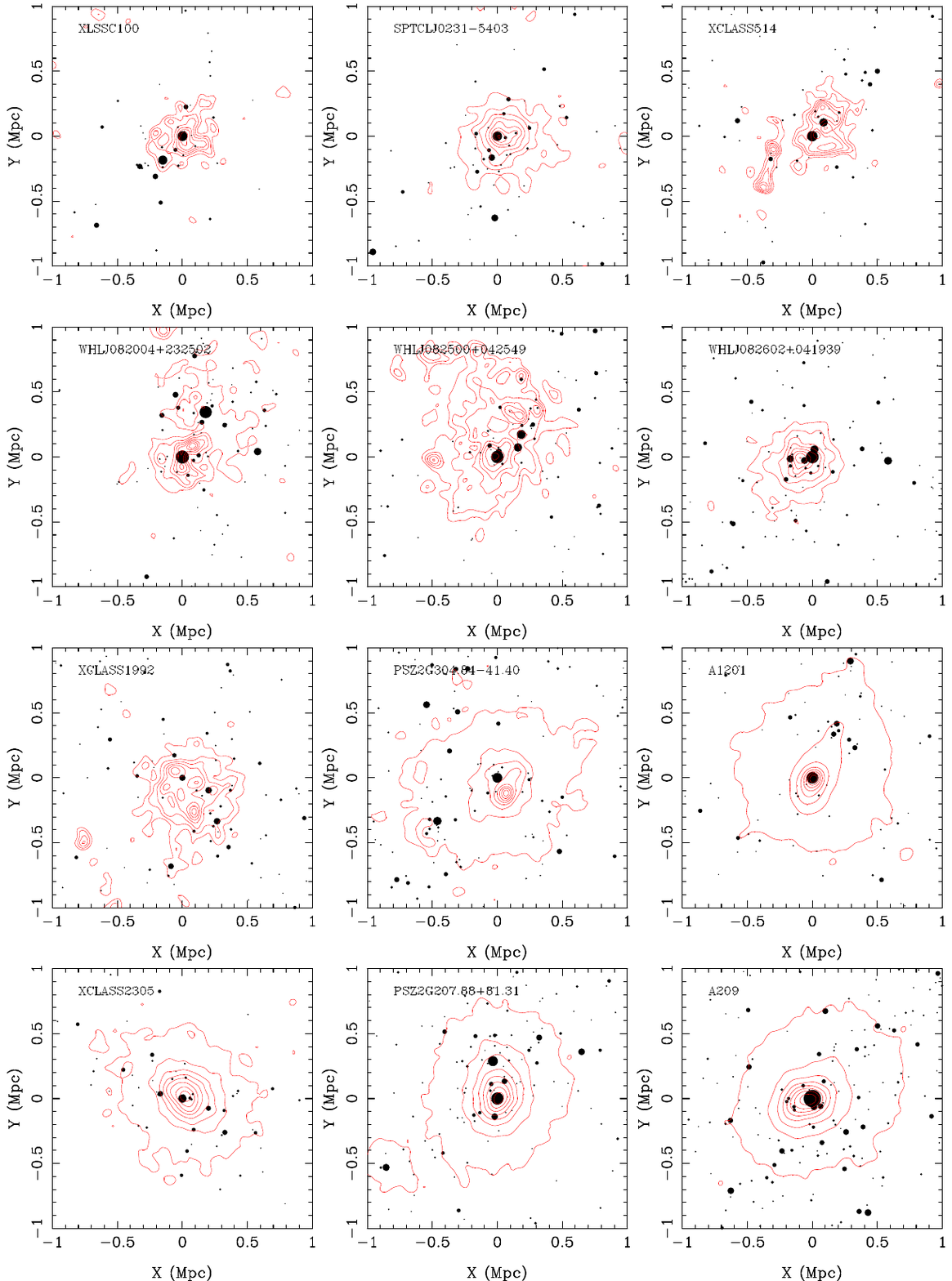}~%
\caption{{\it --- continued}} 
\end{figure*}

\addtocounter{figure}{-1}
\begin{figure*}
\centering
\includegraphics[width=0.95\textwidth]{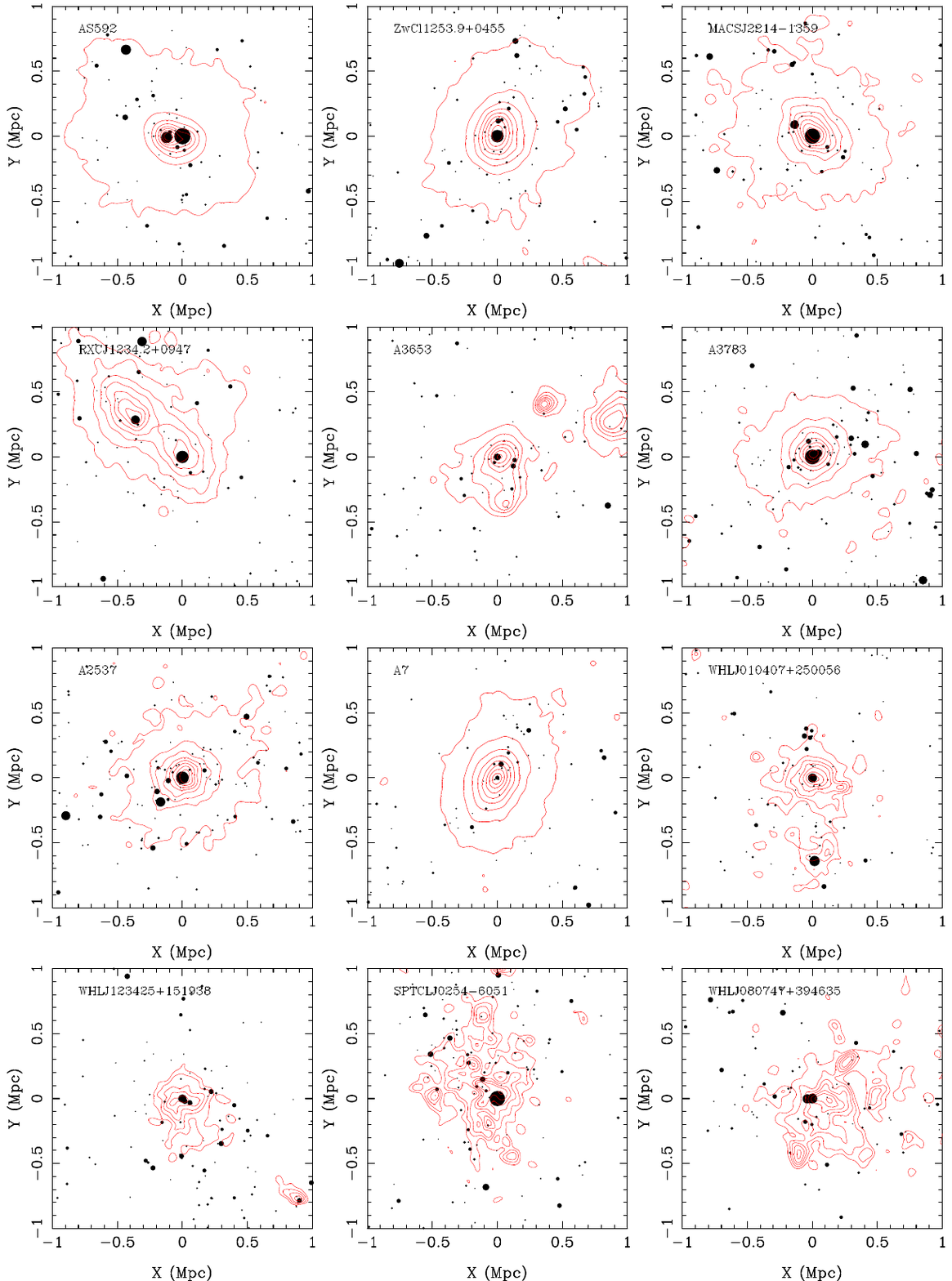}~%
\caption{{\it --- continued}} 
\end{figure*}

\addtocounter{figure}{-1}
\begin{figure*}
\centering
\includegraphics[width=0.95\textwidth]{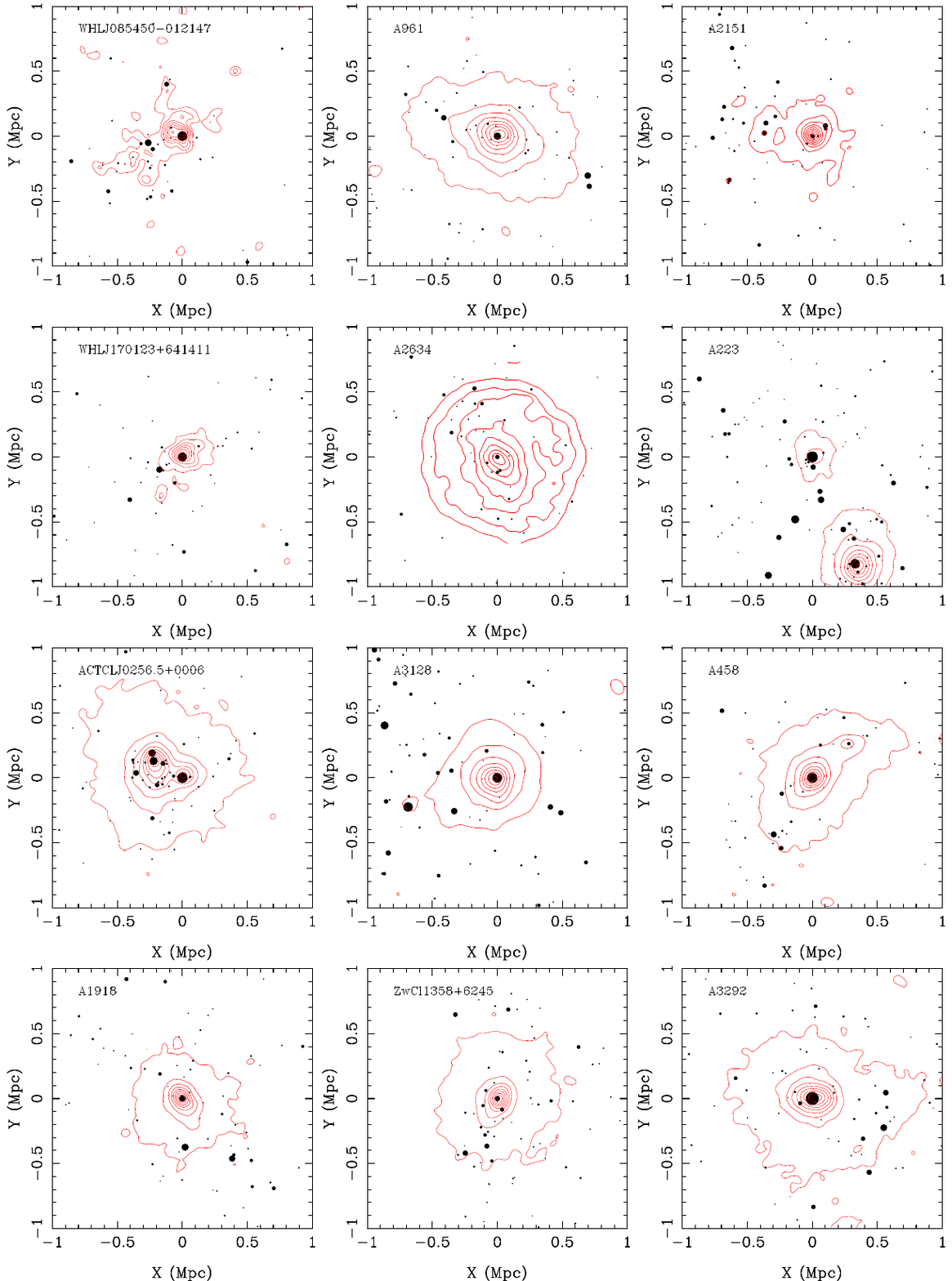}~%
\caption{{\it --- continued}} 
\end{figure*}

\addtocounter{figure}{-1}
\begin{figure*}
\centering
\includegraphics[width=0.95\textwidth]{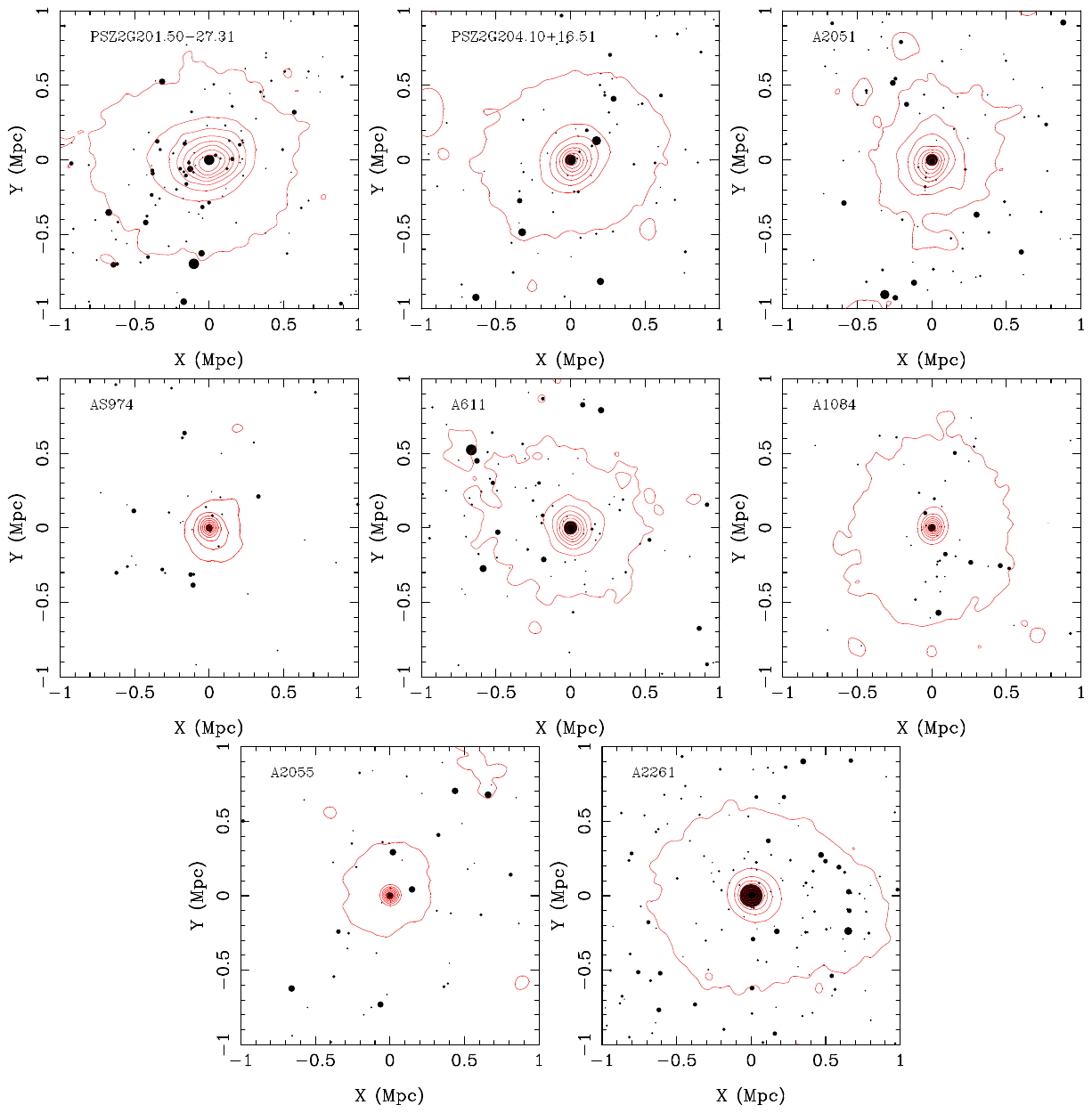}~%
\caption{{\it --- continued}} 
\end{figure*}

%% file: ms.bbl
\begin{thebibliography}{}
\makeatletter
\relax
\def\mn@urlcharsother{\let\do\@makeother \do\$\do\&\do\#\do\^\do\_\do\%\do\~}
\def\mn@doi{\begingroup\mn@urlcharsother \@ifnextchar [ {\mn@doi@}
  {\mn@doi@[]}}
\def\mn@doi@[#1]#2{\def\@tempa{#1}\ifx\@tempa\@empty \href
  {http://dx.doi.org/#2} {doi:#2}\else \href {http://dx.doi.org/#2} {#1}\fi
  \endgroup}
\def\mn@eprint#1#2{\mn@eprint@#1:#2::\@nil}
\def\mn@eprint@arXiv#1{\href {http://arxiv.org/abs/#1} {{\tt arXiv:#1}}}
\def\mn@eprint@dblp#1{\href {http://dblp.uni-trier.de/rec/bibtex/#1.xml}
  {dblp:#1}}
\def\mn@eprint@#1:#2:#3:#4\@nil{\def\@tempa {#1}\def\@tempb {#2}\def\@tempc
  {#3}\ifx \@tempc \@empty \let \@tempc \@tempb \let \@tempb \@tempa \fi \ifx
  \@tempb \@empty \def\@tempb {arXiv}\fi \@ifundefined
  {mn@eprint@\@tempb}{\@tempb:\@tempc}{\expandafter \expandafter \csname
  mn@eprint@\@tempb\endcsname \expandafter{\@tempc}}}

\bibitem[\protect\citeauthoryear{{Akamatsu} et~al.,}{{Akamatsu}
  et~al.}{2016}]{ags+16}
{Akamatsu} H.,  et~al., 2016, \mn@doi [\aap] {10.1051/0004-6361/201629275},
  \href {https://ui.adsabs.harvard.edu/abs/2016A&A...593L...7A} {593, L7}

\bibitem[\protect\citeauthoryear{{Altay}, {Colberg}  \& {Croft}}{{Altay}
  et~al.}{2006}]{acc+06}
{Altay} G.,  {Colberg} J.~M.,   {Croft} R. A.~C.,  2006, \mn@doi [\mnras]
  {10.1111/j.1365-2966.2006.10555.x}, \href
  {https://ui.adsabs.harvard.edu/abs/2006MNRAS.370.1422A} {370, 1422}

\bibitem[\protect\citeauthoryear{{Bagchi}, {Durret}, {Neto}  \&
  {Paul}}{{Bagchi} et~al.}{2006}]{bdn+06}
{Bagchi} J.,  {Durret} F.,  {Neto} G. B.~L.,   {Paul} S.,  2006, \mn@doi
  [Science] {10.1126/science.1131189}, \href
  {https://ui.adsabs.harvard.edu/abs/2006Sci...314..791B} {314, 791}

\bibitem[\protect\citeauthoryear{{Banerjee}, {Szabo}, {Pierpaoli}, {Franco},
  {Ortiz}, {Oramas}  \& {Tornello}}{{Banerjee} et~al.}{2018}]{bsp+18}
{Banerjee} P.,  {Szabo} T.,  {Pierpaoli} E.,  {Franco} G.,  {Ortiz} M.,
  {Oramas} A.,   {Tornello} B.,  2018, \mn@doi [\na]
  {10.1016/j.newast.2017.07.008}, \href
  {http://adsabs.harvard.edu/abs/2018NewA...58...61B} {58, 61}

\bibitem[\protect\citeauthoryear{{Barrena}, {Girardi}, {Boschin}  \&
  {Das{\'{\i}}}}{{Barrena} et~al.}{2009}]{bgb+09}
{Barrena} R.,  {Girardi} M.,  {Boschin} W.,   {Das{\'{\i}}} M.,  2009, \mn@doi
  [\aap] {10.1051/0004-6361/200911788}, \href
  {http://adsabs.harvard.edu/abs/2009A%26A...503..357B} {503, 357}

\bibitem[\protect\citeauthoryear{{Barrena}, {B{\"o}hringer}  \&
  {Chon}}{{Barrena} et~al.}{2022}]{bbc22}
{Barrena} R.,  {B{\"o}hringer} H.,   {Chon} G.,  2022, \mn@doi [\aap]
  {10.1051/0004-6361/202243418}, \href
  {https://ui.adsabs.harvard.edu/abs/2022A&A...663A..78B} {663, A78}

\bibitem[\protect\citeauthoryear{{Beers}, {Geller}  \& {Huchra}}{{Beers}
  et~al.}{1982}]{bgh82}
{Beers} T.~C.,  {Geller} M.~J.,   {Huchra} J.~P.,  1982, \mn@doi [\apj]
  {10.1086/159958}, \href {http://adsabs.harvard.edu/abs/1982ApJ...257...23B}
  {257, 23}

\bibitem[\protect\citeauthoryear{{Belsole}, {Pratt}, {Sauvageot}  \&
  {Bourdin}}{{Belsole} et~al.}{2004}]{bps+04}
{Belsole} E.,  {Pratt} G.~W.,  {Sauvageot} J.-L.,   {Bourdin} H.,  2004,
  \mn@doi [\aap] {10.1051/0004-6361:20034239}, \href
  {http://adsabs.harvard.edu/abs/2004A%26A...415..821B} {415, 821}

\bibitem[\protect\citeauthoryear{{Benson}, {Wittman}, {Golovich}, {Jee}, {van
  Weeren}  \& {Dawson}}{{Benson} et~al.}{2017}]{bwg+17}
{Benson} B.,  {Wittman} D.~M.,  {Golovich} N.,  {Jee} M.~J.,  {van Weeren}
  R.~J.,   {Dawson} W.~A.,  2017, \mn@doi [\apj] {10.3847/1538-4357/aa6d66},
  \href {http://adsabs.harvard.edu/abs/2017ApJ...841....7B} {841, 7}

\bibitem[\protect\citeauthoryear{{Boschin}, {Girardi}, {Barrena}  \&
  {Nonino}}{{Boschin} et~al.}{2012a}]{bgb+12}
{Boschin} W.,  {Girardi} M.,  {Barrena} R.,   {Nonino} M.,  2012a, \mn@doi
  [\aap] {10.1051/0004-6361/201118076}, \href
  {https://ui.adsabs.harvard.edu/abs/2012A%26A...540A..43B} {540, A43}

\bibitem[\protect\citeauthoryear{{Boschin}, {Girardi}  \& {Barrena}}{{Boschin}
  et~al.}{2012b}]{bgb12}
{Boschin} W.,  {Girardi} M.,   {Barrena} R.,  2012b, \mn@doi [\aap]
  {10.1051/0004-6361/201219508}, \href
  {http://adsabs.harvard.edu/abs/2012A%26A...547A..44B} {547, A44}

\bibitem[\protect\citeauthoryear{{Brada{\v{c}}} et~al.,}{{Brada{\v{c}}}
  et~al.}{2006}]{bcg+06}
{Brada{\v{c}}} M.,  et~al., 2006, \mn@doi [\apj] {10.1086/508601}, \href
  {https://ui.adsabs.harvard.edu/abs/2006ApJ...652..937B} {652, 937}

\bibitem[\protect\citeauthoryear{{Bulbul} et~al.,}{{Bulbul}
  et~al.}{2016}]{brb+16}
{Bulbul} E.,  et~al., 2016, \mn@doi [\apj] {10.3847/0004-637X/818/2/131}, \href
  {http://adsabs.harvard.edu/abs/2016ApJ...818..131B} {818, 131}

\bibitem[\protect\citeauthoryear{{Buote} \& {Tsai}}{{Buote} \&
  {Tsai}}{1995}]{bt95}
{Buote} D.~A.,  {Tsai} J.~C.,  1995, \mn@doi [\apj] {10.1086/176326}, \href
  {http://adsabs.harvard.edu/abs/1995ApJ...452..522B} {452, 522}

\bibitem[\protect\citeauthoryear{{Chon} \& {B{\"o}hringer}}{{Chon} \&
  {B{\"o}hringer}}{2015}]{cb15}
{Chon} G.,  {B{\"o}hringer} H.,  2015, \mn@doi [\aap]
  {10.1051/0004-6361/201425143}, \href
  {https://ui.adsabs.harvard.edu/abs/2015A&A...574A.132C} {574, A132}

\bibitem[\protect\citeauthoryear{{Clowe}, {Brada{\v{c}}}, {Gonzalez},
  {Markevitch}, {Randall}, {Jones}  \& {Zaritsky}}{{Clowe}
  et~al.}{2006}]{cbg+06}
{Clowe} D.,  {Brada{\v{c}}} M.,  {Gonzalez} A.~H.,  {Markevitch} M.,  {Randall}
  S.~W.,  {Jones} C.,   {Zaritsky} D.,  2006, \mn@doi [\apjl] {10.1086/508162},
  \href {https://ui.adsabs.harvard.edu/abs/2006ApJ...648L.109C} {648, L109}

\bibitem[\protect\citeauthoryear{{David} \& {Kempner}}{{David} \&
  {Kempner}}{2004}]{dk04}
{David} L.~P.,  {Kempner} J.,  2004, \mn@doi [\apj] {10.1086/423195}, \href
  {https://ui.adsabs.harvard.edu/abs/2004ApJ...613..831D} {613, 831}

\bibitem[\protect\citeauthoryear{{Dawson} et~al.,}{{Dawson}
  et~al.}{2015}]{djs+15}
{Dawson} W.~A.,  et~al., 2015, \mn@doi [\apj] {10.1088/0004-637X/805/2/143},
  \href {http://adsabs.harvard.edu/abs/2015ApJ...805..143D} {805, 143}

\bibitem[\protect\citeauthoryear{{Dey} et~al.,}{{Dey} et~al.}{2019}]{dsl+19}
{Dey} A.,  et~al., 2019, \mn@doi [\aj] {10.3847/1538-3881/ab089d}, \href
  {https://ui.adsabs.harvard.edu/abs/2019AJ....157..168D} {157, 168}

\bibitem[\protect\citeauthoryear{{Dietrich}, {Schneider}, {Clowe},
  {Romano-D{\'\i}az}  \& {Kerp}}{{Dietrich} et~al.}{2005}]{dsc+05}
{Dietrich} J.~P.,  {Schneider} P.,  {Clowe} D.,  {Romano-D{\'\i}az} E.,
  {Kerp} J.,  2005, \mn@doi [\aap] {10.1051/0004-6361:20041523}, \href
  {https://ui.adsabs.harvard.edu/abs/2005A&A...440..453D} {440, 453}

\bibitem[\protect\citeauthoryear{{Dressler} \& {Shectman}}{{Dressler} \&
  {Shectman}}{1988}]{ds88}
{Dressler} A.,  {Shectman} S.~A.,  1988, \mn@doi [\aj] {10.1086/114694}, \href
  {http://adsabs.harvard.edu/abs/1988AJ.....95..985D} {95, 985}

\bibitem[\protect\citeauthoryear{{Driver} et~al.,}{{Driver}
  et~al.}{2022}]{dbr+22}
{Driver} S.~P.,  et~al., 2022, \mn@doi [\mnras] {10.1093/mnras/stac472}, \href
  {https://ui.adsabs.harvard.edu/abs/2022MNRAS.513..439D} {513, 439}

\bibitem[\protect\citeauthoryear{{Duchesne}, {Johnston-Hollitt}, {Bartalucci},
  {Hodgson}  \& {Pratt}}{{Duchesne} et~al.}{2021a}]{djb+21}
{Duchesne} S.~W.,  {Johnston-Hollitt} M.,  {Bartalucci} I.,  {Hodgson} T.,
  {Pratt} G.~W.,  2021a, \mn@doi [\pasa] {10.1017/pasa.2020.51}, \href
  {https://ui.adsabs.harvard.edu/abs/2021PASA...38....5D} {38, e005}

\bibitem[\protect\citeauthoryear{{Duchesne}, {Johnston-Hollitt}  \&
  {Wilber}}{{Duchesne} et~al.}{2021b}]{djw+21}
{Duchesne} S.~W.,  {Johnston-Hollitt} M.,   {Wilber} A.~G.,  2021b, \mn@doi
  [\pasa] {10.1017/pasa.2021.24}, \href
  {https://ui.adsabs.harvard.edu/abs/2021PASA...38...31D} {38, e031}

\bibitem[\protect\citeauthoryear{{Dwarakanath}, {Parekh}, {Kale}  \&
  {George}}{{Dwarakanath} et~al.}{2018}]{dpk+18}
{Dwarakanath} K.~S.,  {Parekh} V.,  {Kale} R.,   {George} L.~T.,  2018, \mn@doi
  [\mnras] {10.1093/mnras/sty744}, \href
  {https://ui.adsabs.harvard.edu/abs/2018MNRAS.477..957D} {477, 957}

\bibitem[\protect\citeauthoryear{{Feretti}, {Giovannini}, {Govoni}  \&
  {Murgia}}{{Feretti} et~al.}{2012}]{fgg+12}
{Feretti} L.,  {Giovannini} G.,  {Govoni} F.,   {Murgia} M.,  2012, \mn@doi
  [\aapr] {10.1007/s00159-012-0054-z}, \href
  {http://adsabs.harvard.edu/abs/2012A%26ARv..20...54F} {20, 54}

\bibitem[\protect\citeauthoryear{{Finner} et~al.,}{{Finner}
  et~al.}{2017}]{fjg+17}
{Finner} K.,  et~al., 2017, \mn@doi [\apj] {10.3847/1538-4357/aa998c}, \href
  {http://adsabs.harvard.edu/abs/2017ApJ...851...46F} {851, 46}

\bibitem[\protect\citeauthoryear{{Ge}, {Wang}, {Tripp}, {Li}, {Gu}  \&
  {Ji}}{{Ge} et~al.}{2016}]{gwt+16}
{Ge} C.,  {Wang} Q.~D.,  {Tripp} T.~M.,  {Li} Z.,  {Gu} Q.,   {Ji} L.,  2016,
  \mn@doi [\mnras] {10.1093/mnras/stw599}, \href
  {https://ui.adsabs.harvard.edu/abs/2016MNRAS.459..366G} {459, 366}

\bibitem[\protect\citeauthoryear{{Giacintucci}, {Venturi}, {Cassano},
  {Dallacasa}  \& {Brunetti}}{{Giacintucci} et~al.}{2009}]{gvc+09}
{Giacintucci} S.,  {Venturi} T.,  {Cassano} R.,  {Dallacasa} D.,   {Brunetti}
  G.,  2009, \mn@doi [\apjl] {10.1088/0004-637X/704/1/L54}, \href
  {https://ui.adsabs.harvard.edu/abs/2009ApJ...704L..54G} {704, L54}

\bibitem[\protect\citeauthoryear{{Golovich}, {Dawson}, {Wittman}, {Ogrean},
  {van Weeren}  \& {Bonafede}}{{Golovich} et~al.}{2016}]{gdw+16}
{Golovich} N.,  {Dawson} W.~A.,  {Wittman} D.,  {Ogrean} G.,  {van Weeren} R.,
   {Bonafede} A.,  2016, \mn@doi [\apj] {10.3847/0004-637X/831/1/110}, \href
  {http://adsabs.harvard.edu/abs/2016ApJ...831..110G} {831, 110}

\bibitem[\protect\citeauthoryear{{Golovich}, {van Weeren}, {Dawson}, {Jee}  \&
  {Wittman}}{{Golovich} et~al.}{2017}]{gvd+17}
{Golovich} N.,  {van Weeren} R.~J.,  {Dawson} W.~A.,  {Jee} M.~J.,   {Wittman}
  D.,  2017, \mn@doi [\apj] {10.3847/1538-4357/aa667f}, \href
  {http://adsabs.harvard.edu/abs/2017ApJ...838..110G} {838, 110}

\bibitem[\protect\citeauthoryear{{Gonzalez} et~al.,}{{Gonzalez}
  et~al.}{2018}]{gdo+18}
{Gonzalez} E.~J.,  et~al., 2018, \mn@doi [\aap] {10.1051/0004-6361/201732003},
  \href {https://ui.adsabs.harvard.edu/abs/2018A%26A...611A..78G} {611, A78}

\bibitem[\protect\citeauthoryear{{Gray}, {Wolf}, {Meisenheimer}, {Taylor},
  {Dye}, {Borch}  \& {Kleinheinrich}}{{Gray} et~al.}{2004}]{gwm+04}
{Gray} M.~E.,  {Wolf} C.,  {Meisenheimer} K.,  {Taylor} A.,  {Dye} S.,  {Borch}
  A.,   {Kleinheinrich} M.,  2004, \mn@doi [\mnras]
  {10.1111/j.1365-2966.2004.07489.x}, \href
  {https://ui.adsabs.harvard.edu/abs/2004MNRAS.347L..73G} {347, L73}

\bibitem[\protect\citeauthoryear{{Gu} et~al.,}{{Gu} et~al.}{2019}]{gas+19}
{Gu} L.,  et~al., 2019, \mn@doi [Nature Astronomy] {10.1038/s41550-019-0798-8},
  \href {https://ui.adsabs.harvard.edu/abs/2019NatAs...3..838G} {3, 838}

\bibitem[\protect\citeauthoryear{{Hansen}, {Hassani}, {Lombriser}  \&
  {Kunz}}{{Hansen} et~al.}{2020}]{hhl+20}
{Hansen} S.~H.,  {Hassani} F.,  {Lombriser} L.,   {Kunz} M.,  2020, \mn@doi
  [\jcap] {10.1088/1475-7516/2020/01/048}, \href
  {https://ui.adsabs.harvard.edu/abs/2020JCAP...01..048H} {2020, 048}

\bibitem[\protect\citeauthoryear{{Hayashi} \& {White}}{{Hayashi} \&
  {White}}{2006}]{hw06}
{Hayashi} E.,  {White} S. D.~M.,  2006, \mn@doi [\mnras]
  {10.1111/j.1745-3933.2006.00184.x}, \href
  {https://ui.adsabs.harvard.edu/abs/2006MNRAS.370L..38H} {370, L38}

\bibitem[\protect\citeauthoryear{{Hwang} \& {Lee}}{{Hwang} \&
  {Lee}}{2009}]{hl09}
{Hwang} H.~S.,  {Lee} M.~G.,  2009, \mn@doi [\mnras]
  {10.1111/j.1365-2966.2009.15100.x}, \href
  {http://adsabs.harvard.edu/abs/2009MNRAS.397.2111H} {397, 2111}

\bibitem[\protect\citeauthoryear{{HyeongHan} et~al.,}{{HyeongHan}
  et~al.}{2020}]{hjr+20}
{HyeongHan} K.,  et~al., 2020, \mn@doi [\apj] {10.3847/1538-4357/aba742}, \href
  {https://ui.adsabs.harvard.edu/abs/2020ApJ...900..127H} {900, 127}

\bibitem[\protect\citeauthoryear{{Ichinohe}, {Werner}, {Simionescu}, {Allen},
  {Canning}, {Ehlert}, {Mernier}  \& {Takahashi}}{{Ichinohe}
  et~al.}{2015}]{iws+15}
{Ichinohe} Y.,  {Werner} N.,  {Simionescu} A.,  {Allen} S.~W.,  {Canning}
  R.~E.~A.,  {Ehlert} S.,  {Mernier} F.,   {Takahashi} T.,  2015, \mn@doi
  [\mnras] {10.1093/mnras/stv217}, \href
  {https://ui.adsabs.harvard.edu/abs/2015MNRAS.448.2971I} {448, 2971}

\bibitem[\protect\citeauthoryear{{Jee}, {Hughes}, {Menanteau}, {Sif{\'o}n},
  {Mandelbaum}, {Barrientos}, {Infante}  \& {Ng}}{{Jee} et~al.}{2014}]{jhm+14}
{Jee} M.~J.,  {Hughes} J.~P.,  {Menanteau} F.,  {Sif{\'o}n} C.,  {Mandelbaum}
  R.,  {Barrientos} L.~F.,  {Infante} L.,   {Ng} K.~Y.,  2014, \mn@doi [\apj]
  {10.1088/0004-637X/785/1/20}, \href
  {http://adsabs.harvard.edu/abs/2014ApJ...785...20J} {785, 20}

\bibitem[\protect\citeauthoryear{{Jee} et~al.,}{{Jee} et~al.}{2015}]{jsd+15}
{Jee} M.~J.,  et~al., 2015, \mn@doi [\apj] {10.1088/0004-637X/802/1/46}, \href
  {http://adsabs.harvard.edu/abs/2015ApJ...802...46J} {802, 46}

\bibitem[\protect\citeauthoryear{{Jee}, {Dawson}, {Stroe}, {Wittman}, {van
  Weeren}, {Br{\"u}ggen}, {Brada{\v c}}  \& {R{\"o}ttgering}}{{Jee}
  et~al.}{2016}]{jds+16}
{Jee} M.~J.,  {Dawson} W.~A.,  {Stroe} A.,  {Wittman} D.,  {van Weeren} R.~J.,
  {Br{\"u}ggen} M.,  {Brada{\v c}} M.,   {R{\"o}ttgering} H.,  2016, \mn@doi
  [\apj] {10.3847/0004-637X/817/2/179}, \href
  {http://adsabs.harvard.edu/abs/2016ApJ...817..179J} {817, 179}

\bibitem[\protect\citeauthoryear{{Jing}, {Mo}, {Borner}  \& {Fang}}{{Jing}
  et~al.}{1995}]{jmb+95}
{Jing} Y.~P.,  {Mo} H.~J.,  {Borner} G.,   {Fang} L.~Z.,  1995, \mn@doi
  [\mnras] {10.1093/mnras/276.2.417}, \href
  {http://adsabs.harvard.edu/abs/1995MNRAS.276..417J} {276, 417}

\bibitem[\protect\citeauthoryear{{Kato}, {Nakazawa}, {Gu}, {Akahori},
  {Takizawa}, {Fujita}  \& {Makishima}}{{Kato} et~al.}{2015}]{kng+15}
{Kato} Y.,  {Nakazawa} K.,  {Gu} L.,  {Akahori} T.,  {Takizawa} M.,  {Fujita}
  Y.,   {Makishima} K.,  2015, \mn@doi [\pasj] {10.1093/pasj/psv029}, \href
  {https://ui.adsabs.harvard.edu/abs/2015PASJ...67...71K} {67, 71}

\bibitem[\protect\citeauthoryear{{Kempner} \& {David}}{{Kempner} \&
  {David}}{2004}]{kd04}
{Kempner} J.~C.,  {David} L.~P.,  2004, \mn@doi [\mnras]
  {10.1111/j.1365-2966.2004.07534.x}, \href
  {https://ui.adsabs.harvard.edu/abs/2004MNRAS.349..385K} {349, 385}

\bibitem[\protect\citeauthoryear{{Kim}, {Jee}, {Finner}, {Golovich}, {Wittman},
  {van Weeren}  \& {Dawson}}{{Kim} et~al.}{2019}]{kjf+19}
{Kim} M.,  {Jee} M.~J.,  {Finner} K.,  {Golovich} N.,  {Wittman} D.~M.,  {van
  Weeren} R.~J.,   {Dawson} W.~A.,  2019, \mn@doi [\apj]
  {10.3847/1538-4357/ab0d7c}, \href
  {https://ui.adsabs.harvard.edu/abs/2019ApJ...874..143K} {874, 143}

\bibitem[\protect\citeauthoryear{{Kim} et~al.,}{{Kim} et~al.}{2021}]{kjh+21}
{Kim} J.,  et~al., 2021, \mn@doi [\apj] {10.3847/1538-4357/ac294f}, \href
  {https://ui.adsabs.harvard.edu/abs/2021ApJ...923..101K} {923, 101}

\bibitem[\protect\citeauthoryear{{Knowles} et~al.,}{{Knowles}
  et~al.}{2016}]{kib+16}
{Knowles} K.,  et~al., 2016, \mn@doi [\mnras] {10.1093/mnras/stw795}, \href
  {http://adsabs.harvard.edu/abs/2016MNRAS.459.4240K} {459, 4240}

\bibitem[\protect\citeauthoryear{{Knowles} et~al.,}{{Knowles}
  et~al.}{2022}]{kcr+22}
{Knowles} K.,  et~al., 2022, \mn@doi [\aap] {10.1051/0004-6361/202141488},
  \href {https://ui.adsabs.harvard.edu/abs/2022A&A...657A..56K} {657, A56}

\bibitem[\protect\citeauthoryear{{Koester} et~al.,}{{Koester}
  et~al.}{2007}]{kma+07b}
{Koester} B.~P.,  et~al., 2007, \mn@doi [\apj] {10.1086/509599}, \href
  {http://ads.ari.uni-heidelberg.de/abs/2007ApJ...660..239K} {660, 239}

\bibitem[\protect\citeauthoryear{{Kraljic} \& {Sarkar}}{{Kraljic} \&
  {Sarkar}}{2015}]{ks15}
{Kraljic} D.,  {Sarkar} S.,  2015, \mn@doi [\jcap]
  {10.1088/1475-7516/2015/04/050}, \href
  {https://ui.adsabs.harvard.edu/abs/2015JCAP...04..050K} {2015, 050}

\bibitem[\protect\citeauthoryear{{Kravtsov} \& {Borgani}}{{Kravtsov} \&
  {Borgani}}{2012}]{kb12}
{Kravtsov} A.~V.,  {Borgani} S.,  2012, \mn@doi [\araa]
  {10.1146/annurev-astro-081811-125502}, \href
  {http://adsabs.harvard.edu/abs/2012ARA%26A..50..353K} {50, 353}

\bibitem[\protect\citeauthoryear{{Lauer}, {Postman}, {Strauss}, {Graves}  \&
  {Chisari}}{{Lauer} et~al.}{2014}]{lps+14}
{Lauer} T.~R.,  {Postman} M.,  {Strauss} M.~A.,  {Graves} G.~J.,   {Chisari}
  N.~E.,  2014, \mn@doi [\apj] {10.1088/0004-637X/797/2/82}, \href
  {https://ui.adsabs.harvard.edu/abs/2014ApJ...797...82L} {797, 82}

\bibitem[\protect\citeauthoryear{{Locatelli} et~al.,}{{Locatelli}
  et~al.}{2020}]{lrv+20}
{Locatelli} N.~T.,  et~al., 2020, \mn@doi [\mnras] {10.1093/mnrasl/slaa074},
  \href {https://ui.adsabs.harvard.edu/abs/2020MNRAS.496L..48L} {496, L48}

\bibitem[\protect\citeauthoryear{{Lopes}, {Trevisan}, {Lagan{\'a}}, {Durret},
  {Ribeiro}  \& {Rembold}}{{Lopes} et~al.}{2018}]{ltl+18}
{Lopes} P. A.~A.,  {Trevisan} M.,  {Lagan{\'a}} T.~F.,  {Durret} F.,  {Ribeiro}
  A.~L.~B.,   {Rembold} S.~B.,  2018, \mn@doi [\mnras] {10.1093/mnras/sty1374},
  \href {https://ui.adsabs.harvard.edu/abs/2018MNRAS.478.5473L} {478, 5473}

\bibitem[\protect\citeauthoryear{{Macario}, {Markevitch}, {Giacintucci},
  {Brunetti}, {Venturi}  \& {Murray}}{{Macario} et~al.}{2011}]{mmg+11}
{Macario} G.,  {Markevitch} M.,  {Giacintucci} S.,  {Brunetti} G.,  {Venturi}
  T.,   {Murray} S.~S.,  2011, \mn@doi [\apj] {10.1088/0004-637X/728/2/82},
  \href {http://adsabs.harvard.edu/abs/2011ApJ...728...82M} {728, 82}

\bibitem[\protect\citeauthoryear{{Mann} \& {Ebeling}}{{Mann} \&
  {Ebeling}}{2012}]{me12}
{Mann} A.~W.,  {Ebeling} H.,  2012, \mn@doi [\mnras]
  {10.1111/j.1365-2966.2011.20170.x}, \href
  {http://adsabs.harvard.edu/abs/2012MNRAS.420.2120M} {420, 2120}

\bibitem[\protect\citeauthoryear{{Markevitch} \& {Vikhlinin}}{{Markevitch} \&
  {Vikhlinin}}{2007}]{mv07}
{Markevitch} M.,  {Vikhlinin} A.,  2007, \mn@doi [\physrep]
  {10.1016/j.physrep.2007.01.001}, \href
  {https://ui.adsabs.harvard.edu/abs/2007PhR...443....1M} {443, 1}

\bibitem[\protect\citeauthoryear{{Markevitch}, {Gonzalez}, {David},
  {Vikhlinin}, {Murray}, {Forman}, {Jones}  \& {Tucker}}{{Markevitch}
  et~al.}{2002}]{mgd+02}
{Markevitch} M.,  {Gonzalez} A.~H.,  {David} L.,  {Vikhlinin} A.,  {Murray} S.,
   {Forman} W.,  {Jones} C.,   {Tucker} W.,  2002, \mn@doi [\apjl]
  {10.1086/339619}, \href
  {https://ui.adsabs.harvard.edu/abs/2002ApJ...567L..27M} {567, L27}

\bibitem[\protect\citeauthoryear{{Markevitch}, {Gonzalez}, {Clowe},
  {Vikhlinin}, {Forman}, {Jones}, {Murray}  \& {Tucker}}{{Markevitch}
  et~al.}{2004}]{mgc+04}
{Markevitch} M.,  {Gonzalez} A.~H.,  {Clowe} D.,  {Vikhlinin} A.,  {Forman} W.,
   {Jones} C.,  {Murray} S.,   {Tucker} W.,  2004, \mn@doi [\apj]
  {10.1086/383178}, \href {http://adsabs.harvard.edu/abs/2004ApJ...606..819M}
  {606, 819}

\bibitem[\protect\citeauthoryear{{Markevitch}, {Govoni}, {Brunetti}  \&
  {Jerius}}{{Markevitch} et~al.}{2005}]{mgb+05}
{Markevitch} M.,  {Govoni} F.,  {Brunetti} G.,   {Jerius} D.,  2005, \mn@doi
  [\apj] {10.1086/430695}, \href
  {https://ui.adsabs.harvard.edu/abs/2005ApJ...627..733M} {627, 733}

\bibitem[\protect\citeauthoryear{{Maurogordato} et~al.,}{{Maurogordato}
  et~al.}{2008}]{mcf+08}
{Maurogordato} S.,  et~al., 2008, \mn@doi [\aap] {10.1051/0004-6361:20077614},
  \href {https://ui.adsabs.harvard.edu/abs/2008A&A...481..593M} {481, 593}

\bibitem[\protect\citeauthoryear{{Menanteau} et~al.,}{{Menanteau}
  et~al.}{2012}]{mhs+12}
{Menanteau} F.,  et~al., 2012, \mn@doi [\apj] {10.1088/0004-637X/748/1/7},
  \href {http://adsabs.harvard.edu/abs/2012ApJ...748....7M} {748, 7}

\bibitem[\protect\citeauthoryear{{Mirakhor}, {Walker}  \& {Runge}}{{Mirakhor}
  et~al.}{2022}]{mwr22}
{Mirakhor} M.~S.,  {Walker} S.~A.,   {Runge} J.,  2022, \mn@doi [\mnras]
  {10.1093/mnras/stab2979}, \href
  {https://ui.adsabs.harvard.edu/abs/2022MNRAS.509.1109M} {509, 1109}

\bibitem[\protect\citeauthoryear{{Mohr}, {Evrard}, {Fabricant}  \&
  {Geller}}{{Mohr} et~al.}{1995}]{mef+95}
{Mohr} J.~J.,  {Evrard} A.~E.,  {Fabricant} D.~G.,   {Geller} M.~J.,  1995,
  \mn@doi [\apj] {10.1086/175852}, \href
  {http://adsabs.harvard.edu/abs/1995ApJ...447....8M} {447, 8}

\bibitem[\protect\citeauthoryear{{Mohr}, {Geller}  \& {Wegner}}{{Mohr}
  et~al.}{1996}]{mgw96}
{Mohr} J.~J.,  {Geller} M.~J.,   {Wegner} G.,  1996, \mn@doi [\aj]
  {10.1086/118144}, \href {http://adsabs.harvard.edu/abs/1996AJ....112.1816M}
  {112, 1816}

\bibitem[\protect\citeauthoryear{{Molino} et~al.,}{{Molino}
  et~al.}{2019}]{mcm+19}
{Molino} A.,  et~al., 2019, \mn@doi [\aap] {10.1051/0004-6361/201731348}, \href
  {http://adsabs.harvard.edu/abs/2019A%26A...622A.178M} {622, A178}

\bibitem[\protect\citeauthoryear{{Monteiro-Oliveira}, {Cypriano}, {Machado},
  {Lima Neto}, {Ribeiro}, {Sodr{\'e}}  \& {Dupke}}{{Monteiro-Oliveira}
  et~al.}{2017}]{mcm+17}
{Monteiro-Oliveira} R.,  {Cypriano} E.~S.,  {Machado} R.~E.~G.,  {Lima Neto}
  G.~B.,  {Ribeiro} A.~L.~B.,  {Sodr{\'e}} L.,   {Dupke} R.,  2017, \mn@doi
  [\mnras] {10.1093/mnras/stw3238}, \href
  {http://adsabs.harvard.edu/abs/2017MNRAS.466.2614M} {466, 2614}

\bibitem[\protect\citeauthoryear{{Monteiro-Oliveira}, {Cypriano}, {Vitorelli},
  {Ribeiro}, {Sodr{\'e}}, {Dupke}  \& {Mendes de Oliveira}}{{Monteiro-Oliveira}
  et~al.}{2018}]{mcv+18}
{Monteiro-Oliveira} R.,  {Cypriano} E.~S.,  {Vitorelli} A.~Z.,  {Ribeiro}
  A.~L.~B.,  {Sodr{\'e}} L.,  {Dupke} R.,   {Mendes de Oliveira} C.,  2018,
  \mn@doi [\mnras] {10.1093/mnras/sty2349}, \href
  {http://adsabs.harvard.edu/abs/2018MNRAS.481.1097M} {481, 1097}

\bibitem[\protect\citeauthoryear{{O'Hara}, {Mohr}  \& {Guerrero}}{{O'Hara}
  et~al.}{2004}]{omg04}
{O'Hara} T.~B.,  {Mohr} J.~J.,   {Guerrero} M.~A.,  2004, \mn@doi [\apj]
  {10.1086/382063}, \href
  {https://ui.adsabs.harvard.edu/abs/2004ApJ...604..604O} {604, 604}

\bibitem[\protect\citeauthoryear{{Oak} \& {Paul}}{{Oak} \& {Paul}}{2024}]{op24}
{Oak} T.,  {Paul} S.,  2024, \mn@doi [\mnras] {10.1093/mnras/stae200}, \href
  {https://ui.adsabs.harvard.edu/abs/2024MNRAS.tmp..185O} {}

\bibitem[\protect\citeauthoryear{{Oguri}}{{Oguri}}{2014}]{ogu14}
{Oguri} M.,  2014, \mn@doi [\mnras] {10.1093/mnras/stu1446}, \href
  {http://adsabs.harvard.edu/abs/2014MNRAS.444..147O} {444, 147}

\bibitem[\protect\citeauthoryear{{Okabe}, {Bourdin}, {Mazzotta}  \&
  {Maurogordato}}{{Okabe} et~al.}{2011}]{obm+11}
{Okabe} N.,  {Bourdin} H.,  {Mazzotta} P.,   {Maurogordato} S.,  2011, \mn@doi
  [\apj] {10.1088/0004-637X/741/2/116}, \href
  {http://adsabs.harvard.edu/abs/2011ApJ...741..116O} {741, 116}

\bibitem[\protect\citeauthoryear{{Owers} et~al.,}{{Owers}
  et~al.}{2013}]{obb+13}
{Owers} M.~S.,  et~al., 2013, \mn@doi [\apj] {10.1088/0004-637X/772/2/104},
  \href {https://ui.adsabs.harvard.edu/abs/2013ApJ...772..104O} {772, 104}

\bibitem[\protect\citeauthoryear{{Paterno-Mahler}, {Randall}, {Bulbul},
  {Andrade-Santos}, {Blanton}, {Jones}, {Murray}  \&
  {Johnson}}{{Paterno-Mahler} et~al.}{2014}]{prb+14}
{Paterno-Mahler} R.,  {Randall} S.~W.,  {Bulbul} E.,  {Andrade-Santos} F.,
  {Blanton} E.~L.,  {Jones} C.,  {Murray} S.,   {Johnson} R.~E.,  2014, \mn@doi
  [\apj] {10.1088/0004-637X/791/2/104}, \href
  {https://ui.adsabs.harvard.edu/abs/2014ApJ...791..104P} {791, 104}

\bibitem[\protect\citeauthoryear{{Pearson}, {Batiste}  \& {Batuski}}{{Pearson}
  et~al.}{2014}]{pbb+14}
{Pearson} D.~W.,  {Batiste} M.,   {Batuski} D.~J.,  2014, \mn@doi [\mnras]
  {10.1093/mnras/stu693}, \href
  {https://ui.adsabs.harvard.edu/abs/2014MNRAS.441.1601P} {441, 1601}

\bibitem[\protect\citeauthoryear{{Piffaretti}, {Arnaud}, {Pratt},
  {Pointecouteau}  \& {Melin}}{{Piffaretti} et~al.}{2011}]{pap+11}
{Piffaretti} R.,  {Arnaud} M.,  {Pratt} G.~W.,  {Pointecouteau} E.,   {Melin}
  J.-B.,  2011, \mn@doi [\aap] {10.1051/0004-6361/201015377}, \href
  {http://adsabs.harvard.edu/abs/2011A%26A...534A.109P} {534, A109}

\bibitem[\protect\citeauthoryear{{Piraino-Cerda} et~al.,}{{Piraino-Cerda}
  et~al.}{2024}]{pjl+24}
{Piraino-Cerda} F.,  et~al., 2024, \mn@doi [\mnras] {10.1093/mnras/stad3957},
  \href {https://ui.adsabs.harvard.edu/abs/2024MNRAS.528..919P} {528, 919}

\bibitem[\protect\citeauthoryear{{Planck Collaboration} et~al.,}{{Planck
  Collaboration} et~al.}{2013}]{planck13b}
{Planck Collaboration} et~al., 2013, \mn@doi [\aap]
  {10.1051/0004-6361/201220194}, \href
  {https://ui.adsabs.harvard.edu/abs/2013A%26A...550A.134P} {550, A134}

\bibitem[\protect\citeauthoryear{{Porter}, {Raychaudhury}, {Pimbblet}  \&
  {Drinkwater}}{{Porter} et~al.}{2008}]{prp+08}
{Porter} S.~C.,  {Raychaudhury} S.,  {Pimbblet} K.~A.,   {Drinkwater} M.~J.,
  2008, \mn@doi [\mnras] {10.1111/j.1365-2966.2008.13388.x}, \href
  {http://adsabs.harvard.edu/abs/2008MNRAS.388.1152P} {388, 1152}

\bibitem[\protect\citeauthoryear{{Rahaman}, {Raja}, {Datta}, {Burns}, {Alden}
  \& {Rapetti}}{{Rahaman} et~al.}{2021}]{rrd+21}
{Rahaman} M.,  {Raja} R.,  {Datta} A.,  {Burns} J.~O.,  {Alden} B.,   {Rapetti}
  D.,  2021, \mn@doi [\mnras] {10.1093/mnras/stab1225}, \href
  {https://ui.adsabs.harvard.edu/abs/2021MNRAS.505..480R} {505, 480}

\bibitem[\protect\citeauthoryear{{Regos} \& {Geller}}{{Regos} \&
  {Geller}}{1989}]{rg89}
{Regos} E.,  {Geller} M.~J.,  1989, \mn@doi [\aj] {10.1086/115177}, \href
  {https://ui.adsabs.harvard.edu/abs/1989AJ.....98..755R} {98, 755}

\bibitem[\protect\citeauthoryear{{Rines} \& {Diaferio}}{{Rines} \&
  {Diaferio}}{2006}]{rd06}
{Rines} K.,  {Diaferio} A.,  2006, \mn@doi [\aj] {10.1086/506017}, \href
  {https://ui.adsabs.harvard.edu/abs/2006AJ....132.1275R} {132, 1275}

\bibitem[\protect\citeauthoryear{{Russell}, {Sanders}, {Fabian}, {Baum},
  {Donahue}, {Edge}, {McNamara}  \& {O'Dea}}{{Russell} et~al.}{2010}]{rsf+10}
{Russell} H.~R.,  {Sanders} J.~S.,  {Fabian} A.~C.,  {Baum} S.~A.,  {Donahue}
  M.,  {Edge} A.~C.,  {McNamara} B.~R.,   {O'Dea} C.~P.,  2010, \mn@doi
  [\mnras] {10.1111/j.1365-2966.2010.16822.x}, \href
  {https://ui.adsabs.harvard.edu/abs/2010MNRAS.406.1721R} {406, 1721}

\bibitem[\protect\citeauthoryear{{Russell} et~al.,}{{Russell}
  et~al.}{2012}]{rms+12}
{Russell} H.~R.,  et~al., 2012, \mn@doi [\mnras]
  {10.1111/j.1365-2966.2012.20808.x}, \href
  {https://ui.adsabs.harvard.edu/abs/2012MNRAS.423..236R} {423, 236}

\bibitem[\protect\citeauthoryear{{Rykoff} et~al.,}{{Rykoff}
  et~al.}{2014}]{rrb+14}
{Rykoff} E.~S.,  et~al., 2014, \mn@doi [\apj] {10.1088/0004-637X/785/2/104},
  \href {http://adsabs.harvard.edu/abs/2014ApJ...785..104R} {785, 104}

\bibitem[\protect\citeauthoryear{{Rykoff} et~al.,}{{Rykoff}
  et~al.}{2016}]{rrh+16}
{Rykoff} E.~S.,  et~al., 2016, \mn@doi [\apjs] {10.3847/0067-0049/224/1/1},
  \href {http://adsabs.harvard.edu/abs/2016ApJS..224....1R} {224, 1}

\bibitem[\protect\citeauthoryear{{Sakelliou} \& {Ponman}}{{Sakelliou} \&
  {Ponman}}{2004}]{sp04}
{Sakelliou} I.,  {Ponman} T.~J.,  2004, \mn@doi [\mnras]
  {10.1111/j.1365-2966.2004.07889.x}, \href
  {https://ui.adsabs.harvard.edu/abs/2004MNRAS.351.1439S} {351, 1439}

\bibitem[\protect\citeauthoryear{{Sakelliou} \& {Ponman}}{{Sakelliou} \&
  {Ponman}}{2006}]{sp06}
{Sakelliou} I.,  {Ponman} T.~J.,  2006, \mn@doi [\mnras]
  {10.1111/j.1365-2966.2006.10080.x}, \href
  {https://ui.adsabs.harvard.edu/abs/2006MNRAS.367.1409S} {367, 1409}

\bibitem[\protect\citeauthoryear{{Santos}, {Rosati}, {Tozzi}, {B{\"o}hringer},
  {Ettori}  \& {Bignamini}}{{Santos} et~al.}{2008}]{srt+08}
{Santos} J.~S.,  {Rosati} P.,  {Tozzi} P.,  {B{\"o}hringer} H.,  {Ettori} S.,
  {Bignamini} A.,  2008, \mn@doi [\aap] {10.1051/0004-6361:20078815}, \href
  {http://adsabs.harvard.edu/abs/2008A%26A...483...35S} {483, 35}

\bibitem[\protect\citeauthoryear{{Schuecker}, {B{\"o}hringer}, {Reiprich}  \&
  {Feretti}}{{Schuecker} et~al.}{2001}]{sbr+01}
{Schuecker} P.,  {B{\"o}hringer} H.,  {Reiprich} T.~H.,   {Feretti} L.,  2001,
  \mn@doi [\aap] {10.1051/0004-6361:20011215}, \href
  {http://adsabs.harvard.edu/abs/2001A%26A...378..408S} {378, 408}

\bibitem[\protect\citeauthoryear{{Smargon}, {Mandelbaum}, {Bahcall}  \&
  {Niederste-Ostholt}}{{Smargon} et~al.}{2012}]{smb+12}
{Smargon} A.,  {Mandelbaum} R.,  {Bahcall} N.,   {Niederste-Ostholt} M.,  2012,
  \mn@doi [\mnras] {10.1111/j.1365-2966.2012.20923.x}, \href
  {https://ui.adsabs.harvard.edu/abs/2012MNRAS.423..856S} {423, 856}

\bibitem[\protect\citeauthoryear{{Smith}, {Kneib}, {Smail}, {Mazzotta},
  {Ebeling}  \& {Czoske}}{{Smith} et~al.}{2005}]{sks+05}
{Smith} G.~P.,  {Kneib} J.-P.,  {Smail} I.,  {Mazzotta} P.,  {Ebeling} H.,
  {Czoske} O.,  2005, \mn@doi [\mnras] {10.1111/j.1365-2966.2005.08911.x},
  \href {http://adsabs.harvard.edu/abs/2005MNRAS.359..417S} {359, 417}

\bibitem[\protect\citeauthoryear{{Sommer}, {Basu}, {Intema}, {Pacaud},
  {Bonafede}, {Babul}  \& {Bertoldi}}{{Sommer} et~al.}{2017}]{sbi+17}
{Sommer} M.~W.,  {Basu} K.,  {Intema} H.,  {Pacaud} F.,  {Bonafede} A.,
  {Babul} A.,   {Bertoldi} F.,  2017, \mn@doi [\mnras] {10.1093/mnras/stw3015},
  \href {https://ui.adsabs.harvard.edu/abs/2017MNRAS.466..996S} {466, 996}

\bibitem[\protect\citeauthoryear{{Springel} et~al.,}{{Springel}
  et~al.}{2005}]{swj+05}
{Springel} V.,  et~al., 2005, \mn@doi [\nat] {10.1038/nature03597}, \href
  {http://adsabs.harvard.edu/abs/2005Natur.435..629S} {435, 629}

\bibitem[\protect\citeauthoryear{{Stancioli}, {Wittman}, {Finner}  \&
  {Bouhrik}}{{Stancioli} et~al.}{2023}]{swf+23}
{Stancioli} R.,  {Wittman} D.,  {Finner} K.,   {Bouhrik} F.,  2023, \mn@doi
  [arXiv e-prints] {10.48550/arXiv.2307.10174}, \href
  {https://ui.adsabs.harvard.edu/abs/2023arXiv230710174S} {p. arXiv:2307.10174}

\bibitem[\protect\citeauthoryear{{Str{\"u}der} et~al.,}{{Str{\"u}der}
  et~al.}{2001}]{sbd+01}
{Str{\"u}der} L.,  et~al., 2001, \mn@doi [\aap] {10.1051/0004-6361:20000066},
  \href {https://ui.adsabs.harvard.edu/abs/2001A&A...365L..18S} {365, L18}

\bibitem[\protect\citeauthoryear{{Tanaka}, {Fujimoto}, {Okabe}, {Mitsuishi},
  {Akamatsu}, {Ota}, {Oguri}  \& {Nishizawa}}{{Tanaka} et~al.}{2021}]{tfo+21}
{Tanaka} K.,  {Fujimoto} R.,  {Okabe} N.,  {Mitsuishi} I.,  {Akamatsu} H.,
  {Ota} N.,  {Oguri} M.,   {Nishizawa} A.~J.,  2021, \mn@doi [\pasj]
  {10.1093/pasj/psab022}, \href
  {https://ui.adsabs.harvard.edu/abs/2021PASJ...73..584T} {73, 584}

\bibitem[\protect\citeauthoryear{{Tejos} et~al.,}{{Tejos}
  et~al.}{2016}]{tpc+16}
{Tejos} N.,  et~al., 2016, \mn@doi [\mnras] {10.1093/mnras/stv2376}, \href
  {https://ui.adsabs.harvard.edu/abs/2016MNRAS.455.2662T} {455, 2662}

\bibitem[\protect\citeauthoryear{{Tempel}, {Tuvikene}, {Kipper}  \&
  {Libeskind}}{{Tempel} et~al.}{2017}]{ttk+17}
{Tempel} E.,  {Tuvikene} T.,  {Kipper} R.,   {Libeskind} N.~I.,  2017, \mn@doi
  [\aap] {10.1051/0004-6361/201730499}, \href
  {http://adsabs.harvard.edu/abs/2017A%26A...602A.100T} {602, A100}

\bibitem[\protect\citeauthoryear{{Ulmer} \& {Cruddace}}{{Ulmer} \&
  {Cruddace}}{1982}]{uc82}
{Ulmer} M.~P.,  {Cruddace} R.~G.,  1982, \mn@doi [\apj] {10.1086/160096}, \href
  {https://ui.adsabs.harvard.edu/abs/1982ApJ...258..434U} {258, 434}

\bibitem[\protect\citeauthoryear{{Vikhlinin}, {Markevitch}, {Murray}, {Jones},
  {Forman}  \& {Van Speybroeck}}{{Vikhlinin} et~al.}{2005}]{vmm+05}
{Vikhlinin} A.,  {Markevitch} M.,  {Murray} S.~S.,  {Jones} C.,  {Forman} W.,
  {Van Speybroeck} L.,  2005, \mn@doi [\apj] {10.1086/431142}, \href
  {https://ui.adsabs.harvard.edu/abs/2005ApJ...628..655V} {628, 655}

\bibitem[\protect\citeauthoryear{{Weisskopf}, {Tananbaum}, {Van Speybroeck}  \&
  {O'Dell}}{{Weisskopf} et~al.}{2000}]{wtv+00}
{Weisskopf} M.~C.,  {Tananbaum} H.~D.,  {Van Speybroeck} L.~P.,   {O'Dell}
  S.~L.,  2000, in {Truemper} J.~E.,  {Aschenbach} B.,  eds,  Society of
  Photo-Optical Instrumentation Engineers (SPIE) Conference Series Vol. 4012,
  X-Ray Optics, Instruments, and Missions III. pp 2--16 (\mn@eprint {arXiv}
  {astro-ph/0004127}), \mn@doi{10.1117/12.391545}

\bibitem[\protect\citeauthoryear{{Wen} \& {Han}}{{Wen} \& {Han}}{2013}]{wh13}
{Wen} Z.~L.,  {Han} J.~L.,  2013, \mn@doi [\mnras] {10.1093/mnras/stt1581},
  \href {http://adsabs.harvard.edu/abs/2013MNRAS.436..275W} {436, 275}

\bibitem[\protect\citeauthoryear{{Wen} \& {Han}}{{Wen} \& {Han}}{2015a}]{wh15a}
{Wen} Z.~L.,  {Han} J.~L.,  2015a, \mn@doi [\mnras] {10.1093/mnras/stu2722},
  \href {https://ui.adsabs.harvard.edu/abs/2015MNRAS.448....2W} {448, 2}

\bibitem[\protect\citeauthoryear{{Wen} \& {Han}}{{Wen} \& {Han}}{2015b}]{wh15b}
{Wen} Z.~L.,  {Han} J.~L.,  2015b, \mn@doi [\apj]
  {10.1088/0004-637X/807/2/178}, \href
  {http://adsabs.harvard.edu/abs/2015ApJ...807..178W} {807, 178}

\bibitem[\protect\citeauthoryear{{Wen} \& {Han}}{{Wen} \& {Han}}{2021}]{wh21}
{Wen} Z.~L.,  {Han} J.~L.,  2021, \mn@doi [\mnras] {10.1093/mnras/staa3308},
  \href {https://ui.adsabs.harvard.edu/abs/2021MNRAS.500.1003W} {500, 1003}

\bibitem[\protect\citeauthoryear{{Wen} \& {Han}}{{Wen} \& {Han}}{2022}]{wh22}
{Wen} Z.~L.,  {Han} J.~L.,  2022, \mn@doi [\mnras] {10.1093/mnras/stac1149},
  \href {https://ui.adsabs.harvard.edu/abs/2022MNRAS.tmp.1125W} {}

\bibitem[\protect\citeauthoryear{{Wen} \& {Han}}{{Wen} \& {Han}}{2024}]{wh24}
{Wen} Z.~L.,  {Han} J.~L.,  2024, \mn@doi [arXiv e-prints]
  {10.48550/arXiv.2404.02002}, \href
  {https://ui.adsabs.harvard.edu/abs/2024arXiv240402002W} {p. arXiv:2404.02002}

\bibitem[\protect\citeauthoryear{{Wen}, {Han}  \& {Liu}}{{Wen}
  et~al.}{2009}]{whl09}
{Wen} Z.~L.,  {Han} J.~L.,   {Liu} F.~S.,  2009, \mn@doi [\apjs]
  {10.1088/0067-0049/183/2/197}, \href
  {http://adsabs.harvard.edu/abs/2009ApJS..183..197W} {183, 197}

\bibitem[\protect\citeauthoryear{{Wen}, {Han}  \& {Liu}}{{Wen}
  et~al.}{2012}]{whl12}
{Wen} Z.~L.,  {Han} J.~L.,   {Liu} F.~S.,  2012, \mn@doi [\apjs]
  {10.1088/0067-0049/199/2/34}, \href
  {http://adsabs.harvard.edu/abs/2012ApJS..199...34W} {199, 34}

\bibitem[\protect\citeauthoryear{{Werner}, {Finoguenov}, {Kaastra},
  {Simionescu}, {Dietrich}, {Vink}  \& {B{\"o}hringer}}{{Werner}
  et~al.}{2008}]{wfk+08}
{Werner} N.,  {Finoguenov} A.,  {Kaastra} J.~S.,  {Simionescu} A.,  {Dietrich}
  J.~P.,  {Vink} J.,   {B{\"o}hringer} H.,  2008, \mn@doi [\aap]
  {10.1051/0004-6361:200809599}, \href
  {http://adsabs.harvard.edu/abs/2008A%26A...482L..29W} {482, L29}

\bibitem[\protect\citeauthoryear{{West}, {Oemler}  \& {Dekel}}{{West}
  et~al.}{1988}]{wod88}
{West} M.~J.,  {Oemler} Jr. A.,   {Dekel} A.,  1988, \mn@doi [\apj]
  {10.1086/166163}, \href {http://adsabs.harvard.edu/abs/1988ApJ...327....1W}
  {327, 1}

\bibitem[\protect\citeauthoryear{{Wittman}, {Stancioli}, {Finner}, {Bouhrik},
  {van Weeren}  \& {Botteon}}{{Wittman} et~al.}{2023}]{wsf+23}
{Wittman} D.,  {Stancioli} R.,  {Finner} K.,  {Bouhrik} F.,  {van Weeren} R.,
  {Botteon} A.,  2023, \mn@doi [arXiv e-prints] {10.48550/arXiv.2306.01715},
  \href {https://ui.adsabs.harvard.edu/abs/2023arXiv230601715W} {p.
  arXiv:2306.01715}

\bibitem[\protect\citeauthoryear{{Yang} et~al.,}{{Yang} et~al.}{2021}]{yxh+21}
{Yang} X.,  et~al., 2021, \mn@doi [\apj] {10.3847/1538-4357/abddb2}, \href
  {https://ui.adsabs.harvard.edu/abs/2021ApJ...909..143Y} {909, 143}

\bibitem[\protect\citeauthoryear{{Yu}, {Diaferio}, {Serra}  \& {Baldi}}{{Yu}
  et~al.}{2018}]{yds+18}
{Yu} H.,  {Diaferio} A.,  {Serra} A.~L.,   {Baldi} M.,  2018, \mn@doi [\apj]
  {10.3847/1538-4357/aac263}, \href
  {https://ui.adsabs.harvard.edu/abs/2018ApJ...860..118Y} {860, 118}

\bibitem[\protect\citeauthoryear{{Yuan} \& {Han}}{{Yuan} \& {Han}}{2020}]{yh20}
{Yuan} Z.~S.,  {Han} J.~L.,  2020, \mn@doi [\mnras] {10.1093/mnras/staa2363},
  \href {https://ui.adsabs.harvard.edu/abs/2020MNRAS.497.5485Y} {497, 5485}

\bibitem[\protect\citeauthoryear{{Yuan}, {Han}  \& {Wen}}{{Yuan}
  et~al.}{2022}]{yhw22}
{Yuan} Z.~S.,  {Han} J.~L.,   {Wen} Z.~L.,  2022, \mn@doi [\mnras]
  {10.1093/mnras/stac1037}, \href
  {https://ui.adsabs.harvard.edu/abs/2022MNRAS.513.3013Y} {513, 3013}

\bibitem[\protect\citeauthoryear{{Yuan}, {Han}, {B{\"o}hringer}, {Wen}  \&
  {Chon}}{{Yuan} et~al.}{2023}]{yhb+23}
{Yuan} Z.~S.,  {Han} J.~L.,  {B{\"o}hringer} H.,  {Wen} Z.~L.,   {Chon} G.,
  2023, \mn@doi [\mnras] {10.1093/mnras/stad1426}, \href
  {https://ui.adsabs.harvard.edu/abs/2023MNRAS.523.1364Y} {523, 1364}

\bibitem[\protect\citeauthoryear{{Zhang}, {Dietrich}, {McKay}, {Sheldon}  \&
  {Nguyen}}{{Zhang} et~al.}{2013}]{zdm+13}
{Zhang} Y.,  {Dietrich} J.~P.,  {McKay} T.~A.,  {Sheldon} E.~S.,   {Nguyen} A.
  T.~Q.,  2013, \mn@doi [\apj] {10.1088/0004-637X/773/2/115}, \href
  {https://ui.adsabs.harvard.edu/abs/2013ApJ...773..115Z} {773, 115}

\bibitem[\protect\citeauthoryear{{Zhang} et~al.,}{{Zhang}
  et~al.}{2021}]{zss+21}
{Zhang} X.,  et~al., 2021, \mn@doi [\aap] {10.1051/0004-6361/202141540}, \href
  {https://ui.adsabs.harvard.edu/abs/2021A&A...656A..59Z} {656, A59}

\bibitem[\protect\citeauthoryear{{Zou} et~al.,}{{Zou} et~al.}{2022}]{zsx+22}
{Zou} H.,  et~al., 2022, \mn@doi [Research in Astronomy and Astrophysics]
  {10.1088/1674-4527/ac6416}, \href
  {https://ui.adsabs.harvard.edu/abs/2022RAA....22f5001Z} {22, 065001}

\bibitem[\protect\citeauthoryear{{van Weeren}, {R{\"o}ttgering}, {Br{\"u}ggen}
  \& {Hoeft}}{{van Weeren} et~al.}{2010}]{vrb+10}
{van Weeren} R.~J.,  {R{\"o}ttgering} H.~J.~A.,  {Br{\"u}ggen} M.,   {Hoeft}
  M.,  2010, \mn@doi [Science] {10.1126/science.1194293}, \href
  {http://adsabs.harvard.edu/abs/2010Sci...330..347V} {330, 347}

\bibitem[\protect\citeauthoryear{{van Weeren}, {R{\"o}ttgering}, {Intema},
  {Rudnick}, {Br{\"u}ggen}, {Hoeft}  \& {Oonk}}{{van Weeren}
  et~al.}{2012}]{vri+12}
{van Weeren} R.~J.,  {R{\"o}ttgering} H.~J.~A.,  {Intema} H.~T.,  {Rudnick} L.,
   {Br{\"u}ggen} M.,  {Hoeft} M.,   {Oonk} J.~B.~R.,  2012, \mn@doi [\aap]
  {10.1051/0004-6361/201219000}, \href
  {https://ui.adsabs.harvard.edu/abs/2012A&A...546A.124V} {546, A124}

\bibitem[\protect\citeauthoryear{{van Weeren}, {de Gasperin}, {Akamatsu},
  {Br{\"u}ggen}, {Feretti}, {Kang}, {Stroe}  \& {Zandanel}}{{van Weeren}
  et~al.}{2019}]{vda+19}
{van Weeren} R.~J.,  {de Gasperin} F.,  {Akamatsu} H.,  {Br{\"u}ggen} M.,
  {Feretti} L.,  {Kang} H.,  {Stroe} A.,   {Zandanel} F.,  2019, \mn@doi [\ssr]
  {10.1007/s11214-019-0584-z}, \href
  {https://ui.adsabs.harvard.edu/abs/2019SSRv..215...16V} {215, 16}

\makeatother
\end{thebibliography}
